\documentclass[article,twocolumn]{aastex631}

\usepackage{graphicx}	
\usepackage{amsmath}	
\usepackage{amssymb}	
\usepackage{subfigure}
\usepackage{float}
\usepackage{ae,aecompl,amsmath,amsbsy,mathtools, nccmath}
\usepackage{epstopdf}
\usepackage{multirow}
\usepackage{hyperref}
\usepackage{lipsum}
\begin{document}
\renewcommand{\deg}{$^\circ$}
\title[4]{
AB Dor: Coronal imaging and activity cycles  
}
\email{gurpreet@aries.res.in, jeewan@aries.res.in}
\author[0009-0002-6580-3931]{Gurpreet Singh}
\affiliation{Aryabhatta Research Institute of Observational Sciences (ARIES), Manora Peak, Nainital 263001, India}
\affiliation{Department of Physics, Deen Dayal Upadhyaya Gorakhpur University, Gorakhpur 273009, India}
\author[0000-0002-4331-1867]{J. C. Pandey  }
\affiliation{Aryabhatta Research Institute of Observational Sciences (ARIES), Manora Peak, Nainital 263001, India}
\begin{abstract}

Using long-term X-ray observations, we present the short-term and long-term X-ray variability analysis of the ultra-fast rotating active star AB Dor. 
Flaring events are common in X-ray observations of AB Dor and occupy a substantial portion of the total observation time, averaging around 57$\pm$23\%.  The flare-free X-ray light curves show rotational modulation, indicating the presence of highly active regions in its corona.  We have developed a light curve inversion code to image the corona of active fast rotating stars. The results of coronal imaging reveal the presence of two active regions of different brightness that are separated by $\sim$180\deg in longitude. These active regions are also found to migrate along the longitude and also show variation in their brightness. Our analysis of long-term X-ray data spanning from 1979 to 2022 shows multiple periodicities. The existence of a $\sim$19.2 yr cycle and its first harmonic indicates the presence of a Solar-like, long-term pattern. In comparison, the periodicities of $\sim$3.6 and  $\sim$5.4 yr are possibly due to the presence of a flip-flop cycle in X-rays, which is also supported by similar periods findings from the optical data in the earlier studies. Further confirmation of the existence of the X-ray flip-flop cycle requires long-term observations at regular intervals in the quiescent state.%
\end{abstract}

\keywords{ \href{http://astrothesaurus.org/uat/1580}{Stellar activity (1580)} --- \href{http://astrothesaurus.org/uat/305 }{Stellar Coronae (305)} --- 	\href{http://astrothesaurus.org/uat/1823} {X-ray star (1823)} --- \href{http://astrothesaurus.org/uat/1810} {X-ray astronomy (1810)} --- \href{http://astrothesaurus.org/uat/1145}{Stellar imaging (1145)} }

\section{Introduction} \label{sec:intro}

The variable nature of light curves of late-type active stars can be observed over the entire electromagnetic spectrum with a variability time scale ranging from a few minutes to a few decades. Both short-term variability (STV) and long-term variability (LTV) are found to be present in solar-type stars and are due to the different manifestations of magnetic activities.  The STVs last a few minutes to a few days and are generally attributed to flaring activity and rotational modulation due to inhomogeneities.  The LTVs last from a few yr to a few decades and are linked to the stellar activity cycles.

   The STVs due to flares in X-rays have been studied and modelled in the past for a long and helped to understand the extreme physical condition of solar-type stars  \citep[e.g.][]{1991ARA&A..29..275H,2007A&A...471..271R,2008MNRAS.387.1627P,2012MNRAS.419.1219P}.  
Stellar coronae are spatially unresolved; therefore, different techniques have been developed to extract information from the periodic STVs due to the rotational modulations of active regions in the stellar atmosphere. Doppler imaging \citep[][]{2001ApJ...562L..75B},  extrapolating the surface magnetic maps \citep[][etc]{2007MNRAS.377.1488H,2010MNRAS.404..101J,2010ApJ...721...80C}, and light curve inversion techniques \citep[][]{1992MNRAS.259..453S,1996ApJ...473..470S,2014ApJ...783....2D,2022ApJ...934...20S} are main techniques to explain such type of STVs in X-rays. These techniques have their own limitations. Doppler imaging of X-ray data requires high spectral resolution, which is inadequate for most of the stars due to instrumental and observing limitations. Inferring the coronal structures on the basis of magnetic surface maps requires simultaneous observations in optical and X-ray bands. The light-curve inversion techniques (LCITs) are a mathematically ill-posed problem where  3D information is extracted from 1D time series data. However, these techniques have gained interest with time due to the easy availability of time series data.  

The study of LTVs is useful to understand the underlying dynamo mechanisms. The  Sun is the only star for which the LTVs are studied in detail, and theoretical models are constructed to explain it.
The LTVs for the magnetically active stars have been studied extensively over the past several decades in optical bands \citep[e.g.][]{2002A&A...393..225M,2007ApJS..171..260L}.  In a sample of  Mt. Wilson HK program \citep[][]{1968ApJ...153..221W,1978ApJ...226..379W}, about 60\% of the stars have shown periodic and cyclic variations in their chromospheric activity \citep[][]{1995ApJ...438..269B,1998ASPC..154..153B}. \cite{1998ApJS..118..239R}  have shown that the photometric cycle is in phase with the chromospheric cycle for older active stars, whereas the opposite relation was found for younger active stars.  
\cite{1996A&A...305..284H} found that stars with cyclic variations in Ca {\sc ii} H\&K flux tend to show less X-ray activity than the stars, which show irregular variability in Ca {\sc ii} emission. 
These activity cycles serve as proxies for the stellar dynamo.  However, the X-ray activity cycle corresponding to these cycles in the stellar corona remains challenging due to their longer cyclic period and limited X-ray data.  Only six stars have been reported for which X-ray activity cycles so far. These are  
61 Cyg A \citep[][]{2006A&A...460..261H,2012A&A...543A..84R}, HD 81809 \citep[][]{2008A&A...490.1121F,2017A&A...605A..19O}, $\alpha$ Cen A and $\alpha$ Cen B \citep[][]{2012A&A...543A..84R,2017MNRAS.464.3281W}, $\iota$ Horologii \citep[][]{2019A&A...631A..45S}, and $\epsilon$ Eridani \cite[][]{2020A&A...636A..49C}.
 Apart from a longer activity cycle, the Sun and other sun-like stars have been found to have an activity cycle whose period is nearly one-third of the longer activity cycle \citep[][]{2005LRSP....2....8B}, which is linked with the phenomenon of periodic switching between two active longitudes, i.e. flip-flop cycle. Moreover, for the Sun, it has been found that the flip-flop cycle is different in the northern and southern hemispheres, where it was shown that the northern hemisphere flip-flop period is 5\% slower than that of the southern hemisphere\citep[][]{2003A&A...405.1121B}.

For the present work, we have taken the ultra-fast rotating active star  AB Dor A due to the availability of sufficient observations in the X-ray band by the XMM-Newton satellite.  AB Dor A is part of a quintuplet stellar system. It is a K0-type star that has recently reached the main sequence and is located at a distance of 14.85 pc\citep[][]{2020yCat.1350....0G}. 
AB Dor A  being a fast rotator (P$_{rot}\sim$0.51 d), shows violent magnetic activity with average X-ray luminosity in the range of $\sim10^{30}$ erg s$^{-1}$\citep[][]{2013A&A...559A.119L}. In  X-rays, the contribution from other components in the AB Dor system is negligible; thus, AB Dor A (hereafter AB Dor) can be regarded as a single X-ray-emitting star.  

AB Dor has gained the attention of most X-ray missions because of its higher X-ray flux and location advantage of being distant from the galactic plane. Since the first detection in X-rays by \cite{1981A&A...104...33P}, AB Dor is found to show frequent flaring episodes. The STVs due to flares have been extensively studied in the past \citep[e.g.][etc.]{2002franciosini,2013A&A...559A.119L,2024MNRAS.527.1705D}. Moreover, STVs due to rotational modulation is also reported by several authors in the past \citep[][etc]{1988MNRAS.231..131C,1993A&A...278..449J,2007MNRAS.377.1488H,2013A&A...553A..40H}. \cite{1997A&A...320..831K} reported the first LTV X-ray study, where they reported partial rotational modulation with no long-term X-ray activity trends during the 5.5 yr of X-ray observations, but a slight increase in X-ray flux is observed. 
Based on a longer X-ray data set, \cite{2013A&A...559A.119L} have found a probable activity cycle of about 17 years, with X-ray amplitude fluctuation substantially smaller than the Sun.  
The long-term photometric study revealed different types of activity cycles: one with a period of 5 to 7 yr in which activity switches between the two active longitudes, i.e. flip-flop cycle, and another with a period of 19 to 22 years which is similar to the 11-year solar cycle \citep[][]{2005A&A...432..657J,2001ASPC..223..895A}.

Our paper is organized as follows: In section \ref{sec:method}, we explain the coronal imaging method. The section \ref{sec:app_to_abdor} deals with the observation, light curve analysis, and application of the coronal imaging model to the star AB Dor.  The long-term X-ray activity of AB Dor is analyzed in section \ref{LTVabdor}. The results obtained are discussed in section \ref{sec:discussion} whereas we conclude our findings in section \ref{sec:conclusion}.

\section{Coronal Imaging}\label{sec:method}
\subsection{The method} \label{sec:method1}
\cite{1974AJ.....79..745L} and \cite{1975SoPh...45..301W} developed an iterative technique that maximizes the likelihood. We call this method as LW method throughout the paper.  This imaging technique has been used extensively in medical science and astronomy \citep[for example ][etc]{1992MNRAS.259..453S,1996ApJ...473..470S,4307558,2022ApJ...934...20S}. We have applied the LW method to image the stellar coronae of single active stars. Thus, our LCIT is a method to invert light curves to get information on the geometry of emitting plasma.  It is based on the following assumptions.\\ 
(a) corona is optically thin,\\
(b) it rotates rigidly, \\
(c) active region lasts for at least one complete cycle,\\
(d) rotational modulation is due to those regions in the corona which are being eclipsed by the photosphere of the star, and\\
 (e) distance to the star is much larger than its radius, so the shadows can be cylindrical shaped.\\

Here, we explain the algorithm for the method in the following different steps\\
Step (i). Firstly, we generate a uniform corona of the star from the photosphere to the coronal height ($h_{cor}$)  with a resolution of a cubical bin of  0.05$\times$0.05$\times$0.05 $R_\odot^3$ and assign a uniform emission density, $f_{em}(x,y,z)$ to each cubical bin in corona.\\
Step (ii). An occultation matrix (\textbf{M}), which is dependent on the angle of inclination,  is calculated for each observed phase. It assigns weight to each cubical bin as 0 for occulted and 1 for visible.\\
Step (iii). 
Cubical bins with a constant weight throughout the observations are removed from the solution space as they do not contribute to the rotational modulation of the light curve.\\
Step (iv). Since the plasma is optically thin, so, the total flux observed at any phase ($\phi$) is the sum of contributions from all the cubical bins which are visible at the given phase, i.e.
\begin{equation}
F(\phi)=\sum_{x,y,z}f_{em}(x,y,z)\textbf{M}(\phi,x,y,z)dxdydz    
\end{equation}

Step (v). The discrepancy between the model and observed light curve is calculated using standard $\chi^2$ statistics and  updated the emission density of each bin  until the reduced  $\chi^2$ converges to $\sim$1  using the following equation
\begin{equation}
    f_{em}^{n+1}(x,y,z)=f_{em}^n(x,y,z)\frac{\sum_i\frac{F_o(\phi_i)}{F_c(\phi_i)}\textbf{M}(\phi_i,x,y,z)}{\sum_i\textbf{M}(\phi_i,x,y,z)}
\end{equation}
where $F_c(\phi_i)$ and $F_o(\phi_i)$ are modeled and observed flux at phase $\phi_i$.\\

To optimise the adequate grid size, the model was run on the synthetic light curve (see Section \ref{validationsection}) using different grid resolutions of 0.3, 0.2, 0.1, 0.05, 0.03, and 0.02. 
Subsequently, we examined the standard deviation of the residuals and the number of iterations as a function of different grid resolutions.
Making the grid size less than 0.05$\times$0.05$\times$0.05  $R_\odot^3$, does not improve the standard deviation of residual but requires more number of iterations to converge the solution, thus making computation expensive.  However, increasing the grid size does not give the appropriate solutions as a large number of outliers were obtained in modelled coronal images. Therefore, the grid size of  0.05$\times$0.05$\times$0.05  $R_\odot^3$ was found to be the optimal grid size for the adequate solution.

\subsection{Validation of method}\label{validationsection}
\begin{figure*}
\subfigure[]{\raisebox{0.4cm}{\includegraphics[width=0.6\linewidth]{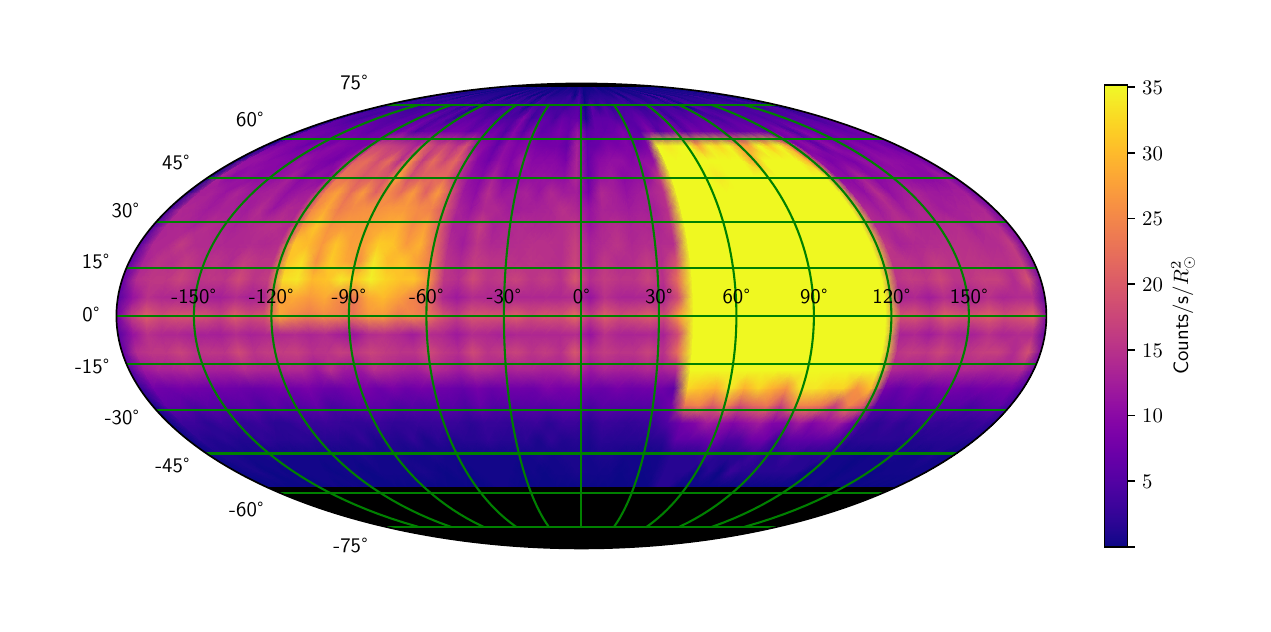}}
    }     
    \subfigure[]{
\raisebox{0.4cm}{\includegraphics[scale=0.25]{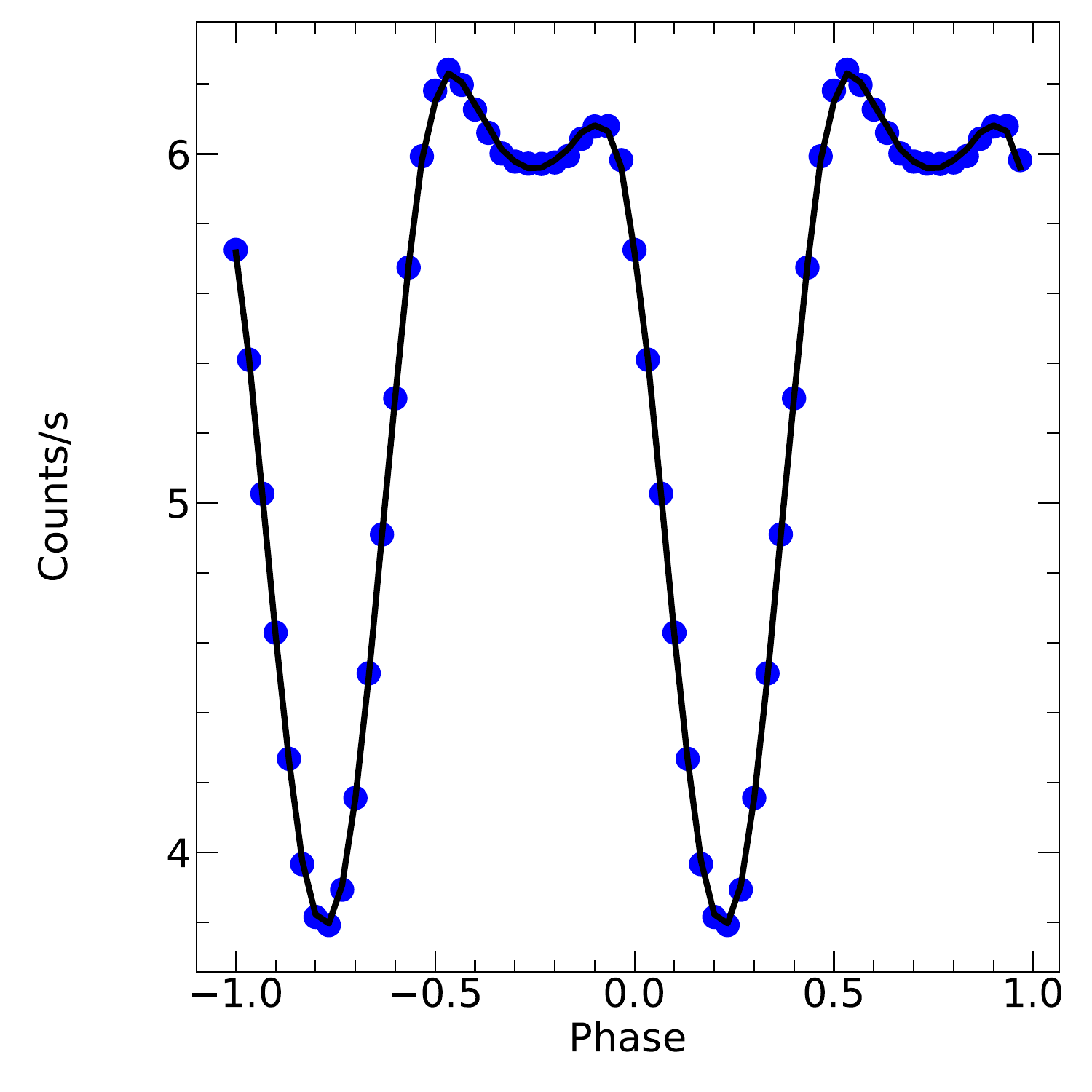}} }\\
    \subfigure[]{
\raisebox{0.4cm}{\includegraphics[width=0.6\linewidth]{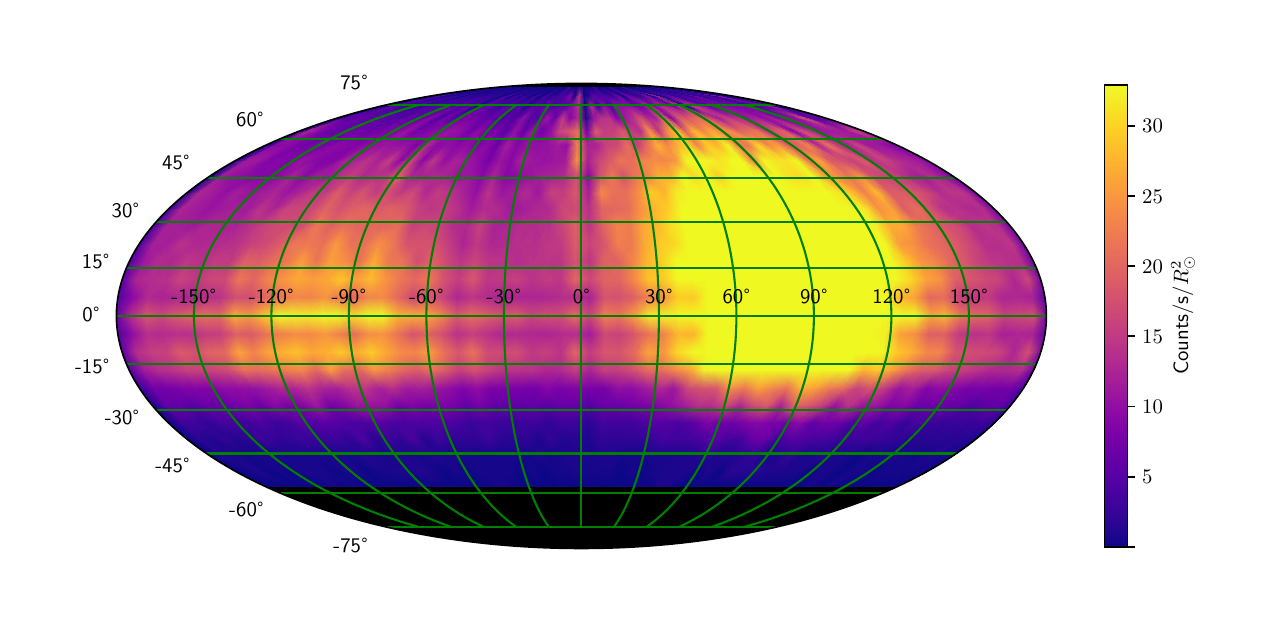}}
}
\subfigure[]{
\raisebox{0.4cm}{\includegraphics[scale=0.23]{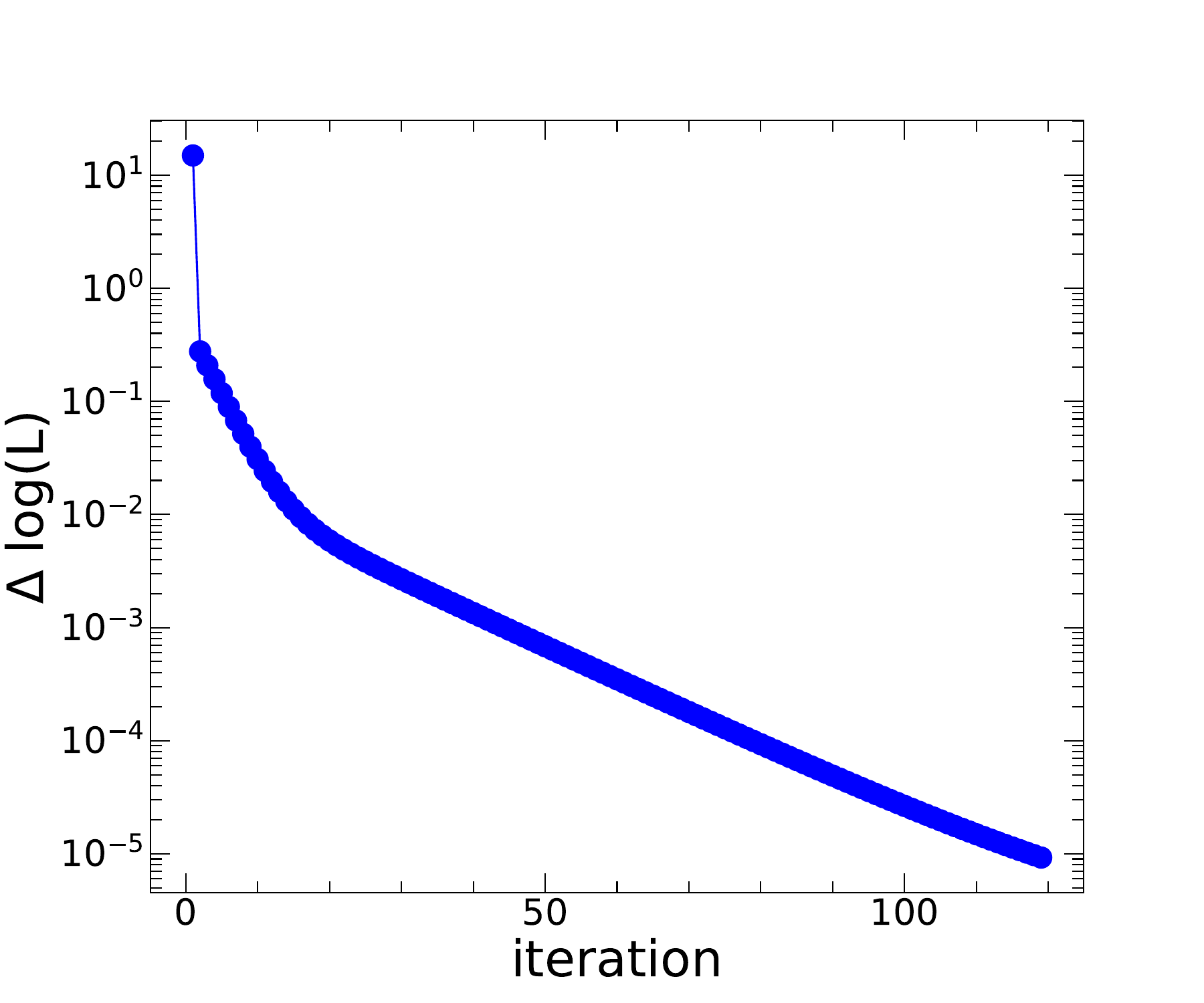}}
}
\caption{(a) Coronal image with injected active regions, (b) Simulated and modelled light curves: Blue dots represent simulated data, while the black solid curve depicts the modelled light curve, (c) Coronal image returned by model, and (d)  $\Delta log(L)$vs iteration plot with $\Delta log(L)$= $log(L_{n})$-$log(L_{n-1})$ and $n$ is the number of iterations.  
}
\label{fig:validation}
\end{figure*}

 To validate our model, we initiated the process by generating simulated light curves, assuming a fully transparent corona with two distinct active regions located at longitudes ranging from 60\deg to 120\deg and 220\deg to 300\deg.
In Figures \ref{fig:validation} (a) and (b), we show an artificial coronal image and the corresponding light curve, respectively. The synthetic light curve was modelled using the above method to construct the coronal image. During each iteration, we computed the conditional probability (P$_i$) of observing counts $F_o(\phi_i)$ at a phase $\phi_i$. 
This computation involved utilizing known emissions in each cubic bin denoted as $f(b)$, their volume element $db=dxdydz$  and the occultation matrix denoted as $M(b, \phi_i)$ as

\begin{equation}
\begin{aligned}
    P_i(F_o(\phi_i)|f,M(\phi_i))=e^{-\lambda}\frac{\lambda^{F_o(\phi_i)}}{F_o(\phi_i)!}; \\
       \lambda=\sum_b f(b)M(b,\phi_i)db  
       \end{aligned}
\end{equation}
 
The corresponding log-likelihood function is defined as

\begin{equation}
\begin{aligned}
    log(L)=\sum_i log(P_i)=\sum_i -\lambda + F_o(\phi_i)log(\lambda)\\
   - log(\Gamma(F_o(\phi_i)+1))
\end{aligned}
\end{equation}

Figure \ref{fig:validation}(c) displays the coronal image generated through modelling the artificial light curve. Additionally, in Figure \ref{fig:validation}(b), we have overlaid the simulated light curve for comparison. Figure \ref{fig:validation} (d) shows $\Delta log(L)$ vs iteration plot, where $\Delta log(L)$= $log(L_{n})$-$log(L_{n-1})$ and $n$ is the number of iterations.  Here, in Figure \ref{fig:validation} (d) $\Delta log(L)$, the difference in likelihood keeps on decreasing, indicating the convergence. Further, the $\Delta log(L) \ge 0$, which shows the method increases likelihood at each successive iteration.


\section{Application to AB Dor}
\label{sec:app_to_abdor}
\subsection{X-ray Observations}
\label{sec:Xrayobs}
AB Dor has been observed by many X-ray missions since its discovery as an active star. It is being monitored regularly by XMM-Newton as it serves as a calibration source.
 The XMM-Newton has five detectors dedicated to X-ray observations: three European photon imaging cameras  (EPIC; one PN \citep[]{2001A&A...365L..18S} and two MOS \citep[]{2001A&A...365L..27T}), two reflection grating spectrometers \citep[RGS;][]{2001A&A...365L...7D}. RGS provides a spectral resolution of 200-800 in the energy range of 0.3 -- 2.5 keV. AB Dor was regularly observed by the RGS detector; therefore, we have used the RGS observations for further analysis. The XMM-Newton has observed AB Dor for 42 epochs from the year 2000 to 2022. The data were reduced using standard {\sc XMM-Newton} Science Analysis System (\textsc{SAS}) v19.0.0 software. The task {\sc rgsproc} was used to reprocess the data. The data was further screened for background flaring events and extraction mask size. The background-subtracted light curves were then extracted using {\sc rgslccorr} task in {\sc SAS}. AB Dor's background-subtracted X-ray light curves from RGS2  are shown in Figure \ref{fig:abdor_flare_qui1} in the appendix \ref{sec:appendix_lcs}.

\begin{figure*}
\vskip -1cm
\setcounter{subfigure}{0}
    \centering
    \subfigure[Obs ID: 0123720201 (2000-05-01)]{\raisebox{0.4cm}{\includegraphics[width=0.27\linewidth]{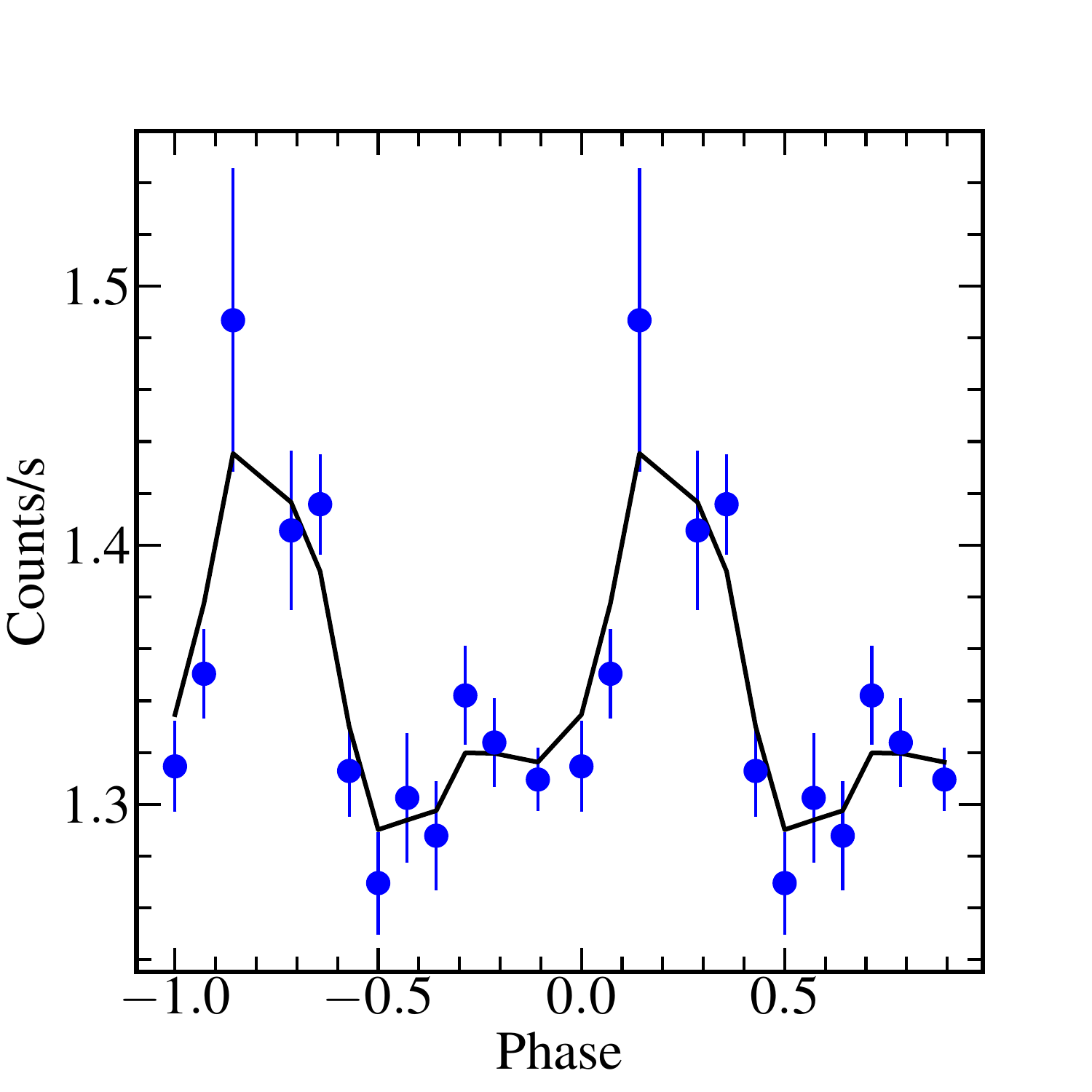}}
    \includegraphics[width=0.60\linewidth]{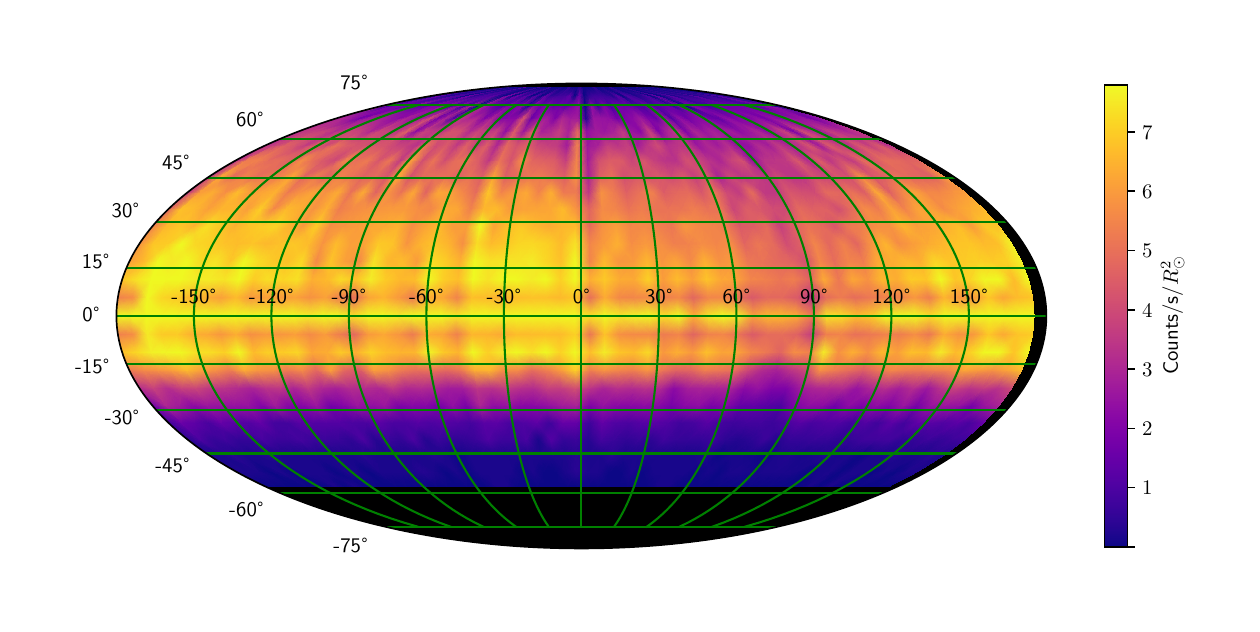}
    }     
    \subfigure[Obs ID: 0134521301 (2001-10-13)]{
\raisebox{0.4cm}{\includegraphics[width=0.25\linewidth]{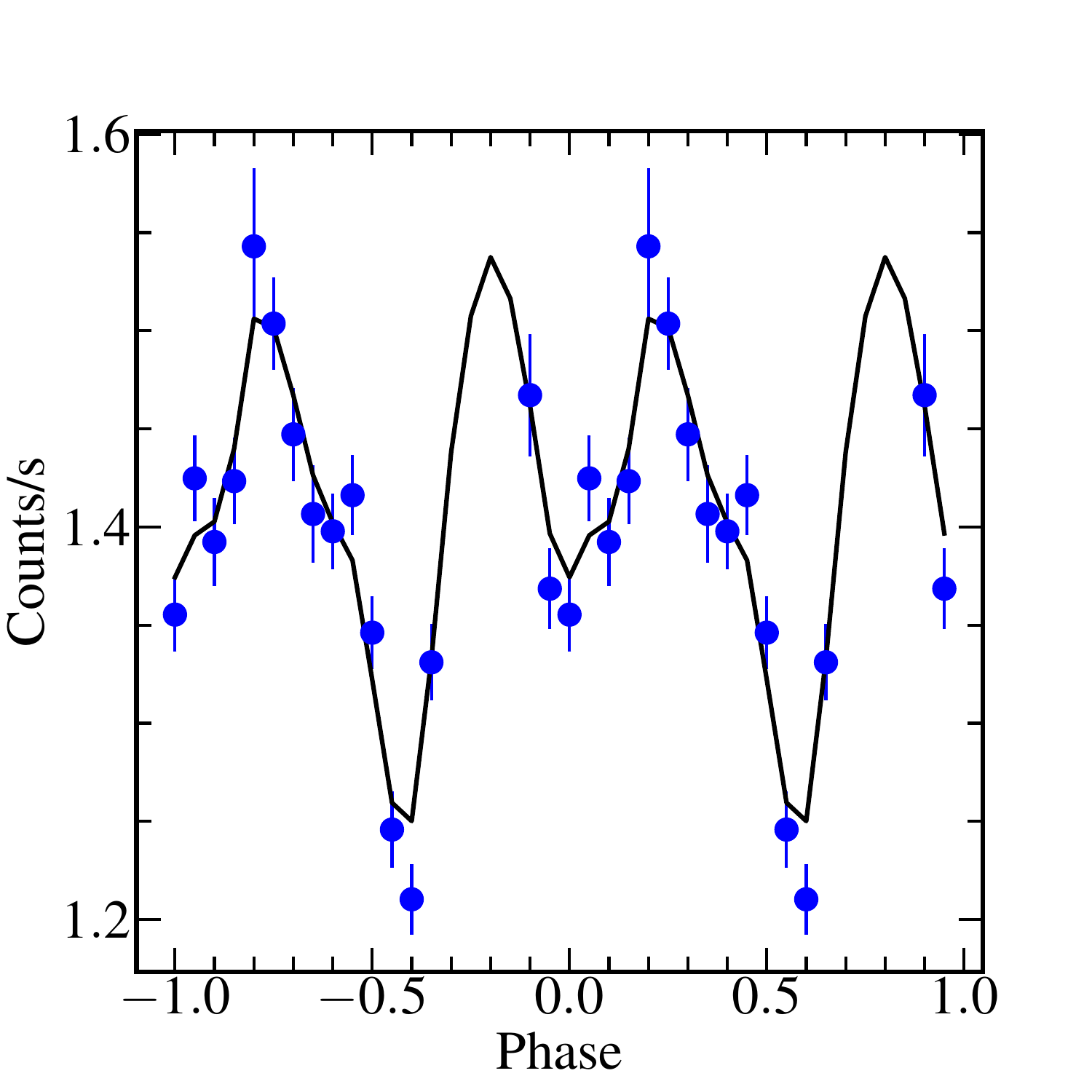}}
\includegraphics[width=0.60\linewidth]{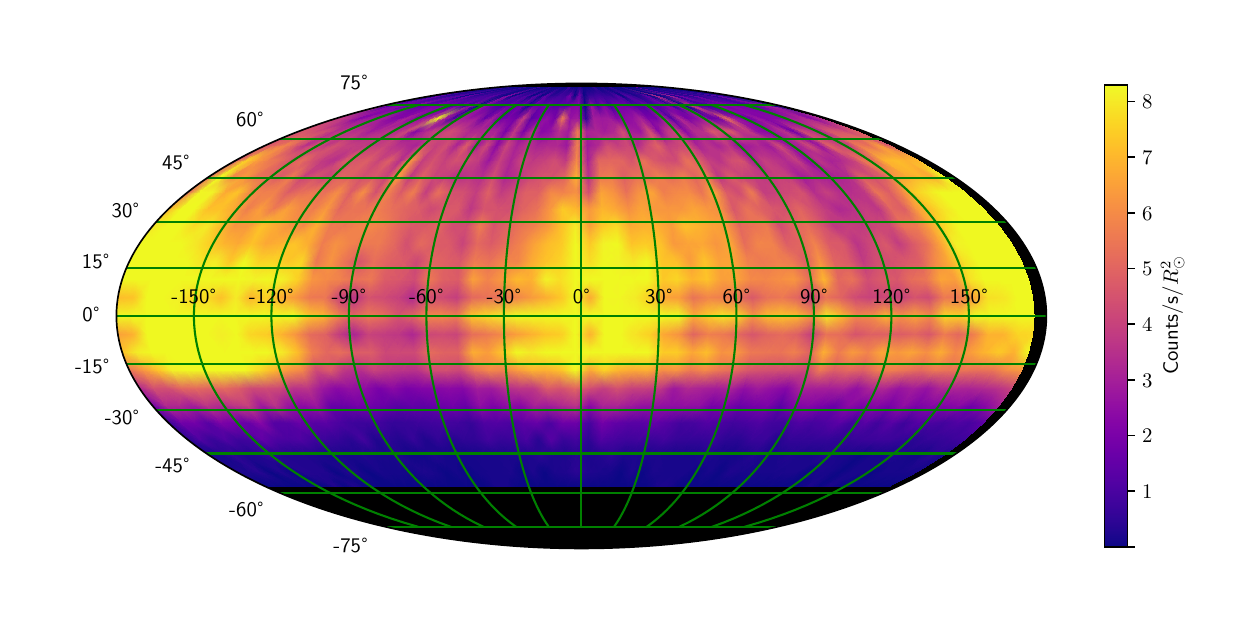} }
    \subfigure[Obs ID: 0134521501 (2002-04-12)]{
\raisebox{0.4cm}{\includegraphics[width=0.25\linewidth]{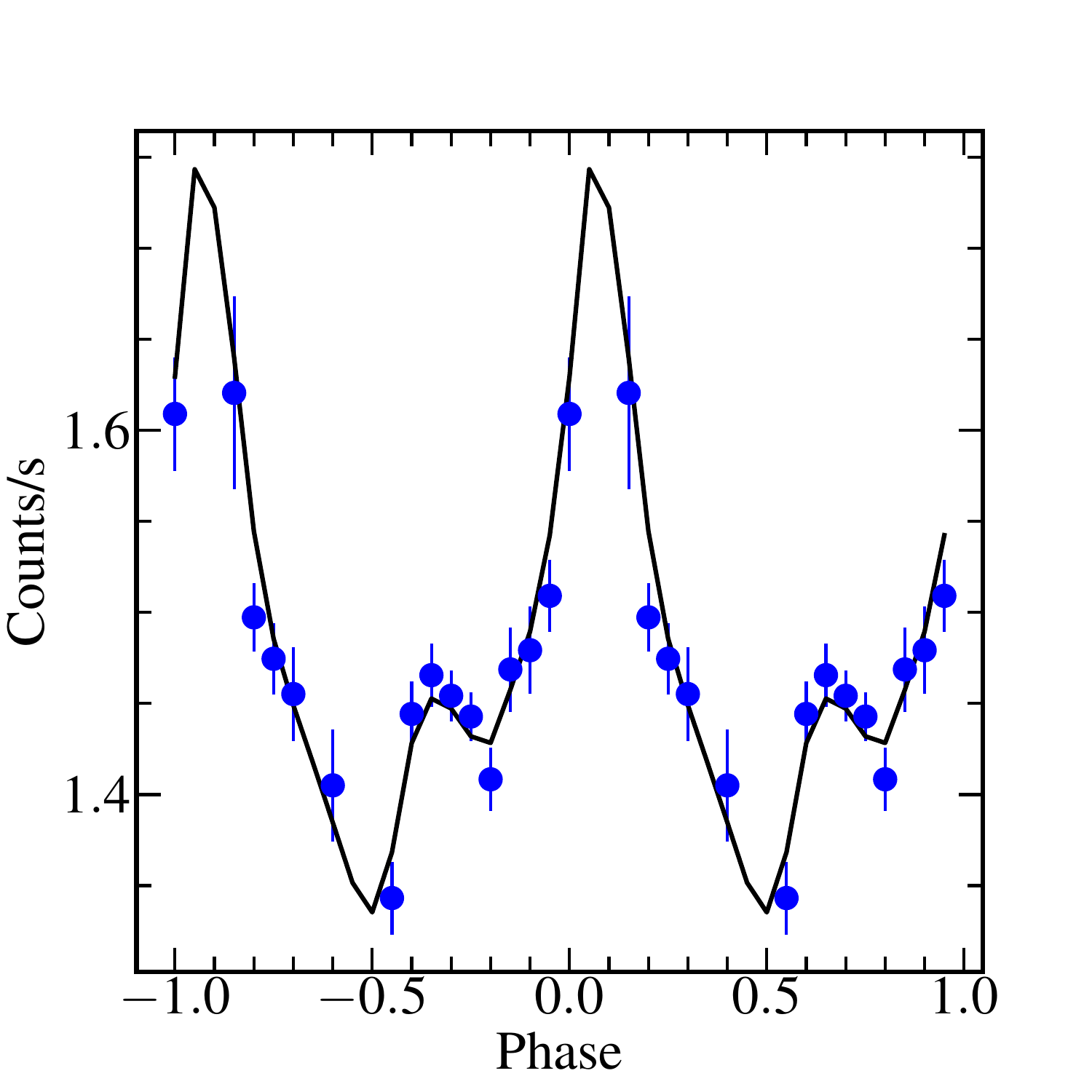}}
\includegraphics[width=0.60\linewidth]{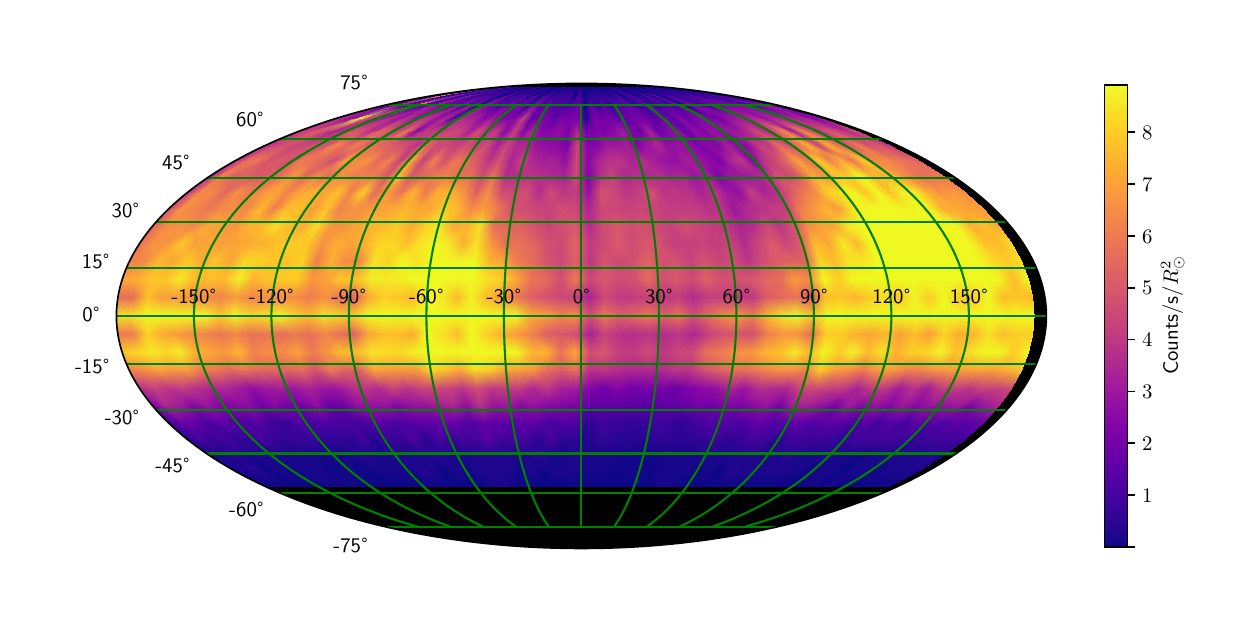} }
\subfigure[Obs ID: 0134522201 (2003-01-23)]{
\raisebox{0.4cm}{\includegraphics[width=0.25\linewidth]{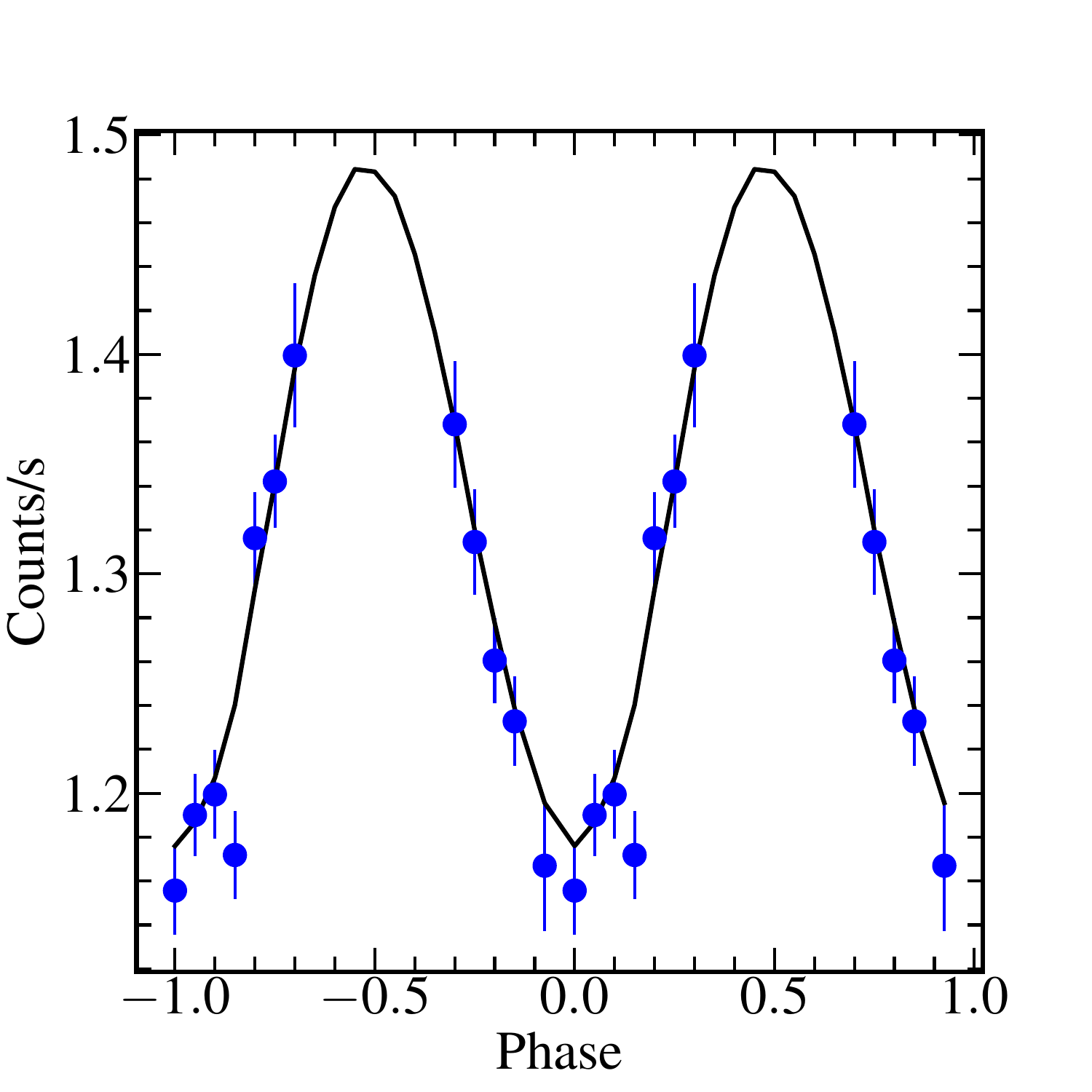}}
\includegraphics[width=0.60\linewidth]{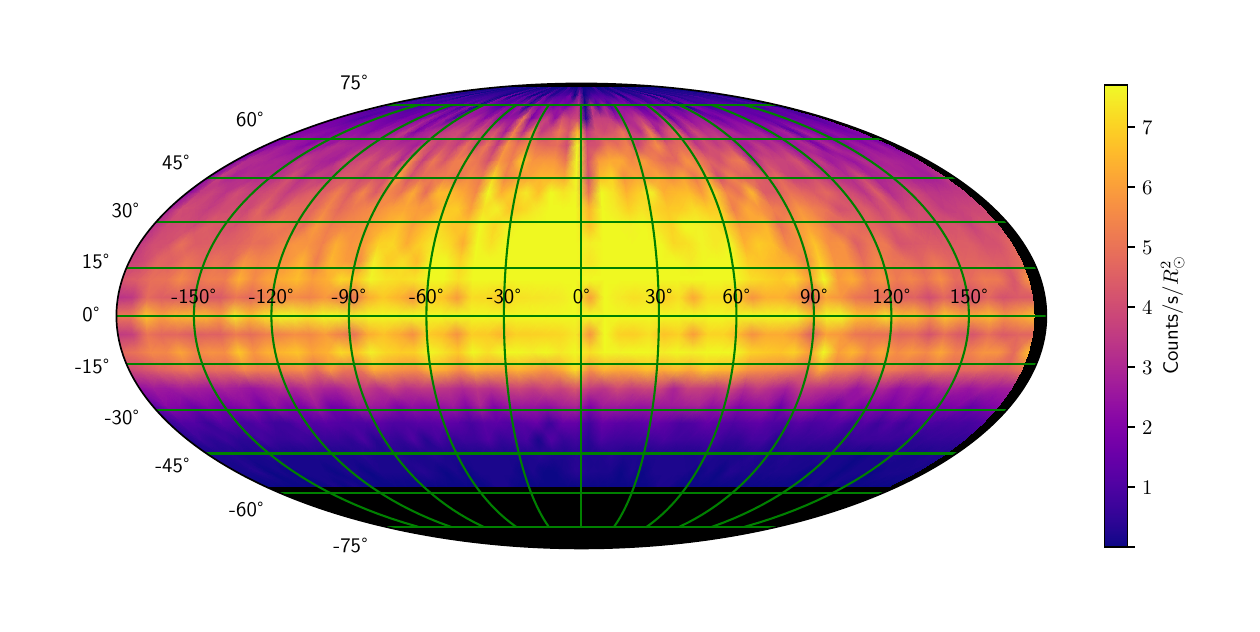} }
     \caption{Left panels: Phase folded X-ray light curves along with the modelled X-ray light curves. Right panels: Coronal images as obtained from the model with phase folded X-ray light curves.} \label{fig:corona1}
\end{figure*}
\setcounter{figure}{1}
\begin{figure*}
\setcounter{subfigure}{4}
    \centering
       \subfigure[Obs ID: 0412580701 (2011-01-02)]{
    \raisebox{0.4cm}{\includegraphics[width=0.25\linewidth]{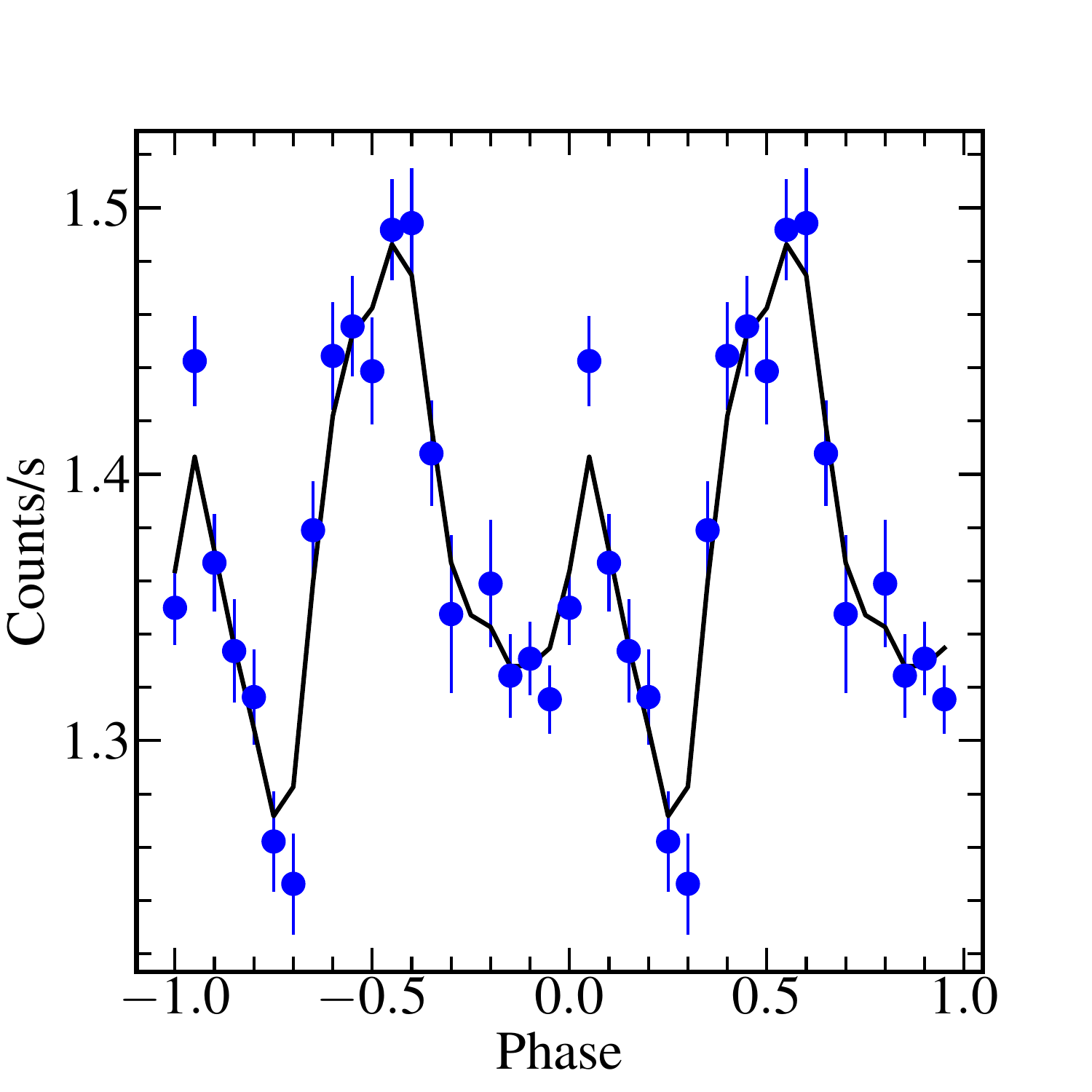}}
    \includegraphics[width=0.60\linewidth]{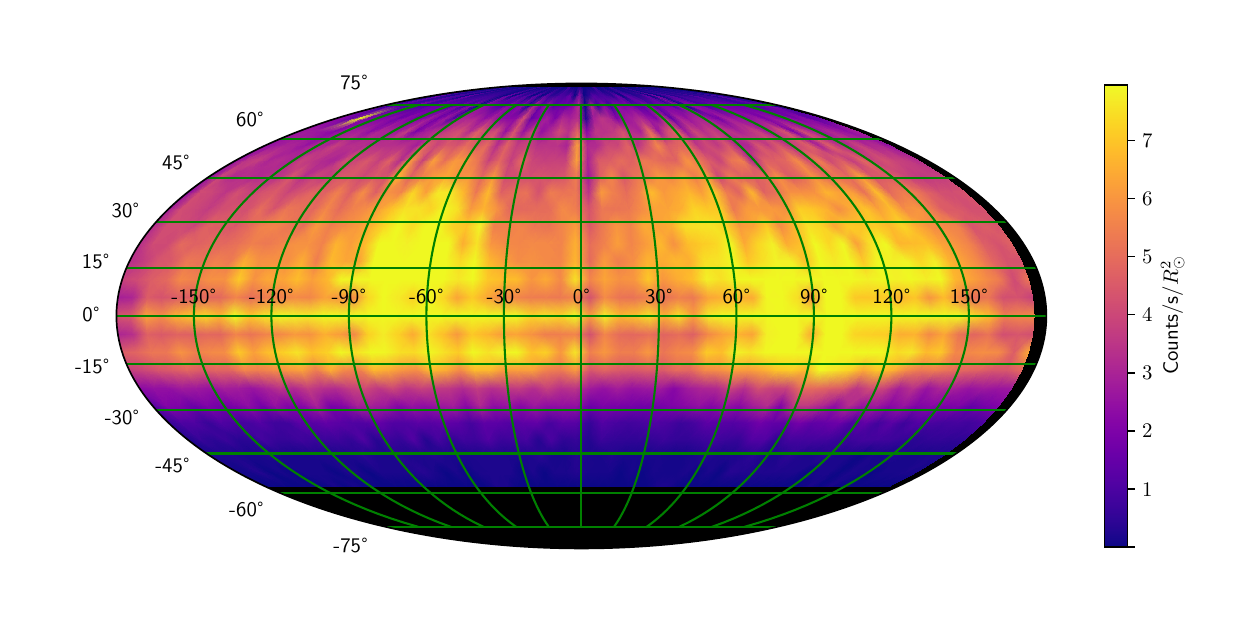} }
        \subfigure[Obs ID: 0791980101 (2016-10-07)]{
\raisebox{0.4cm}{\includegraphics[width=0.25\linewidth]{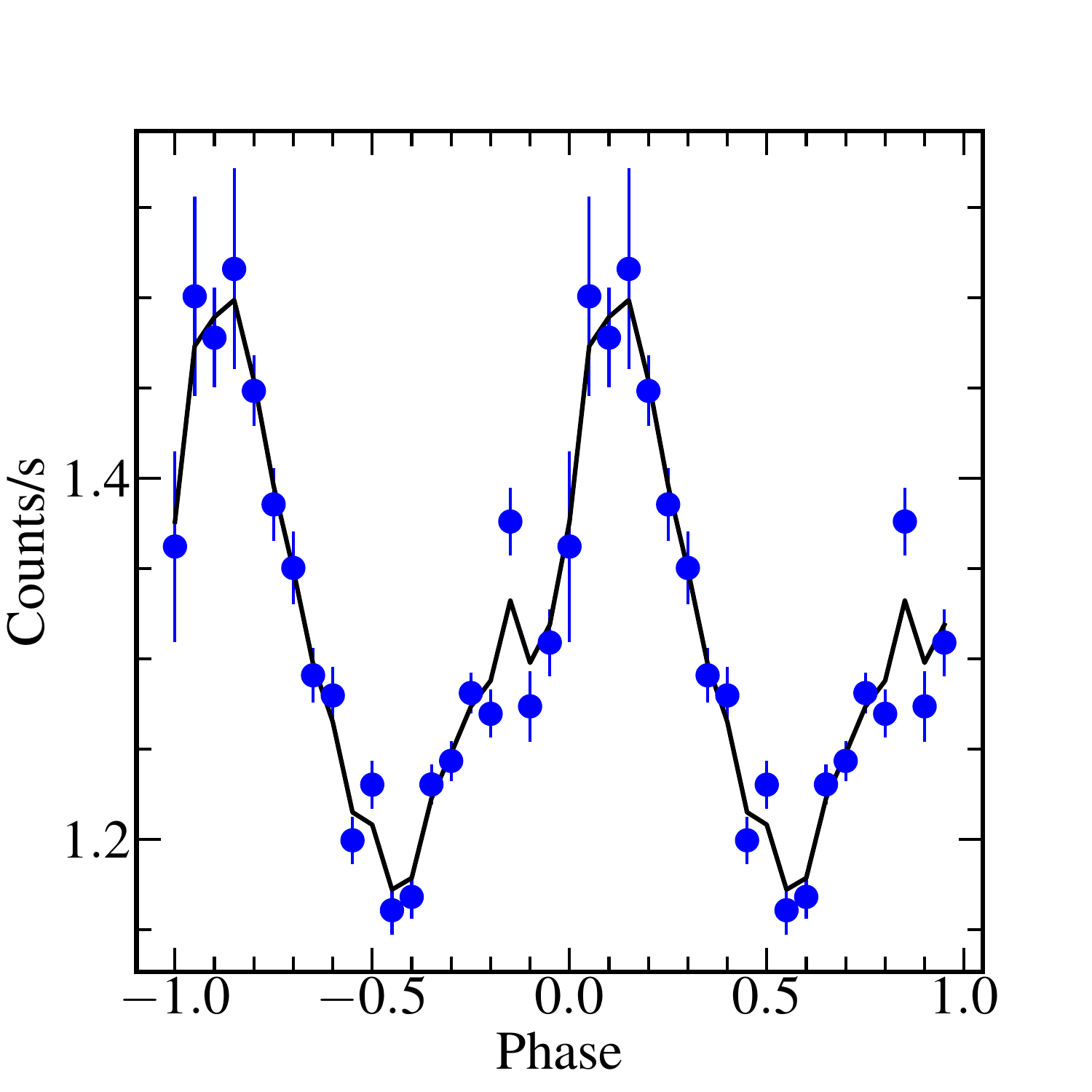}}
\includegraphics[width=0.60\linewidth]{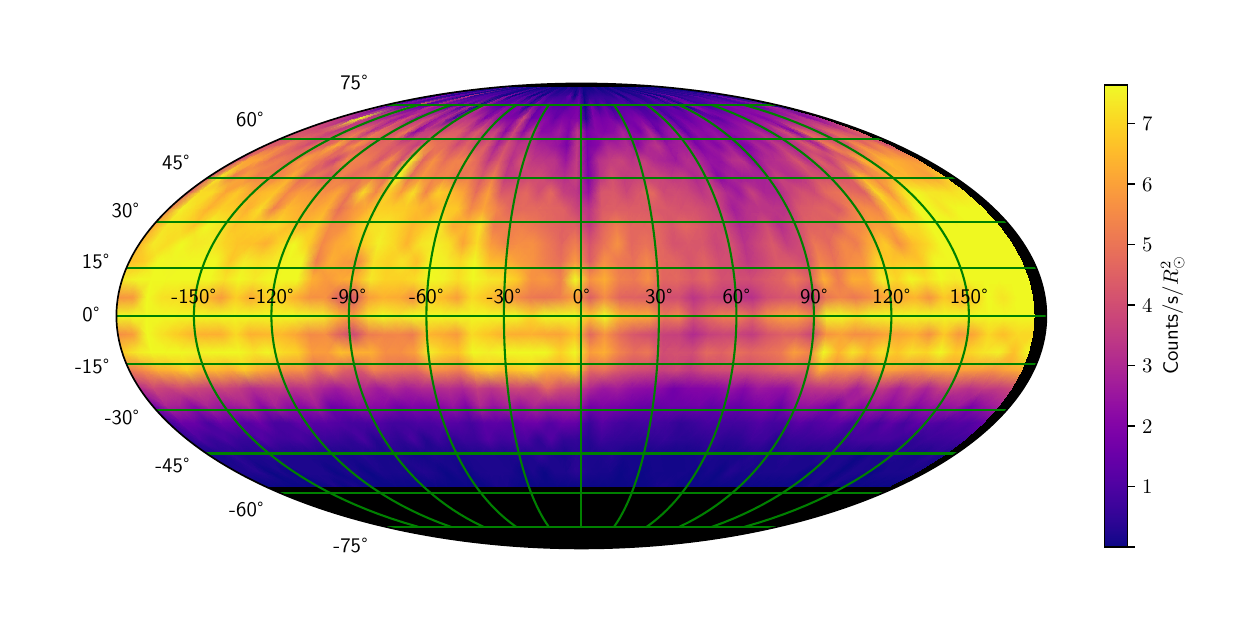} }
    \subfigure[Obs ID: 0791980401 (2017-10-10)]{
\raisebox{0.4cm}{\includegraphics[width=0.25\linewidth]{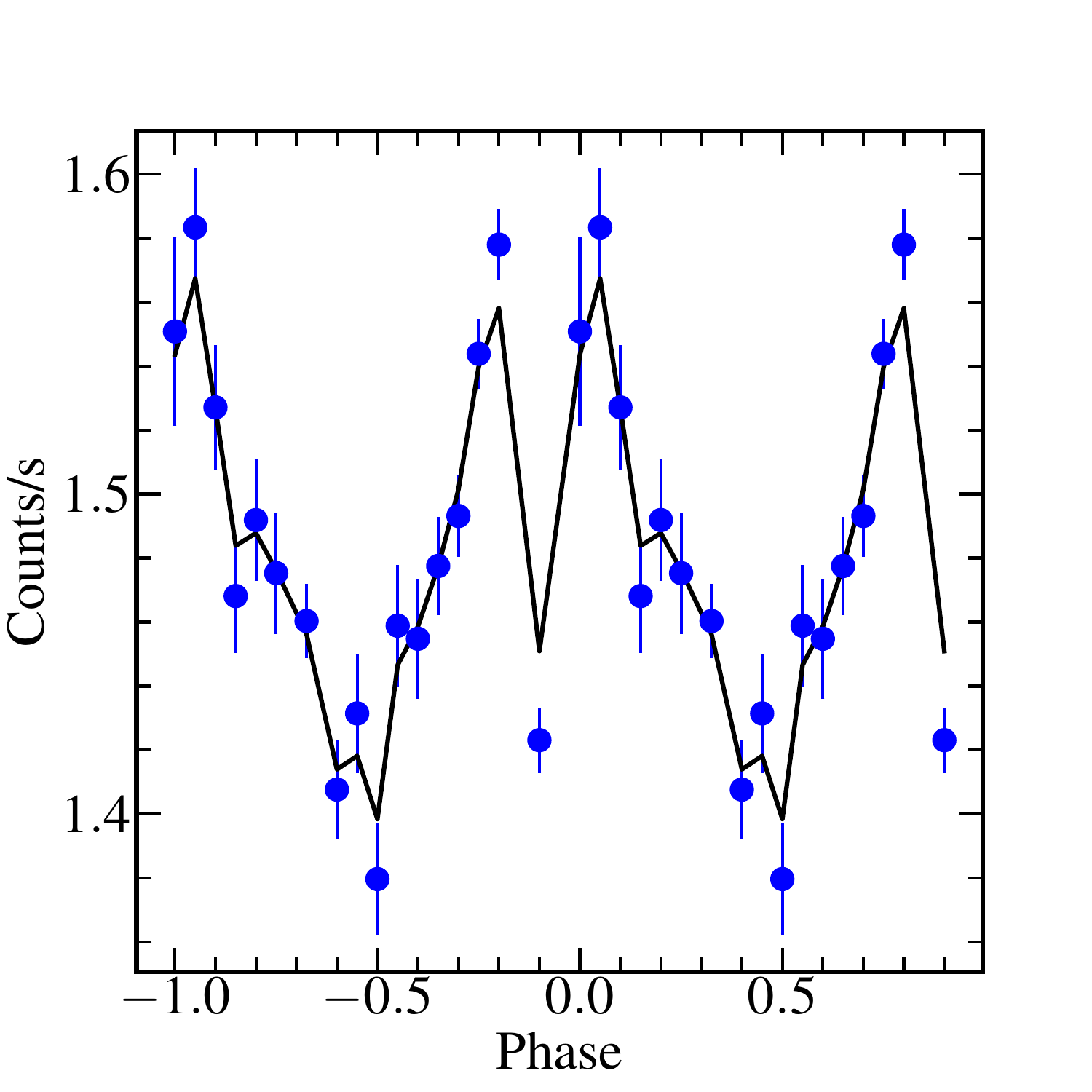}}
\includegraphics[width=0.60\linewidth]{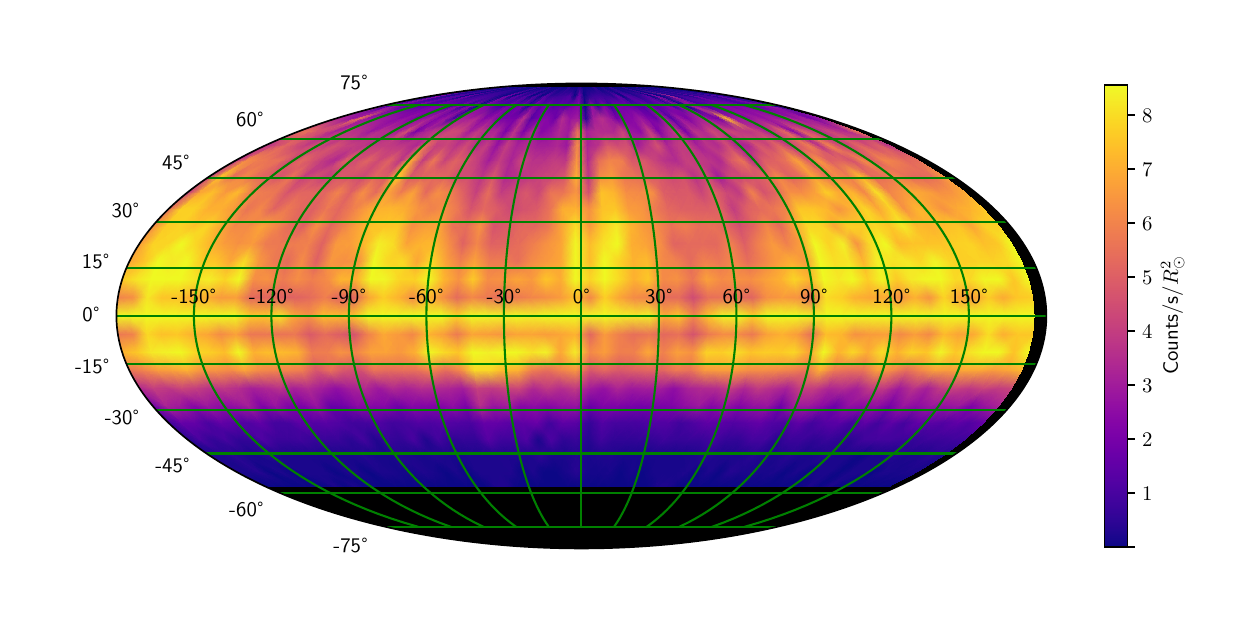} }
    \subfigure[Obs ID: 0810850501 (2019-09-30)]{
\raisebox{0.4cm}{\includegraphics[width=0.25\linewidth]{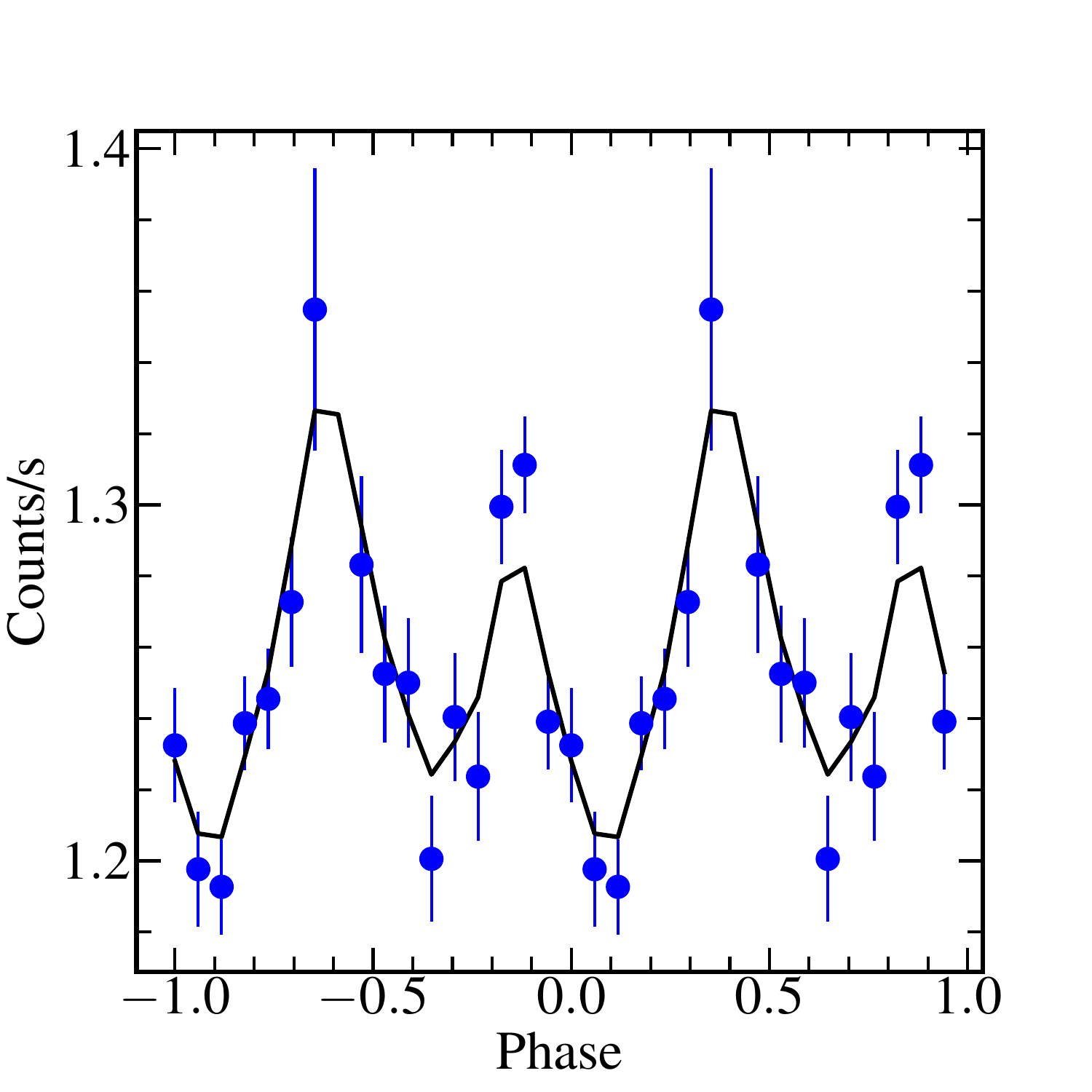}}
\includegraphics[width=0.60\linewidth]{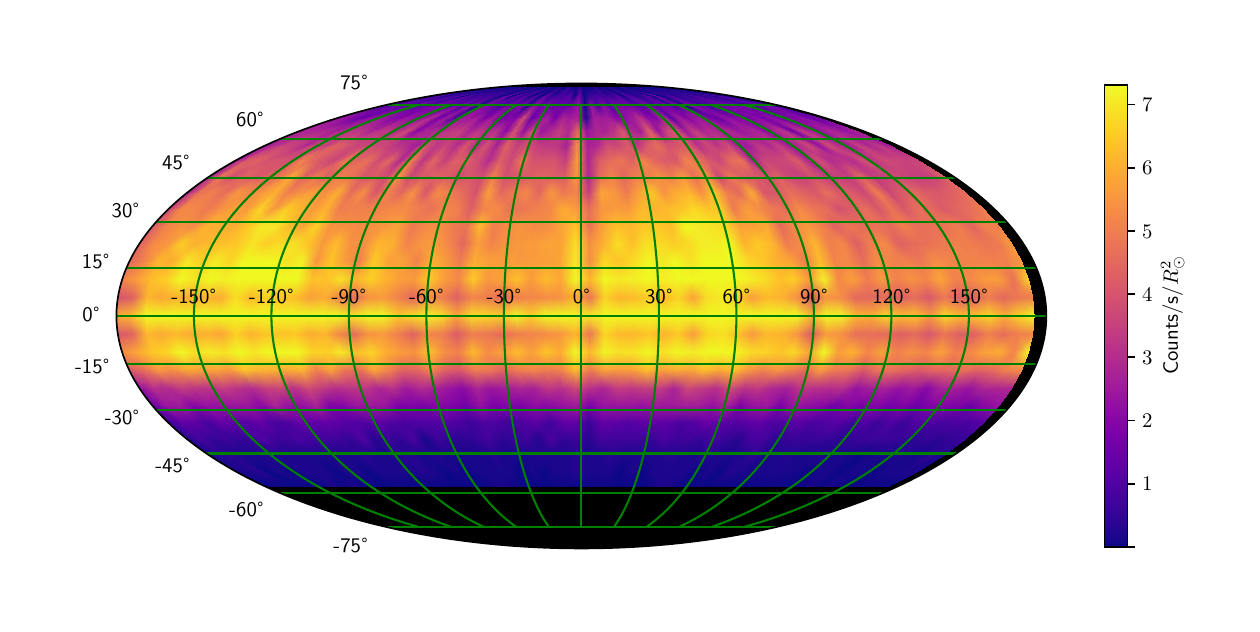} }
    
    \caption{Continued...} \label{fig:corona2}
\end{figure*}
\setcounter{figure}{1}
\begin{figure*}
\setcounter{subfigure}{8}
    \centering
        \subfigure[Obs ID: 0810850601 (2020-09-29)]{
\raisebox{0.4cm}{\includegraphics[width=0.25\linewidth]{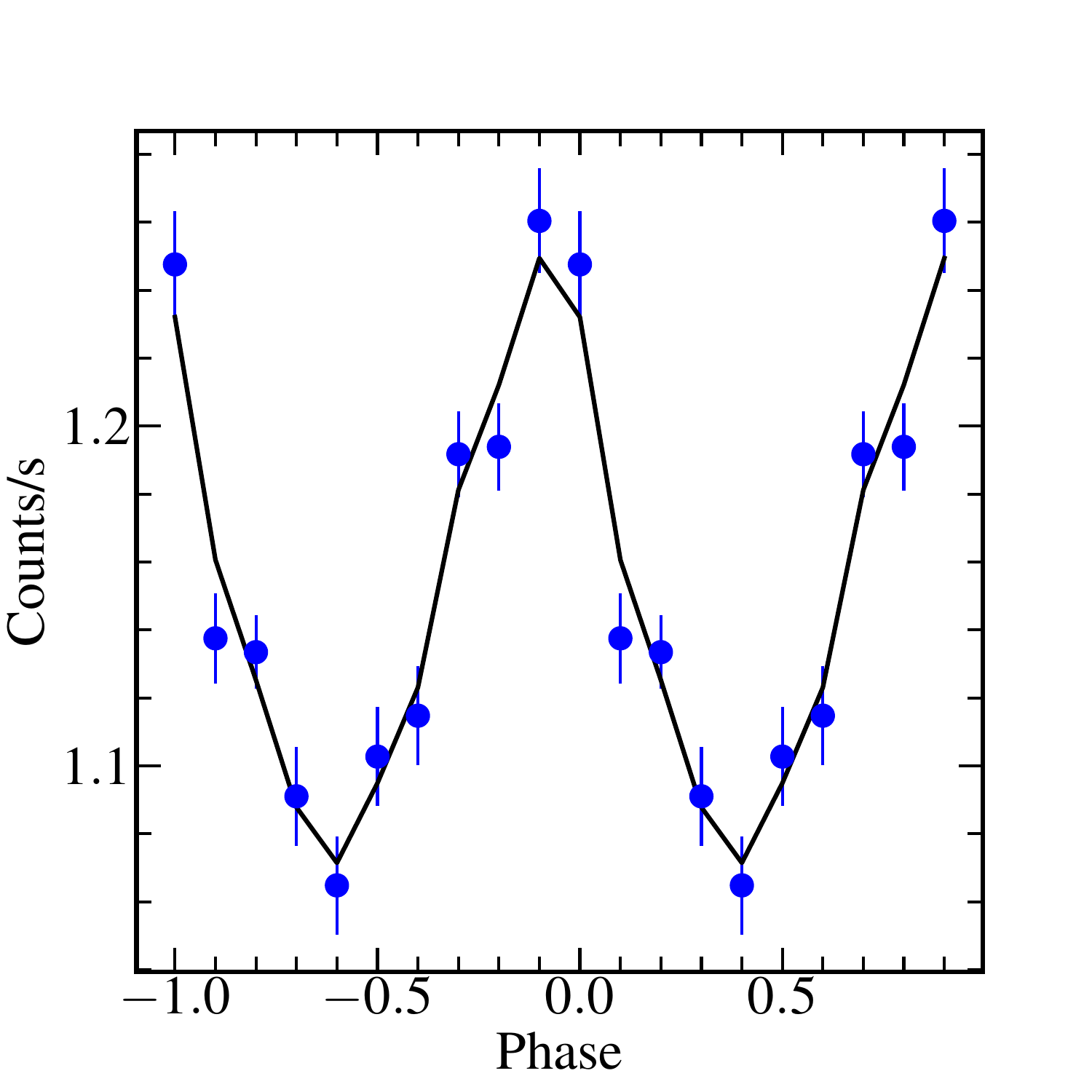}}
\includegraphics[width=0.60\linewidth]{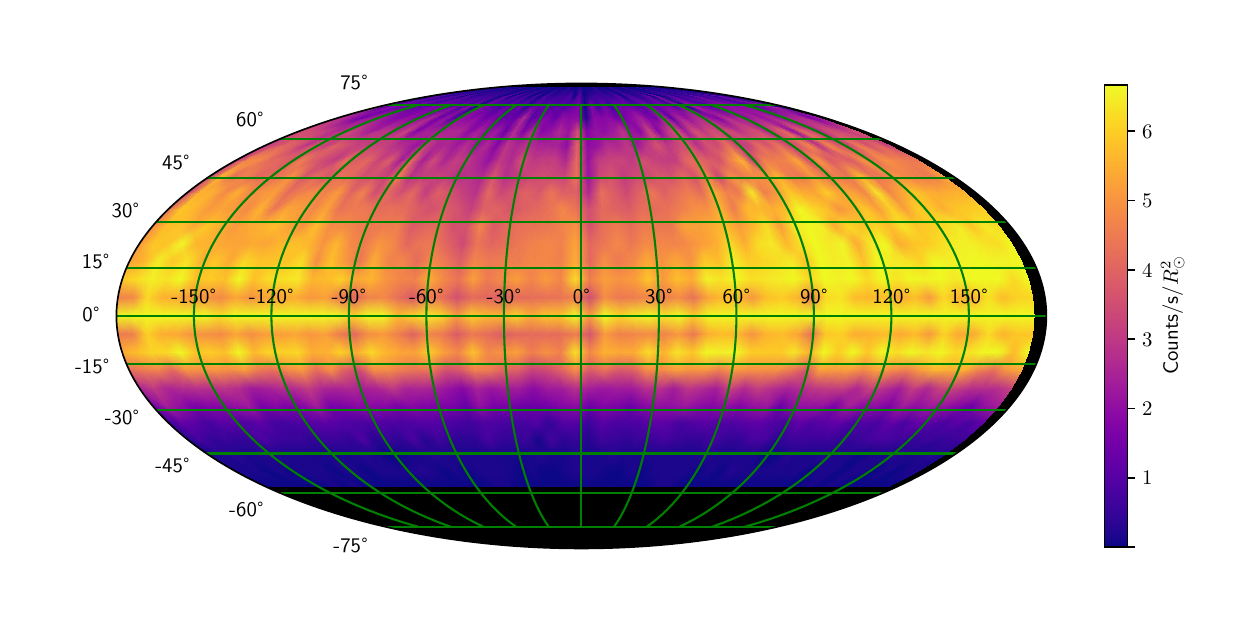} }
     \caption{Continued...} \label{fig:corona3}
\end{figure*}
 
\subsection{Flare detection and Rotational modulation}
\label{sec:rotationalmodulation}
AB Dor showed frequent flaring episodes in almost all the observations. Therefore, it is necessary to remove the flaring events before looking for the rotational modulations.  To remove flaring events from X-ray light curves, we have applied the sigma clipping method with the upper sigma clip to the 2$\sigma$ value of the mean, while the lower sigma clip is set free. By doing so, large flares are removed automatically from the data, which was then screened visually to remove further flaring events left unchecked by the algorithm above. 
The flaring regions are shown by red open circles in Figure \ref{fig:abdor_flare_qui1}.To quantify the flaring duration, we have calculated the flare duty cycle for each observation, which we define as the ratio of the flare duration to the total observing time of that epoch.  AB Dor's average flare duty cycle was found to be  0.57$\pm$0.23, where the error in the flare duty cycle is the standard deviation. This indicates that AB Dor remains in the flaring states for 57$\pm$23 \% of the observation time. Further, there seems to be no preferred rotation phase for the occurrence of flares during all the light curves analyzed in this study. During two epochs, 2016 and 2019, AB Dor showed very strong flares where the count rates during flare peaks were found to be  10-35 times higher than the quiescent count rates. A detailed analysis of these flares is carried out in our other paper \citep[][]{2024MNRAS.527.1705D}.

In order to study the rotational modulation, we removed all the flaring events from the original light curves. The flare-free light curves were then phase-folded using the following ephemeris.
$$MJD=44296.075+0.51479E$$
Where MJD corresponding to the "0" phase is the phase when the dominating active region (say spot A)  is in the centre of the visible disk, and another less active region (say spot B) was 180$^\circ$ apart from spot A \citep[see][]{1993A&A...278..467V}. The quiescent state for at least one rotational phase was observed only for the nine epochs of observations and are shown in the left panels of Figure \ref{fig:corona1}. These phase-folded light curves clearly show the X-ray rotational modulation for all nine epochs. Other light curves also show the signature of the rotational modulation, but complete phase coverage was not observed.

\subsection{Coronal imaging of AB Dor}\label{sec:abdor_corona}
The above-discussed model (Section \ref{sec:method1}) was applied to model the X-ray rotational modulation of AB Dor. We have taken the angle of inclination of 60$^\circ$, the radius of 0.96 $R_\odot$ \citep[][]{2011A&A...533A.106G}, and $h_{cor}$ of 0.4 $R_*$\citep[][]{2007MNRAS.377.1488H}.

Figure \ref{fig:corona1} shows the model results, with the right panels showing the image of the corona and the left panels showing the best fit modelled phase folded light curves along with the observed light curves.  The corona of AB Dor was found to have two active longitudes that were located approximately 180\deg apart. Two active regions were located near the same longitudes during the epochs 2000-05-01 (see Figure \ref{fig:corona1} a) and 2001-10-13 (see Figure \ref{fig:corona1} b). 
The brightest active region was located near the longitude of 0\deg, whereas the less active longitude was near -180\deg. During the observation of the epoch 2002-04-12 (see Figure \ref{fig:corona1} c), both active regions have shifted by 50$^\circ$ towards the decreasing longitude.  After nine months during the epoch 2003-01-23 (see Figure \ref{fig:corona1} d), the corona of AB Dor became dominated by a single active region, spreading over more than half a longitude from -90$^\circ$ to +90$^\circ$.
On the epoch 2011-01-02 (see Figure \ref{fig:corona2} e), AB Dor again showed two active regions near the longitude of -75$^\circ$ and +105$^\circ$. The corona of AB Dor in the yr 2016 (Figure \ref{fig:corona2} f), 2017 (Figure \ref{fig:corona2} g), and 2019  (Figure \ref{fig:corona2} h) appears to be similar to the corona during the periods of 2000-2002. 
It appears that after $\sim$ 17 yr of the observations, AB dor shows a similar activity level, which is close to the photometric activity cycle of $\sim$16.96 years. 
In 2020, the corona of AB Dor exhibits a similar pattern, with one active longitude spreading from +60\deg to -90\deg, but with a 180\deg shift from the active longitude observed in 2003. It appears that the corona of AB Dor also shows a flip-flop like cycle and displays long-term similarities that repeat roughly every $\sim$ 17 years.

\section{Long-term X-ray activity}
\label{LTVabdor}
\begin{figure}
    \centering
    \includegraphics[scale=0.35]{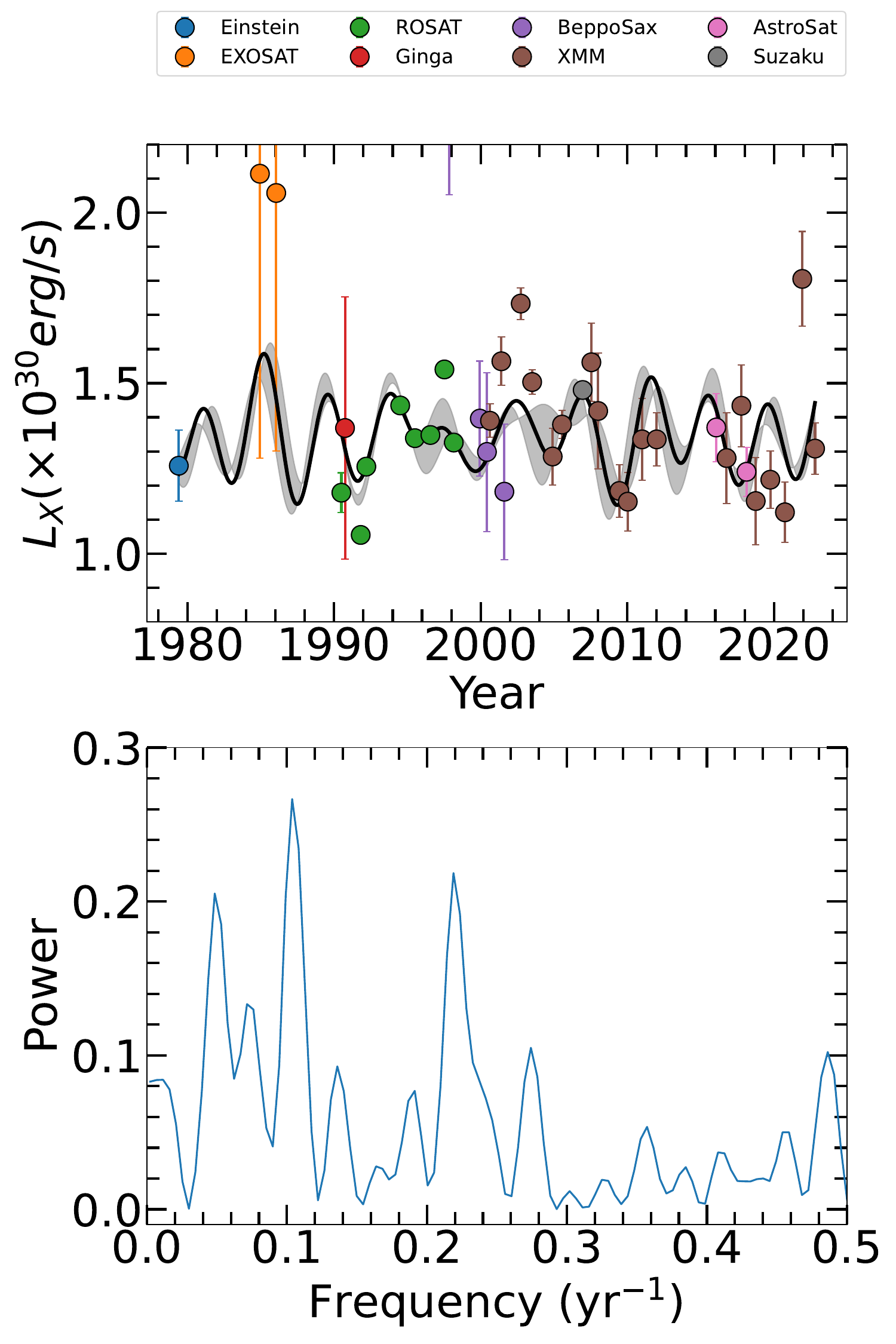}
    \caption{Long-term evolution of corona of AB Dor. The solid black line in the upper panel shows the best-fit line along with 1$\sigma$ confidence limits as grey-shaded regions. The lower panel shows the Lomb-Scargle periodogram of long-term X-ray data of AB Dor. }
    \label{fig:longtermabdor}
\end{figure}

We have compiled a comprehensive dataset of X-ray observations for AB Dor from 1979 to 2022 to analyse the LTV. This dataset was sourced from both X-ray archives and literature. The {\sc Einstein} observatory data was taken from the Einstein slew survey \citep{1992ApJS...80..257E}. 
Additionally, quiescent count rates were obtained from \cite{1993A&A...278..467V} and \cite{2002franciosini} for {\sc Ginga} and {\sc BeppoSax} observations, respectively. We have also included data from the ROSAT HRI observations and discarded those observations that were flagged as potentially variable. The {\sc EXOSAT} observations were obtained from the HEASARC database\footnote{\href{https://heasarc.gsfc.nasa.gov/W3Browse/exosat/me.html}{https://heasarc.gsfc.nasa.gov/W3Browse/exosat/me.html}}.
 The on-source background-subtracted light curves from {\sc EXOSAT} were obtained in 1-8 keV energy band, which were combined using the {\sc fmerge} task of {\sc FTOOLS}. The combined light curve was binned to 300 s, and the flare detection method was applied. Figure \ref{fig:abdor_flare_qui3} ll and \ref{fig:abdor_flare_qui3} mm show the resulting light curves. 
 We have averaged the quiescent state count rates over yearly bins to investigate long-term variability. The count rates were then calibrated in the energy range of 0.3--10.0 keV using the {\sc WebPIMMS}\footnote{\href{https://heasarc.gsfc.nasa.gov/cgi-bin/Tools/w3pimms/w3pimms.pl}{https://heasarc.gsfc.nasa.gov/cgi-bin/Tools/w3pimms/w3pimms.pl}} tool, with a single {\sc apec} model and an assumed coronal temperature of 0.87 keV \citep[][]{2024MNRAS.527.1705D}. The resulting fluxes were converted to luminosity using the distance of 14.85 pc. The quiescent fluxes in 0.3-10.0 keV using Suzaku data during the years 2006 and 2007 are taken from \cite{2014PASA...31...21S}.
 The evolution of quiescent luminosity is summarized in Table \ref{tab:abdorlongterm} and is plotted in Figure \ref{fig:longtermabdor}. 
 
We have performed the Lomb-Scargle periodogram analysis on the long-term quiescent binned data, revealing significant peaks at periods of 3.6$\pm$0.1, 4.3$\pm$0.1, 5.4$\pm$0.2, 9.5$\pm$0.2, and 19.2$\pm$2.1 yrs.  The periods of 3.6 and 5.4 yrs closely align with the period identified in previous studies from optical data\citep{2001ASPC..223..895A, 2005A&A...432..657J}.
 The peak at 19.2 yr bears a strong resemblance to the periods of 16.96 and 20 yr found in studies by \cite{2013A&A...559A.119L} and \cite{2005A&A...432..657J}, respectively.
 The 4.3 yr peak appears to be the beat period of 3.6 and 19.2 yr, indicating the presence of a long-term activity cycle. Furthermore, the 9.5 yr periodicity seems to be the first harmonic of the long-term activity cycle with a period of 19.2 yr.
 
 We first fit the constant L$_X$ model to the data to further investigate the long-term variability and computed the $\chi^2$ as 976.  This $\chi^2$ was then compared with the critical value of $\chi^2$ of 67.9 for the dof of 36 and the confidence limit of 99.9\%.  The derived $\chi^2$ was found to be much higher than the critical value of $\chi^2$, indicating that the L$_X$ is undoubtedly variable.  As mentioned above, we fit the sine curves with all four periods. This fit gives better $\chi^2$ (= 197) than that from the constant model fitting. Excluding the outliers at epochs 1997.53, 2002.73, and 2021.93, we found a much better fit with $\chi^2$ of 55.3. Adding the period of 19.2 yr does not improve the fit significantly. However, sinusoidal behaviours are unreliable when there are fewer than three cycles, as the available dataset was only for 44 years.
 The best-fit curve is overplotted in the upper panel of Figure \ref{fig:longtermabdor} as a solid black line with grey shaded regions corresponding to 1$\sigma$ confidence level. 

\section{Discussion }\label{sec:discussion}

We have carried out the STVs and LTVs observed in the ultra-fast rotating star AB Dor. The X-ray light curves of AB Dor showed a very dynamic nature with at least one flaring episode per epoch and the presence of rotational modulation.   We found that AB Dor flares $\sim$ 60 \% of its observing time, which is high among other active stars \citep[e.g.][]{2012MNRAS.419.1219P,2000A&A...356..949S}. 

\begin{figure*}
\centering
\vspace{-20pt}
 
     \subfigure[Evolution of eastern and western hemisphere with time]{\includegraphics[width=0.95\textwidth,trim={2.0cm 0cm 3.5cm 2cm},clip] {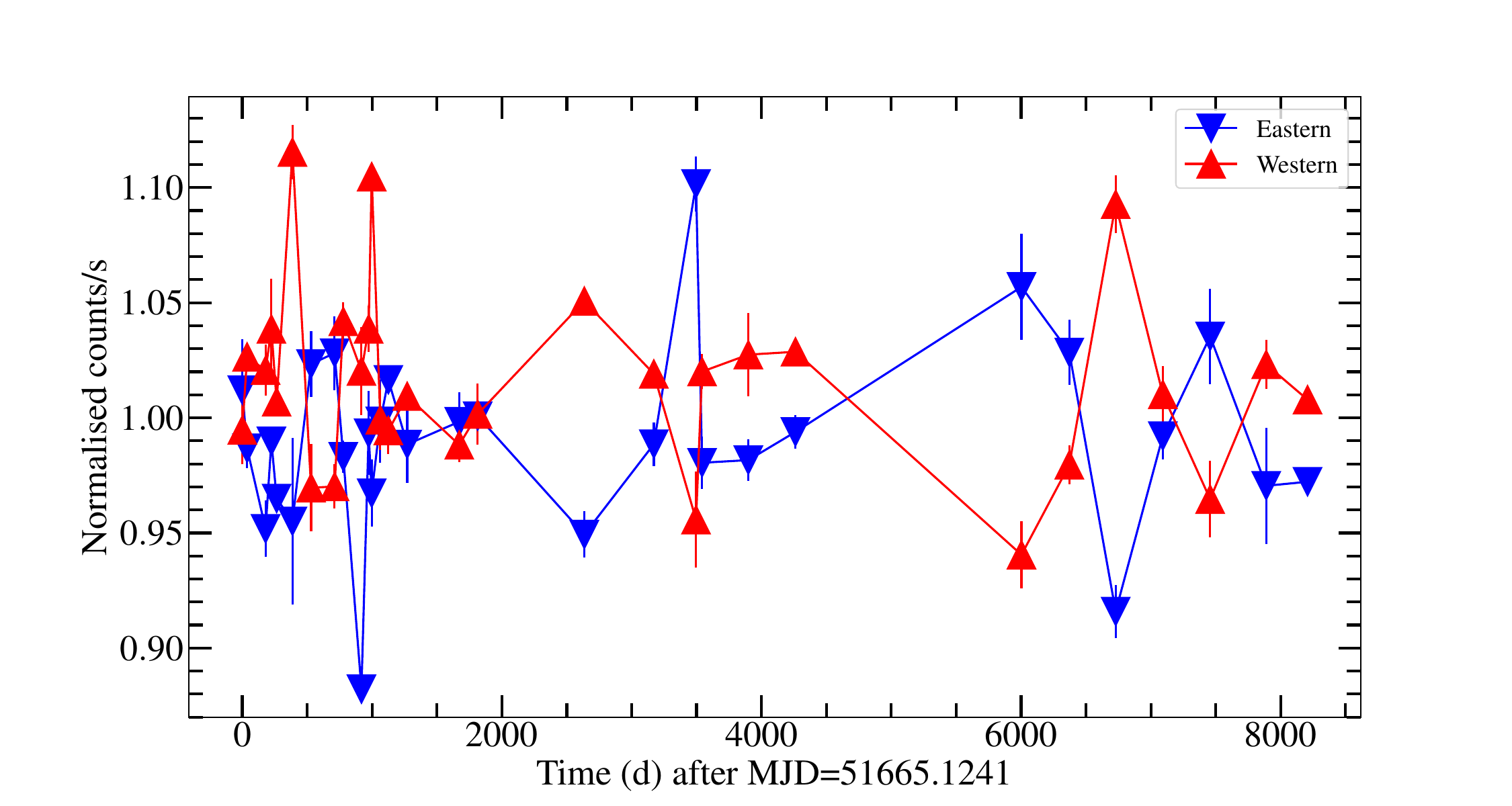}\label{fig:evolu}}
     
    \subfigure[Eastern vs Western hemisphere]{\includegraphics[height=0.46\textwidth]{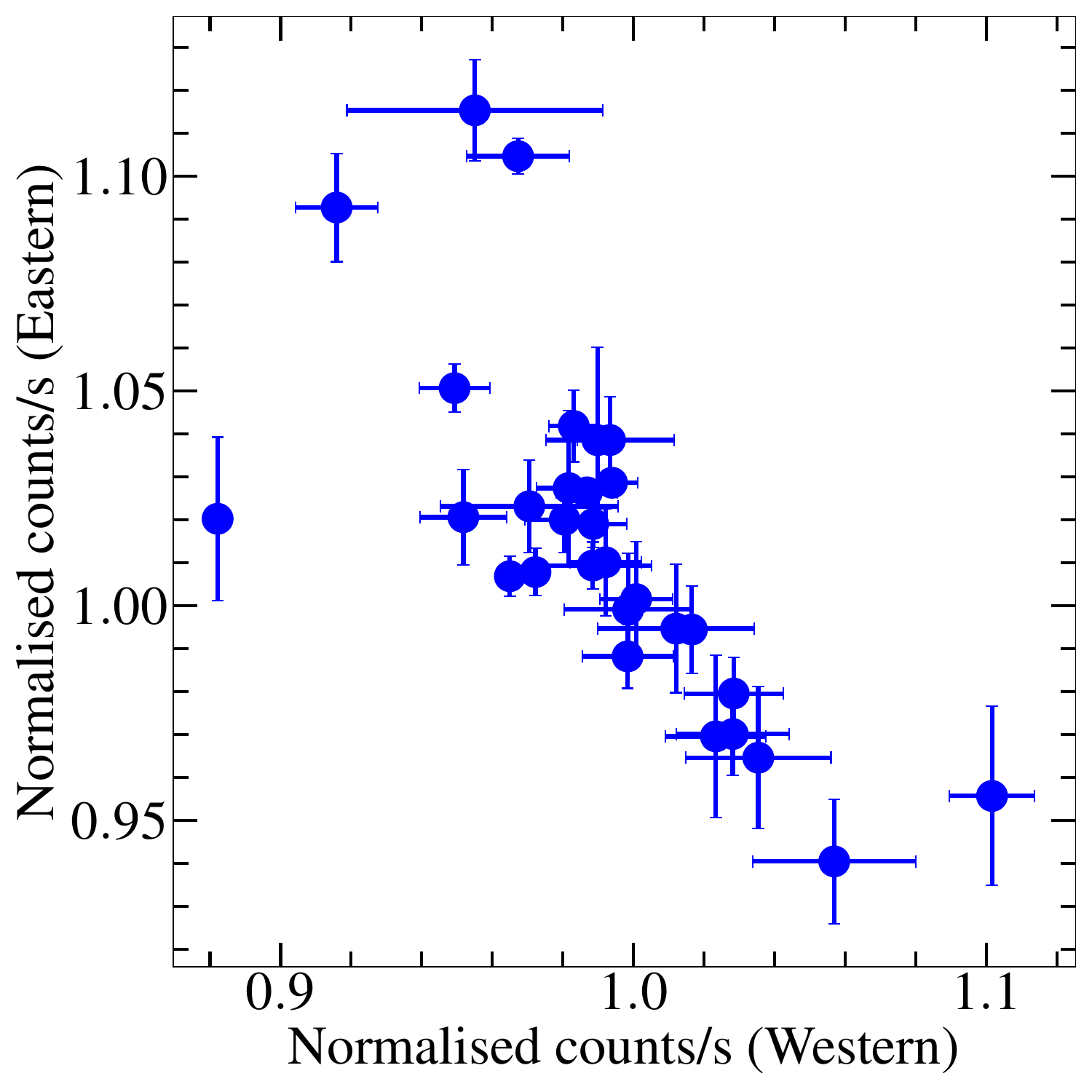} \label{fig:corr}}
   \subfigure[LS power spectra of Eastern and western hemispheres]{\includegraphics[height=0.46\textwidth]{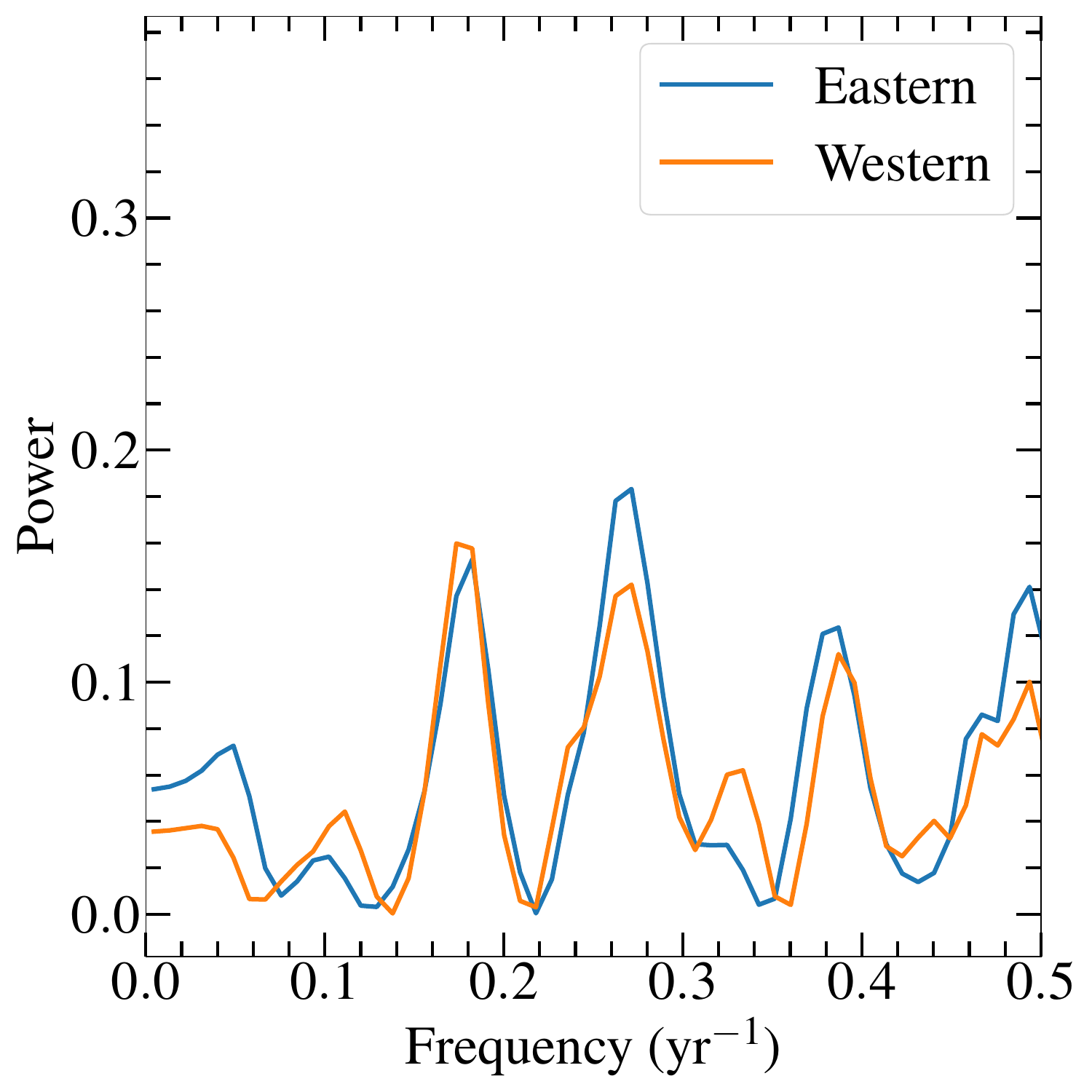} \label{fig:GLSD}}
    \caption{(a) Plot of average count rate of two opposite hemispheres with time as obtained by phase folded quiescent state light curves. (b)  The plot between average X-ray count rate from the eastern hemisphere and average X-ray count rates from the western hemisphere,(c) LS  power spectra Eastern and Western hemispheres' light curves.}  
    \label{fig:abdorspots}
\end{figure*}

The X-ray emission from AB Dor is found to be rotationally modulated. The X-ray rotational modulation in AB Dor has been reported many times in the past \citep[e.g.][etc]{1993A&A...278..467V,1997A&A...320..831K,2003MNRAS.345..601M,2013A&A...559A.119L}.  For a total of nine epochs of observations show complete rotational modulation, whereas, for other epochs of observations, rotational modulation could not be seen for a complete cycle due to a higher flare duty cycle. In order to analyze rotational modulation in AB Dor, we have developed an LCIT to image stellar coronae of single active fast rotators.  The coronal images obtained from X-ray light curve modelling show the bimodal distribution of active regions across the longitudes for most of the epochs. Due to the geometrical constraints of the inclination angle of the star, only a part of the latitudes can be modelled. This fact can be seen in Figure \ref{fig:corona1}, where we could see active regions between -30\deg to +60\deg of latitude.

We have also carried out a long-term X-ray study of AB Dor using the X-ray data obtained from various X-ray missions from the year 1979 to 2022. 
The LS analysis of the data reveals four periods at 3.6,  4.3,  5.4 and 9.5 yr in the power spectra of X-ray light curves. In the earlier studies, the periods at 3.4  and 5.4 ys were also obtained and explained in terms of the flip-flop cycle. The phenomenon of a flip-flop is also evidenced by the coronal images obtained from LCIT, as explained in Section \ref{sec:method}. For example, the coronal image of the epoch 2017 and 2019 show opposite coronal brightness. However, the limited number of coronal images does not allow us to establish firmly the presence of a flip-flop cycle.


 As mentioned above, most of the X-ray light curves suffer from flaring events, and we do not have a complete quiescent state for a single rotation to know the flux of active regions. Further, the unavailability of simultaneous/quasi-simultaneous optical and X-ray images does not allow us to compare the active longitudes. Therefore, we have compared the average X-ray count rates of the first half of the folded light curve with that of the next half for all the quiescent data.  This fixed hemisphere method is equivalent to comparing the X-ray flux from one hemisphere (say eastern) to that of the next hemisphere (say western). The eastern hemisphere is around $\pm$90\deg (or $\pm$0.25 phase) of spot A, whereas the western hemisphere is around the  $\pm$90\deg of spot B. 
  This method may be susceptible to the migration of active regions in and out of the fixed hemispheres. 
 
 To explore the impact of migration on the light curves of a star with two active regions in its corona, we simulated synthetic light curves over 20 years for every 100th rotation, with the brightness of the active regions varying according to a 5.5-year flip-flop cycle, as found in AB Dor from observations in the optical band. Employing the fixed hemisphere method, which calculates the normalized flux of Eastern and Western hemispheres for each rotation and a Lomb-Scargle periodogram analysis, we first confirmed that our method accurately recovers the 5.5-year flip-flop period without migration. We tested various migration rates by introducing synchronous linear migration in the longitude of both active regions. We found that a flip-flop cycle can be detected with synchronous migration of active regions below 0.1 degrees per rotation. However, the method did not recover peaks in the periodogram for migration rates exceeding or equal to 1 degree per rotation. Even at a relatively high migration rate of 0.1 degrees per rotation, our method still captured signatures of the flip-flop cycle. This method gains a significant advantage in the case of AB Dor as active regions steadily migrate throughout a considerable portion of the starspot cycle.

 

We have taken only those light curves for which at least  25 \% of the phased light curve is available for each hemisphere.
    The resulting evolution of X-ray emission from both the eastern and western hemispheres is depicted in Figure \ref{fig:evolu}. Here,  the error in each data point is the weighted standard deviation of X-ray counts of that hemisphere.  Figure \ref{fig:corr} shows the plot of count rates from the eastern hemisphere to count rates from the western hemisphere. The count rates from both hemispheres are found to be anti-correlated with each other. The Spearman correlation coefficient between the count rates of the eastern and western hemispheres is found to be -0.78 with a null hypothesis probability of $3\times10^{-7}$. This suggests that the hemispheres interchange their activity with time.
   
    We conducted periodogram analysis using the LS method for both the flux of the eastern and western hemispheres. The results, depicted in Figure \ref{fig:GLSD}, reveal prominent peaks at periods of 3.6$\pm$0.2 yr and 5.6$\pm$0.4 yr. These frequencies are also present in the power spectra of the long-term X-ray light curve.
   The existence of the 3.4 and 5.4 yr periods in optical data has been previously attributed to the flip-flop cycle  \citep{2005A&A...432..657J}. Thus, the presence of common periodicities of $\sim$ 5.5 yr in optical and X-ray data may also suggest the presence of the flip-flop cycle in X-rays.
   

Since the first discovery of flip-flops on a G-type giant FK Com \citep[][]{1991LNP...380..381J,1993A&A...278..449J}, the flip-flop cycles have been reported for many stars in single as well as binary systems (RS CVn \citep[][]{1998A&A...336L..25B,2009A&A...495..287B}, young solar analogues \citep[][]{2002A&A...393..225M,2005AN....326..283B,2013MNRAS.430.2154P}, active M dwarfs \citep[][]{2010AN....331..250V}, short-period active binaries \citep[][]{2013AN....334..625O}, W UMa systems \citep[][]{2015ApJ...805...22W,2017AJ....154...99L,2018A&A...612A..91M}, and CABS \citep[][]{2017ApJ...838..122J}). However, no flip-flops have been reported in the X-ray band for any active star in the past. 

 Previous studies, such as those by \cite{2002A&A...394..505B}, have put forth theoretical models to explain the flip-flop cycles observed in solar analogues and the Sun. They suggest that flip-flop cycles are due to the existence of two distinct magnetic dynamo modes. One of these modes is an oscillating axisymmetric mode that corresponds to sunspot-like cycles, while the other is a non-axisymmetric mode linked to active longitudes. \cite{2004MNRAS.352L..17M} and \cite{2004SoPh..224..153F} further explored the possible combinations of these modes resulting in flip-flops and active longitudes.
Comparisons between the Sun and young dwarf stars reveal that the relative strength of these modes can change over time. Currently, the Sun is primarily dominated by an axisymmetric dipole-like mode, while both modes are prominent in younger active stars. 
AB Dor shows periodicities at 3.6 and 5.4 yr in long-term X-ray data that were linked to a flip-flop cycle in previous studies, while the signatures for the longer period is seen at 19.2 yr and its first harmonic at period 9.5 yr, which suggests that the non-axisymmetric mode coexist with the axisymmetric mode.

\section{Conclusions}\label{sec:conclusion}
The present study indicates that the AB Dor exhibits frequent flare events, accounting for an average of 57 $\pm$ 23 \% of the total observation time occupied by these flares. Analysis of quiescent light curves revealed the presence of rotational modulation in most observational epochs.  In addition, the coronal imaging of  AB Dor has shown the presence of two distinct active longitudes, each situated 180 degrees apart. These active longitudes exhibit migration and variations in their relative brightness, with one active longitude dominating the other. The periodogram analysis of long-term X-ray lightcurve exhibits multiple periodic signals.  Notably, periods of approximately $\sim$3.6,  $\sim$5.4, and $\sim$19.2 yr manifest, resembling findings from previous studies of optical observations. It is plausible that the $\sim$5.4 yr period is linked to an X-ray flip-flop cycle, whereas the 19.2 yr may correspond to a long-term cycle. The X-ray flip-flop cycle needs to be confirmed using long-term X-ray observations with adequate cadence.


\section{Acknowledgements}
We wish to extend our sincere appreciation to the designated referee and scientific editor of this paper for their insightful comments and valuable suggestions.
The EXOSAT, ROSAT, and XMM-Newton data have been retrieved from the HEASARC archive database. The majority of this work is based on observations obtained with XMM-Newton, an ESA science mission with instruments and contributions directly funded by ESA Member States and NASA.
 XMM-Newton data is reduced using Science Analysis System (SAS), available at https://www.cosmos.esa.int/web/xmm-newton/sas-download.
 HEASOFT used for light-curve analysis is available at \\https://heasarc.gsfc.nasa.gov/lheasoft/download.html.

\bibliography{sample631}{}
\bibliographystyle{aasjournal}

\newpage
\appendix
\renewcommand\thefigure{\thesection.\arabic{figure}}    
\section{X-ray light curves}\label{sec:appendix_lcs}
\setcounter{figure}{0} 
\begin{figure*}
\centering
\subfigure[Obs ID: 0123720201 (2000-05-01)]{\includegraphics[scale=0.15]{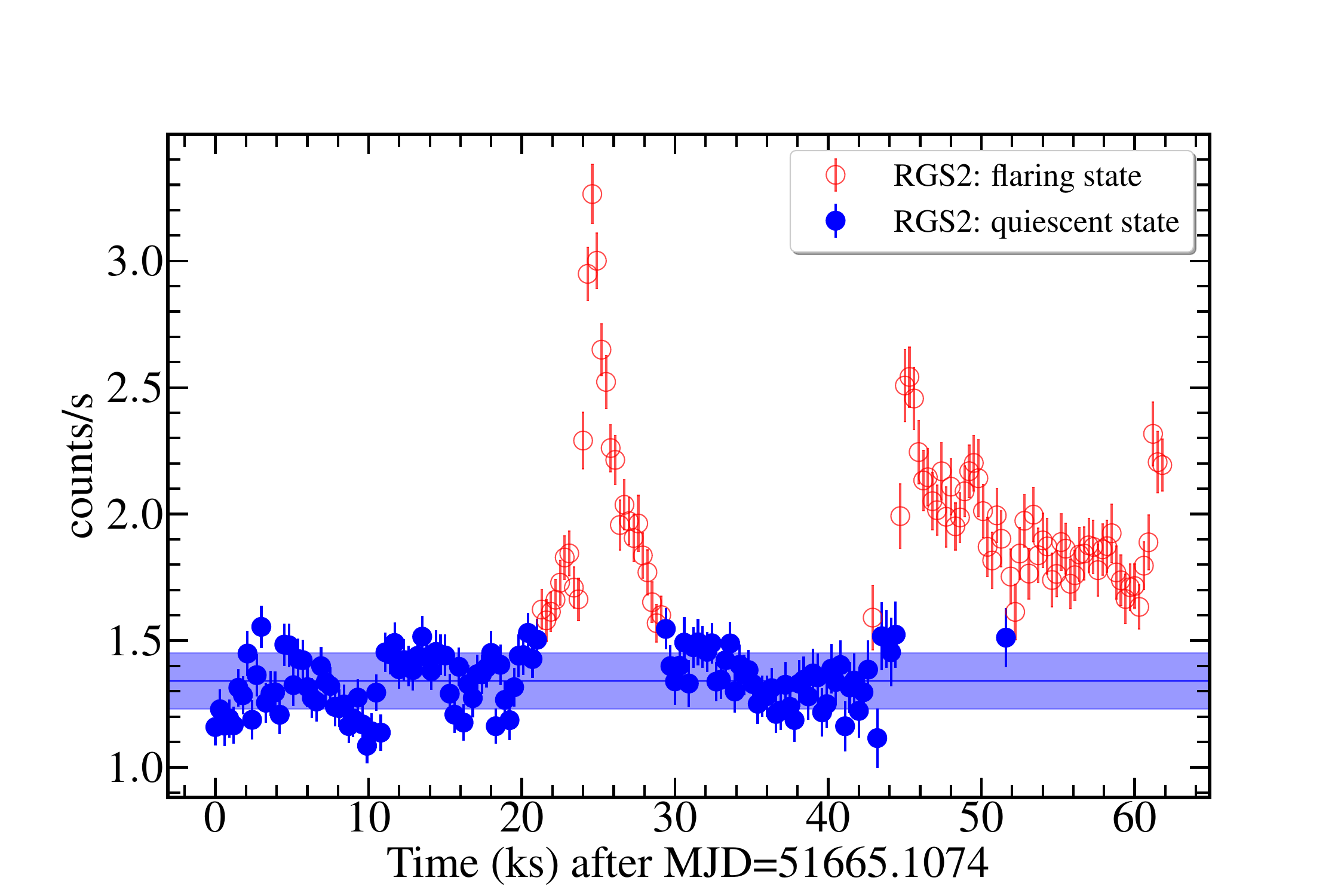}}
\subfigure[Obs ID: 0126130201 (2000-06-07)]{\includegraphics[scale=0.15]{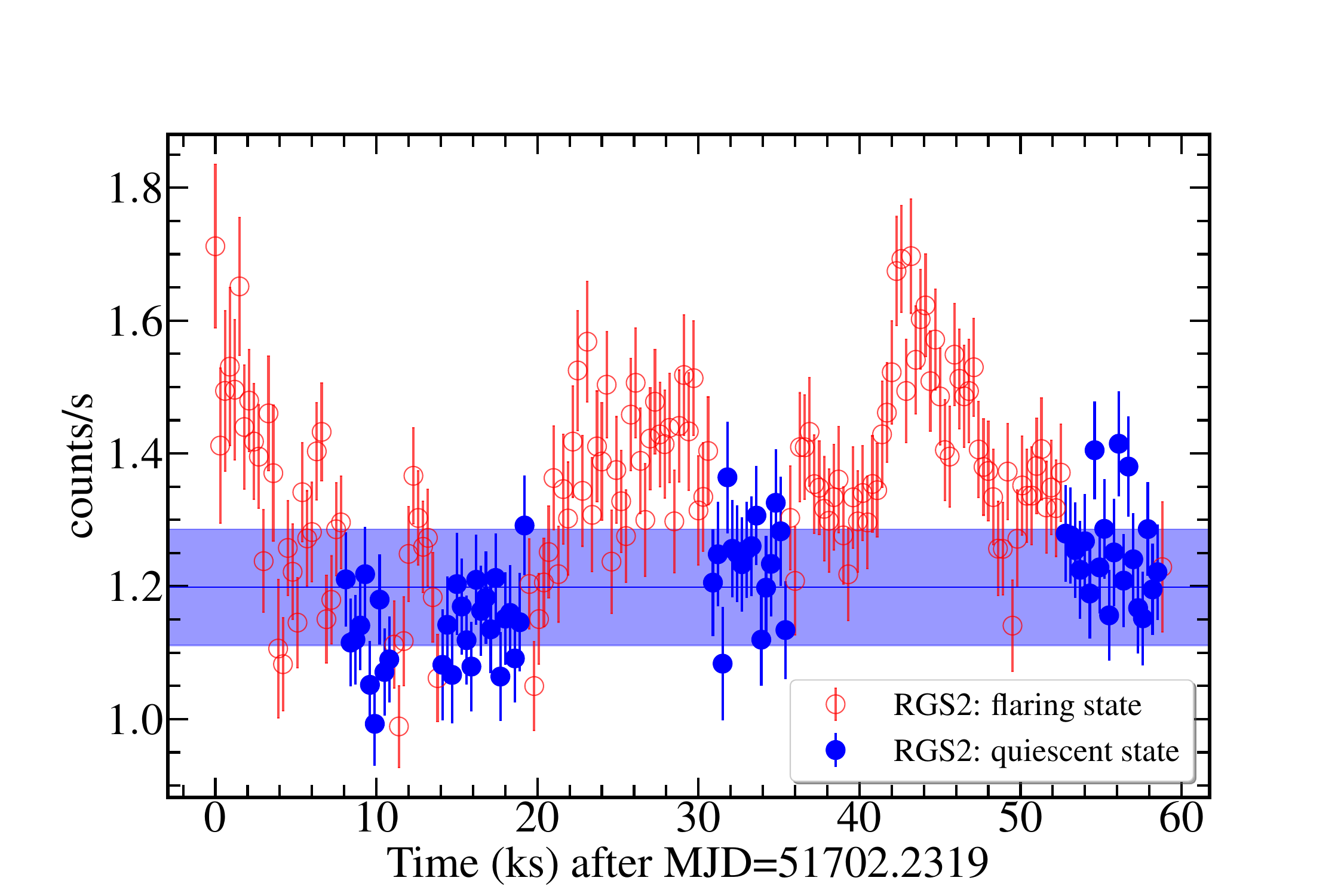}}
\subfigure[Obs ID: 0123720301 (2000-10-27)]{\includegraphics[scale=0.15]{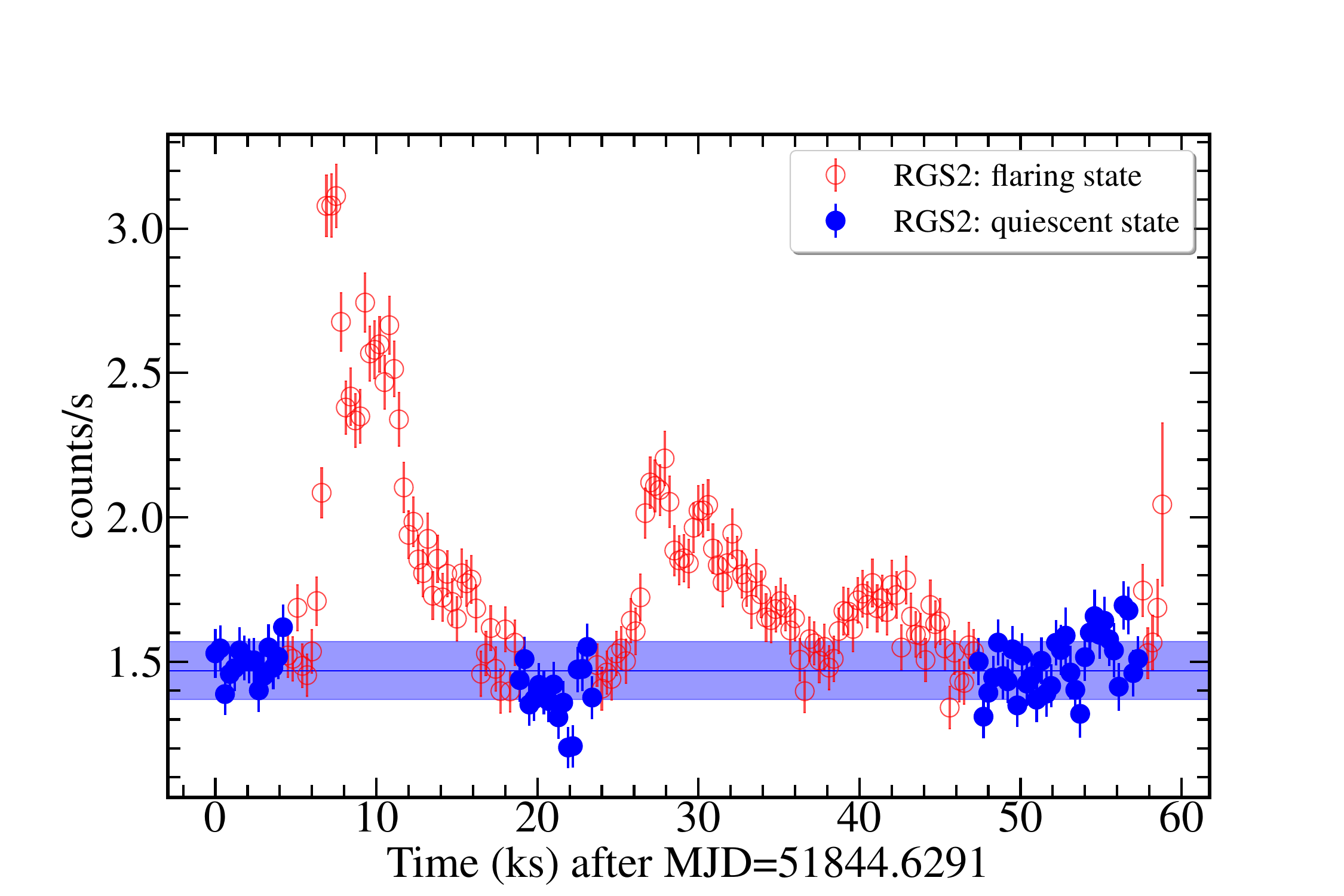}}
\subfigure[Obs ID:0133120701+0133120101+ 0133120201 (2000-12-11)]{\includegraphics[scale=0.15]{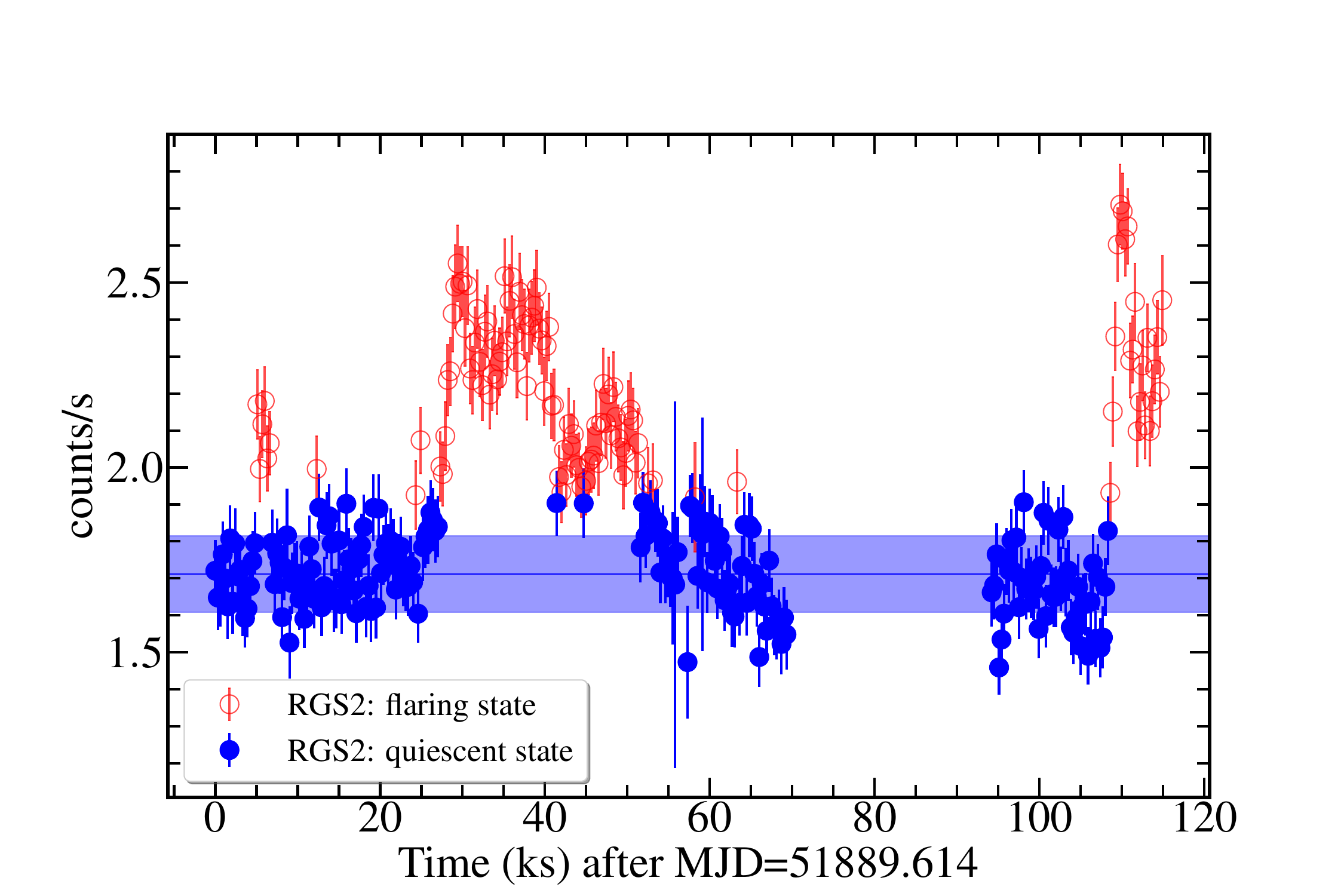}}
\subfigure[Obs ID: 0134520301 (2001-01-20)]{\includegraphics[scale=0.15]{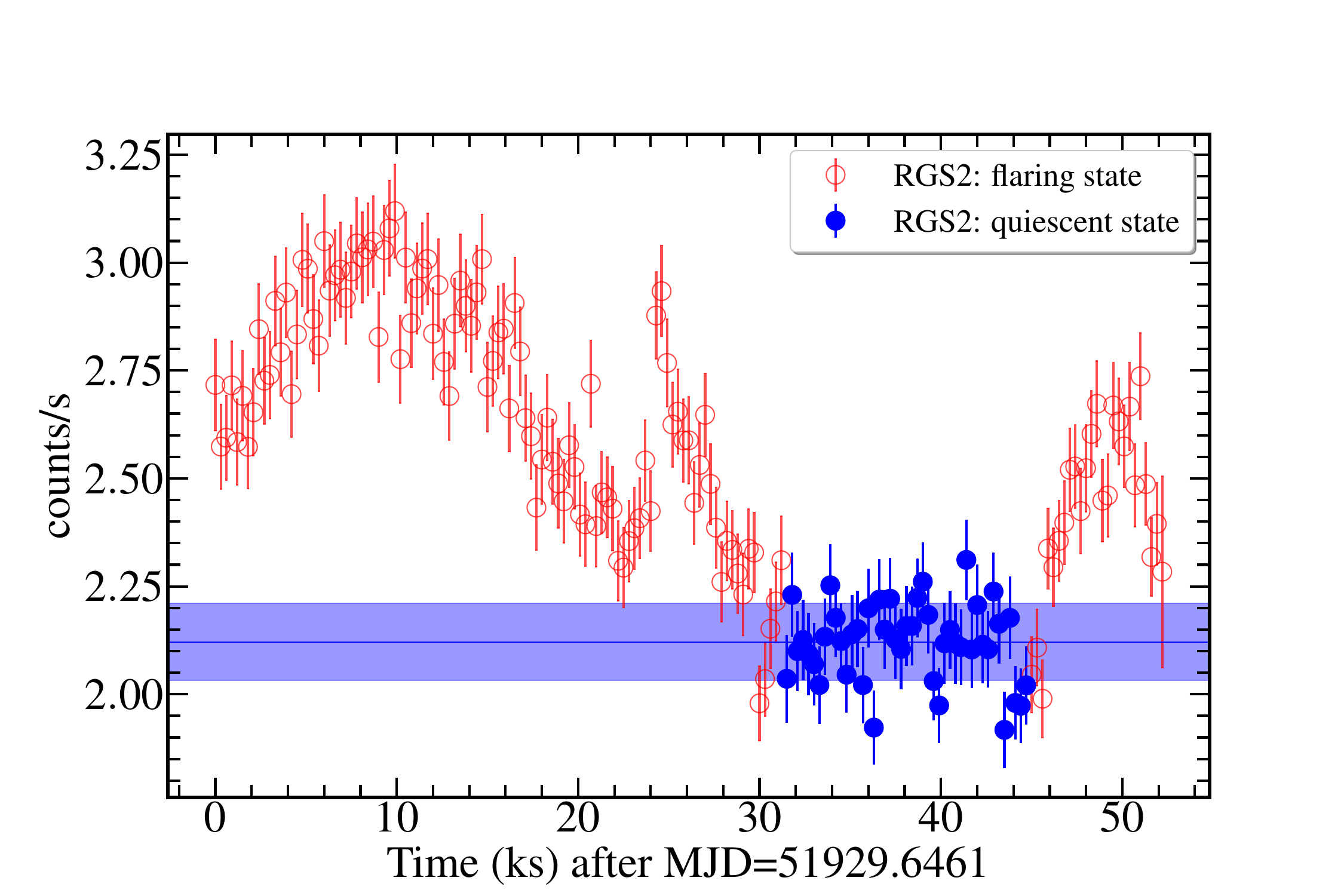}}
\subfigure[Obs ID: 0134520701 (2001-05-22)]{\includegraphics[scale=0.15]{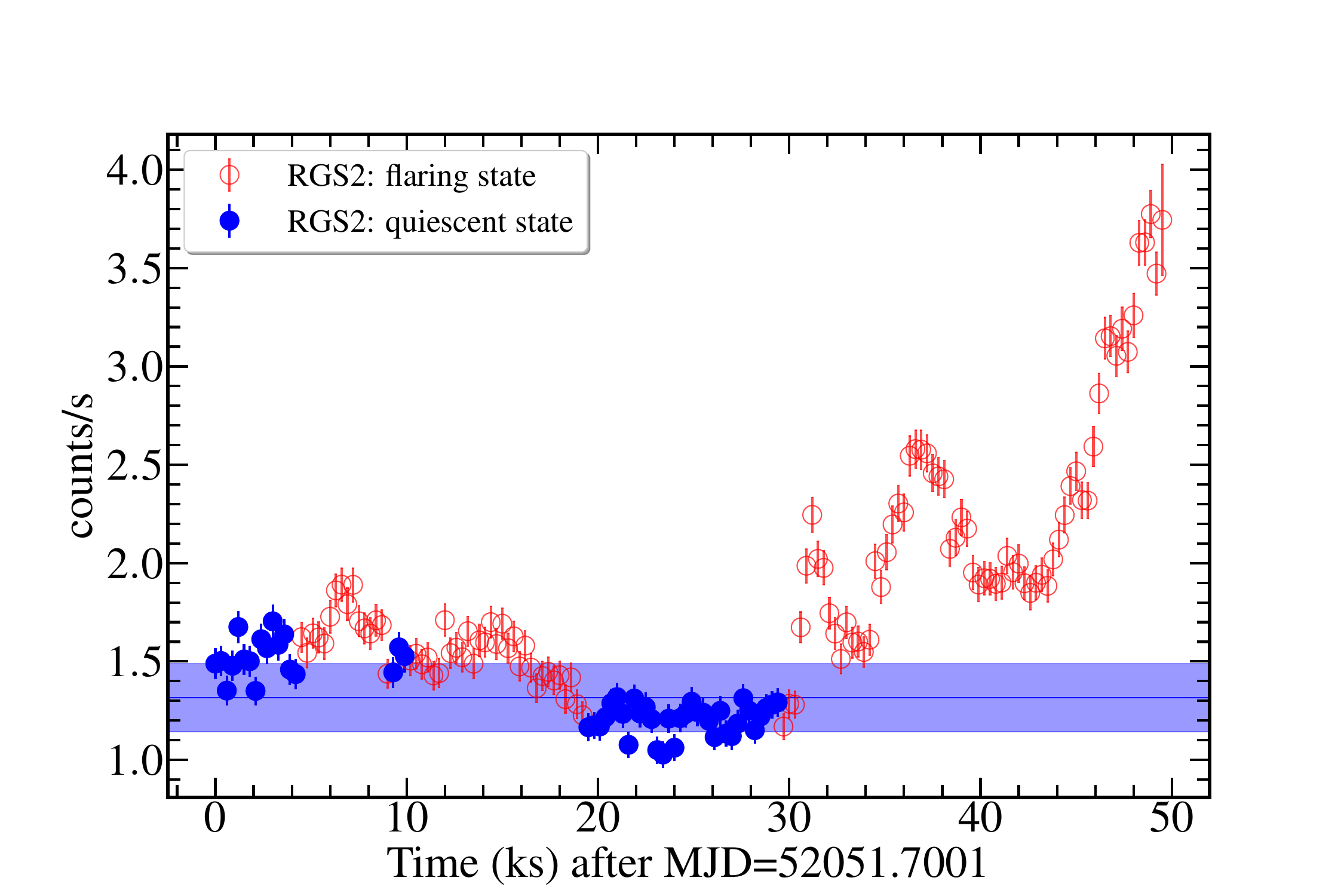}}
\subfigure[Obs ID: 0134521301 (2001-10-13)]{\includegraphics[scale=0.15]{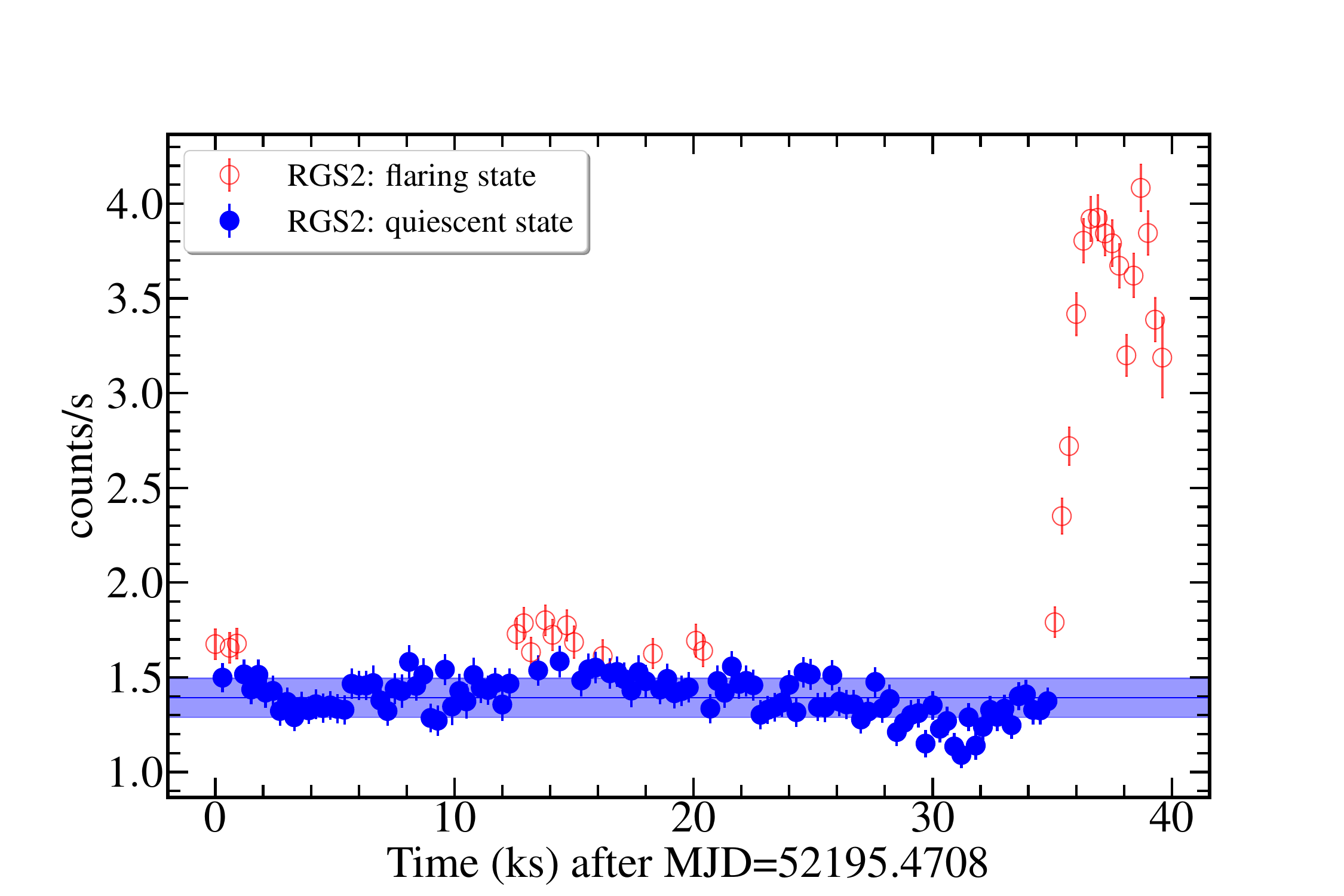}}
\subfigure[Obs ID: 0134521401 (2001-12-26)]{\includegraphics[scale=0.15]{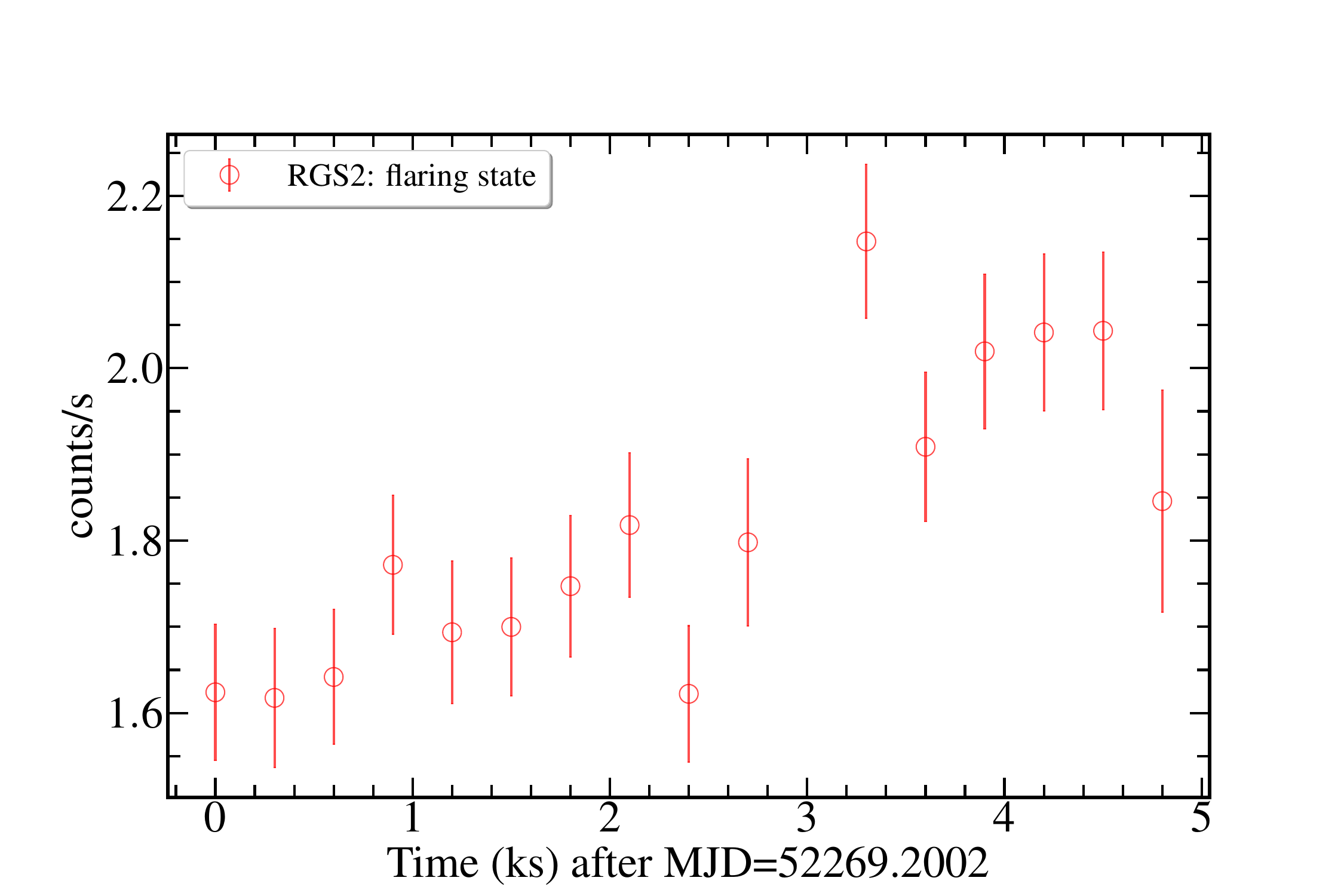}}
\subfigure[Obs ID: 0134521501 (2002-04-12)]{\includegraphics[scale=0.15]{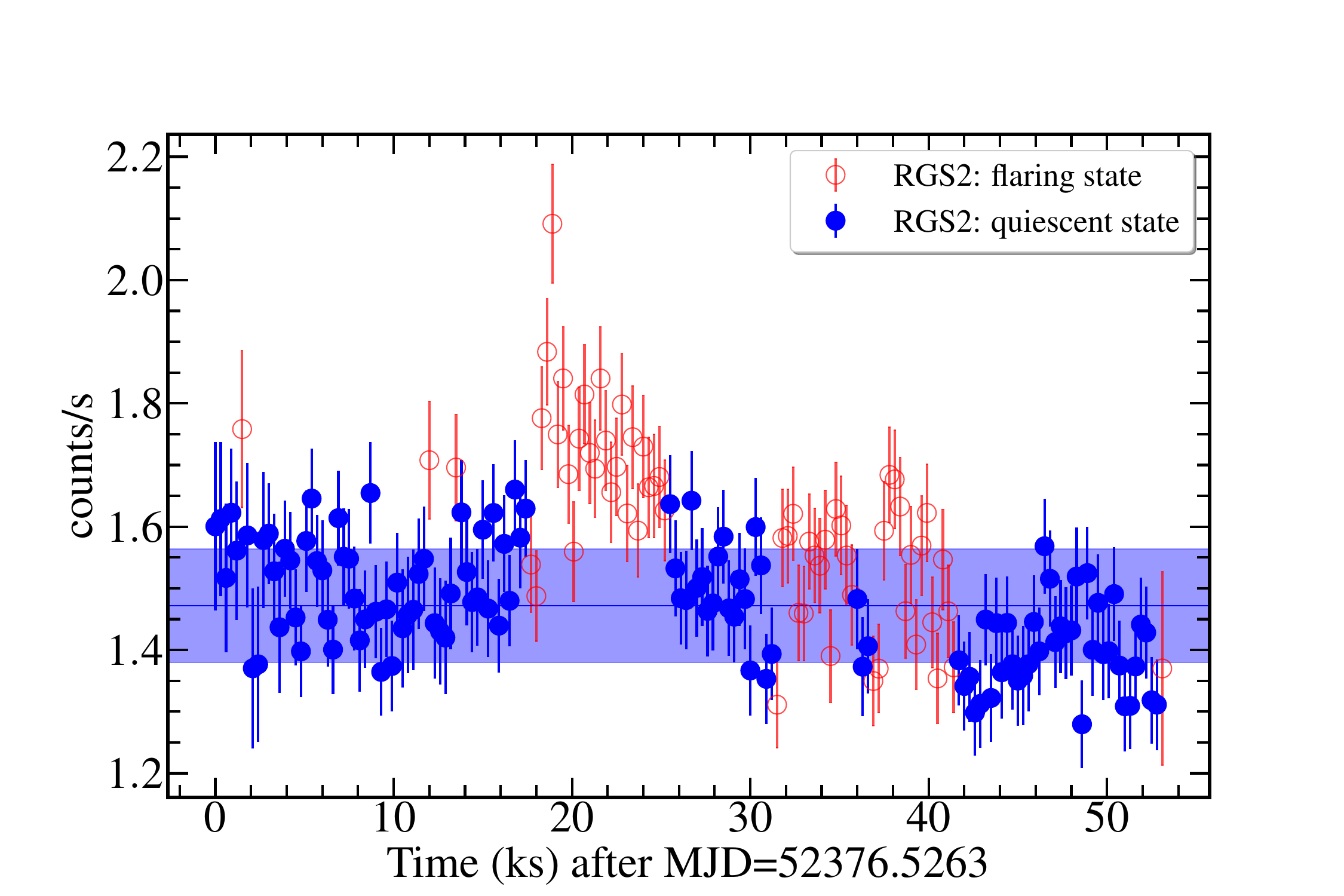}}
\subfigure[Obs ID: 0155150101+0134521601 (2002-06-18)]{\includegraphics[scale=0.15]{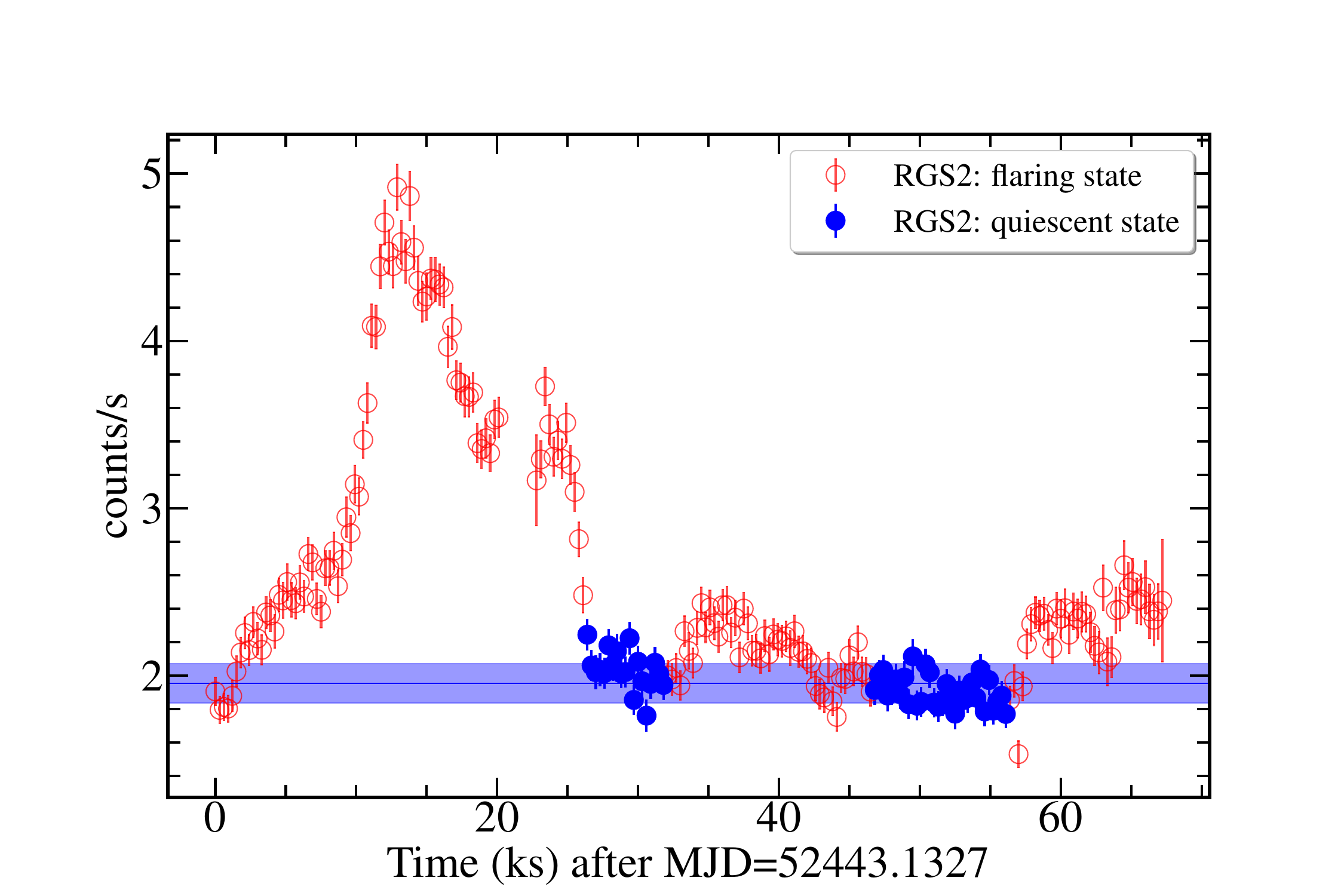}}
\subfigure[Obs ID: 0134521701 (2002-11-15)]{\includegraphics[scale=0.15]{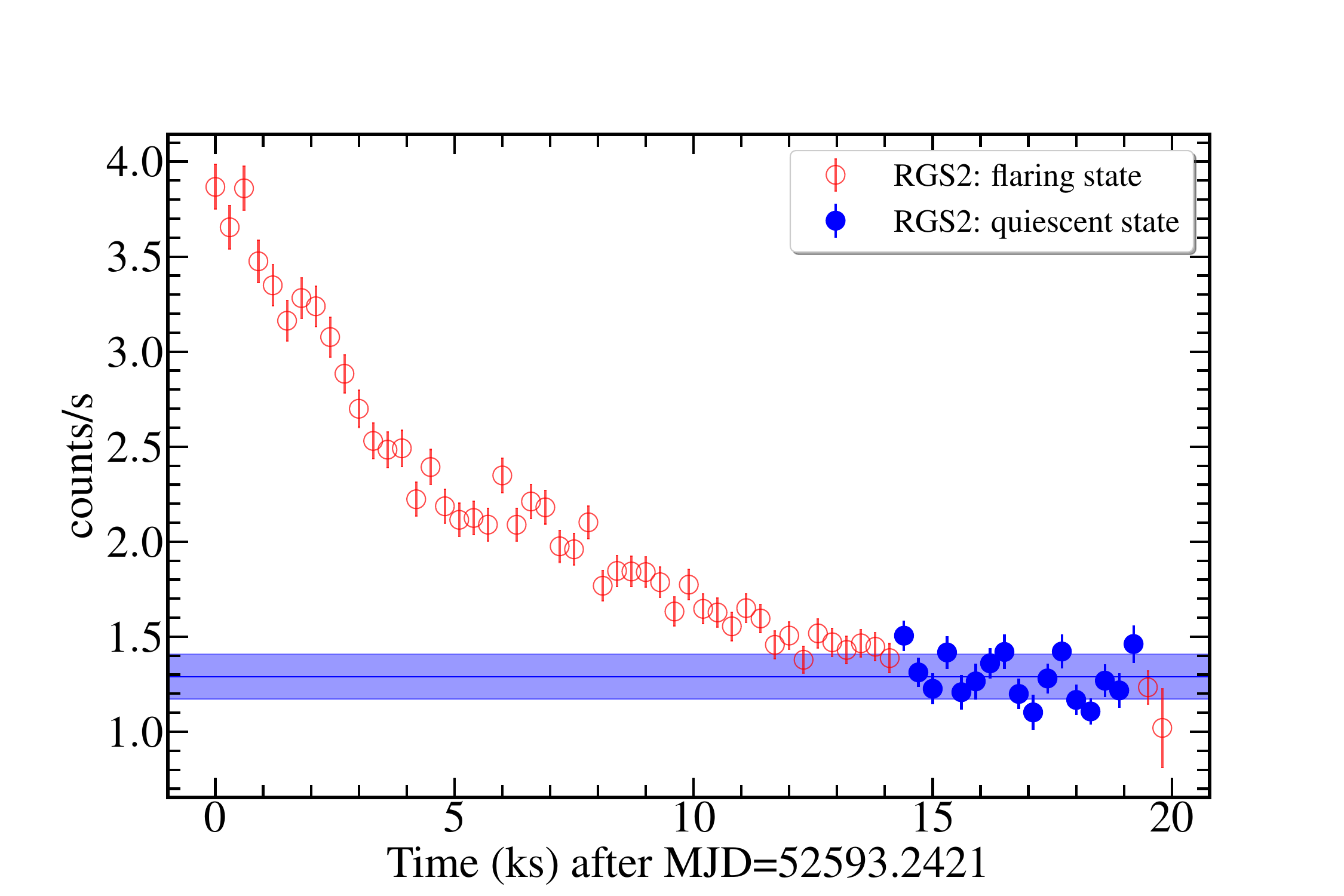}}
\subfigure[Obs ID: 0134521801 (2002-11-05)]{\includegraphics[scale=0.15]{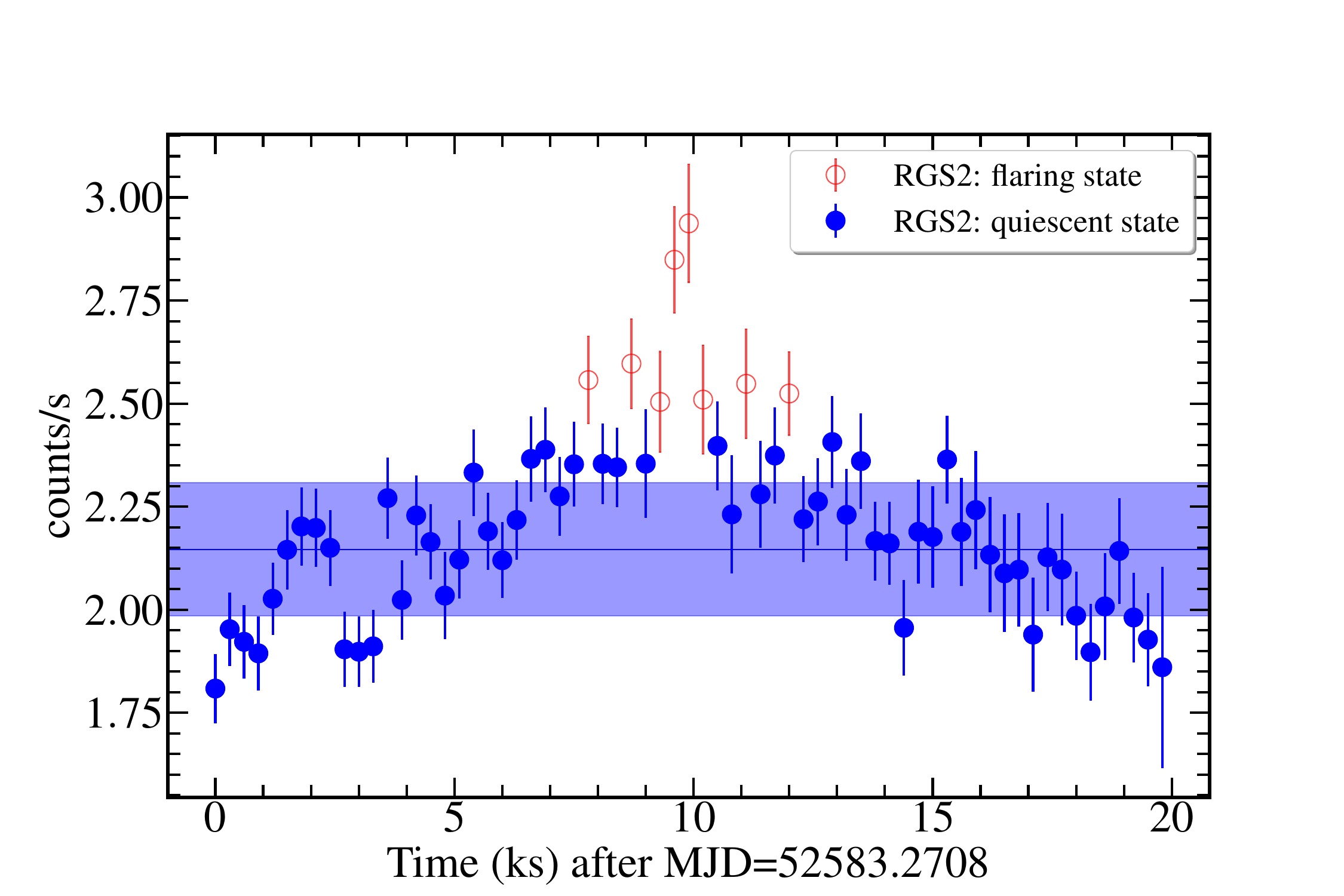}}

\caption{XMM-Newton RGS2 X-ray lightcurves of AB Dor with a 300-sec bin size. The red open circles represent the flaring state, while the blue-filled circles show the quiescent state. The blue solid line and shaded region correspond to the mean quiescent state and standard deviation, respectively.}
    \label{fig:abdor_flare_qui1}
\end{figure*}
\setcounter{figure}{0}  
\begin{figure*}
\setcounter{subfigure}{12}
\centering
\subfigure[Obs ID: 0134522001 (2002-12-03)]{\includegraphics[scale=0.15]{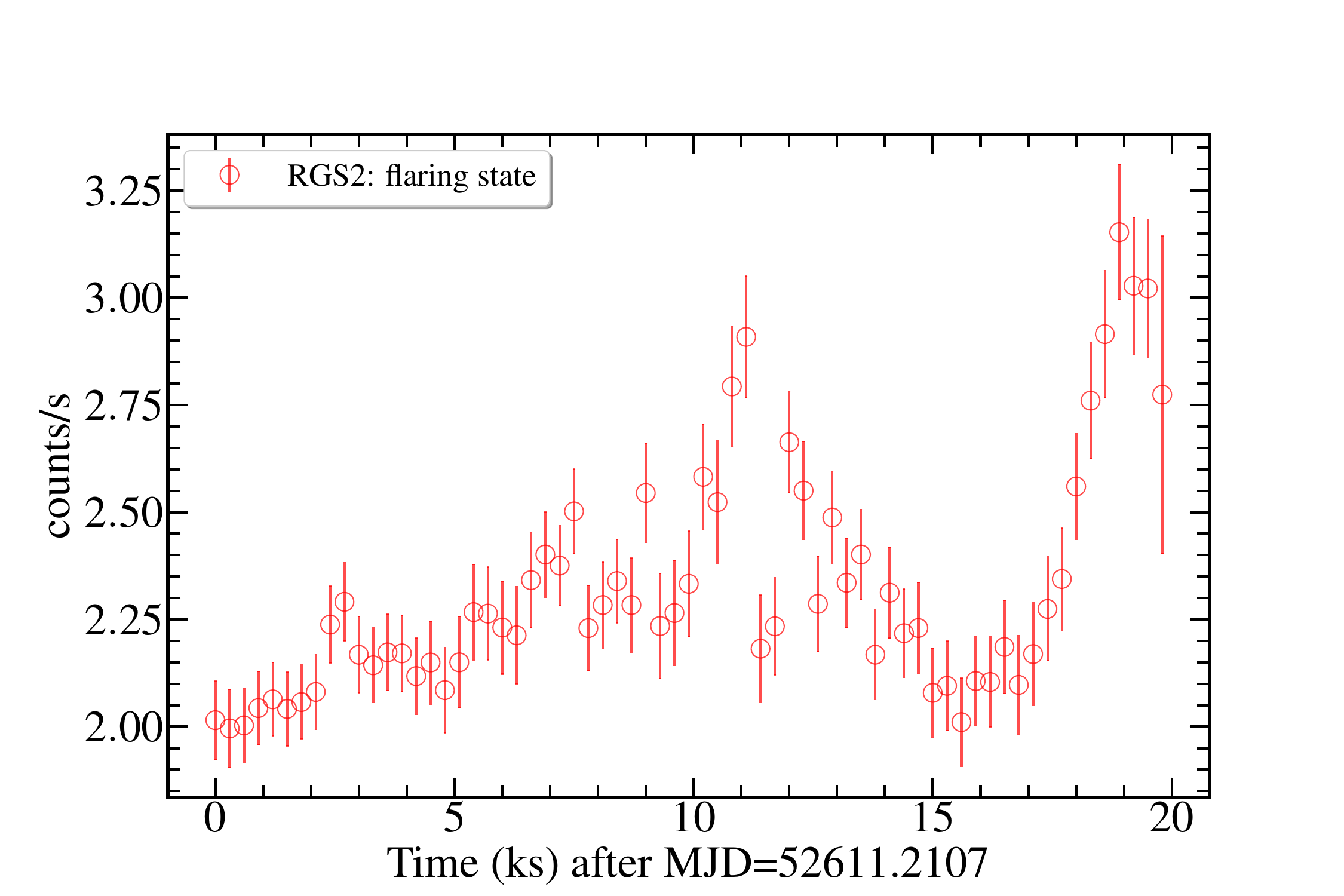}}
\subfigure[Obs ID: 0134522101 (2002-12-30)]{\includegraphics[scale=0.15]{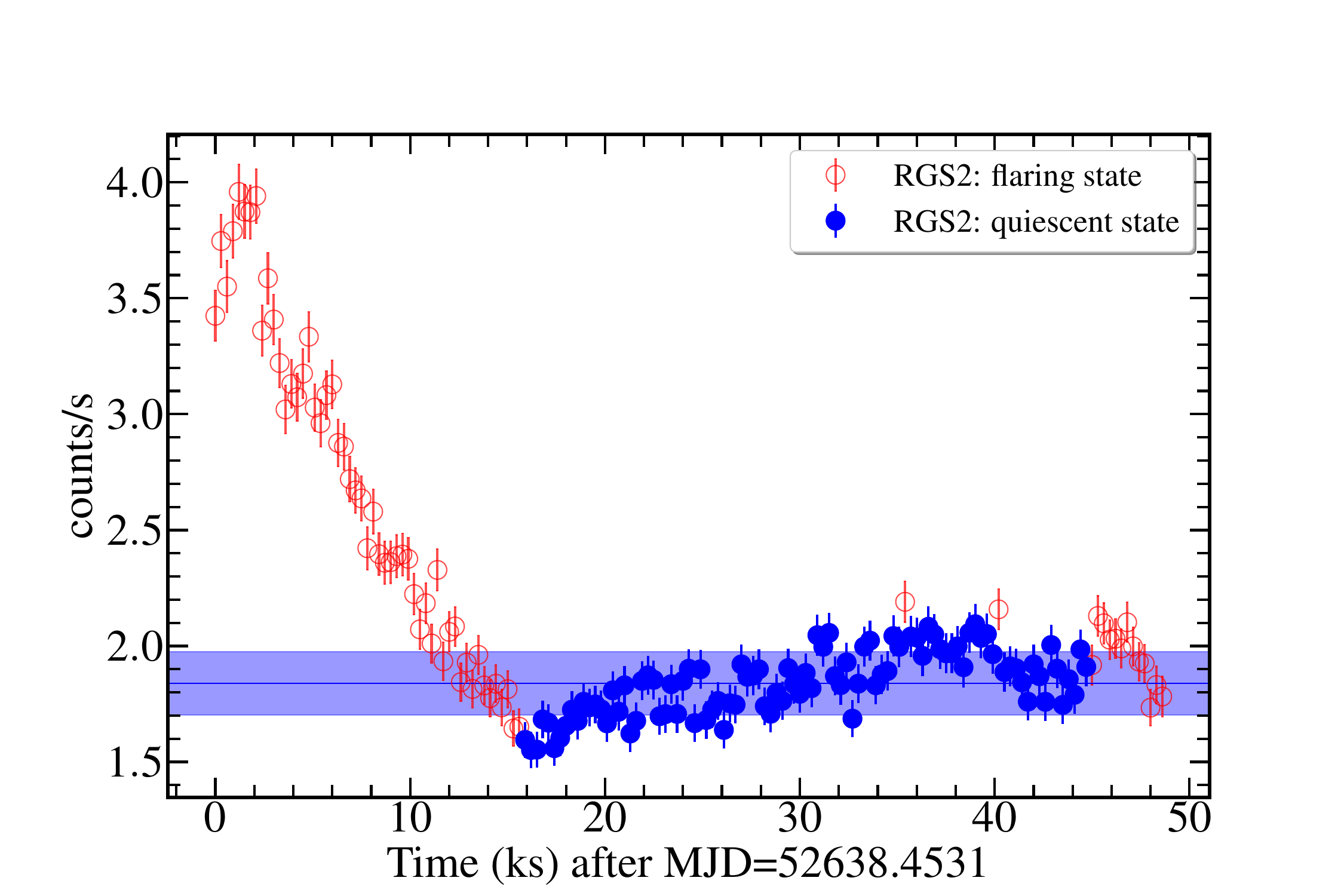}}
\subfigure[Obs ID: 0134522201 (2003-01-23)]{\includegraphics[scale=0.15]{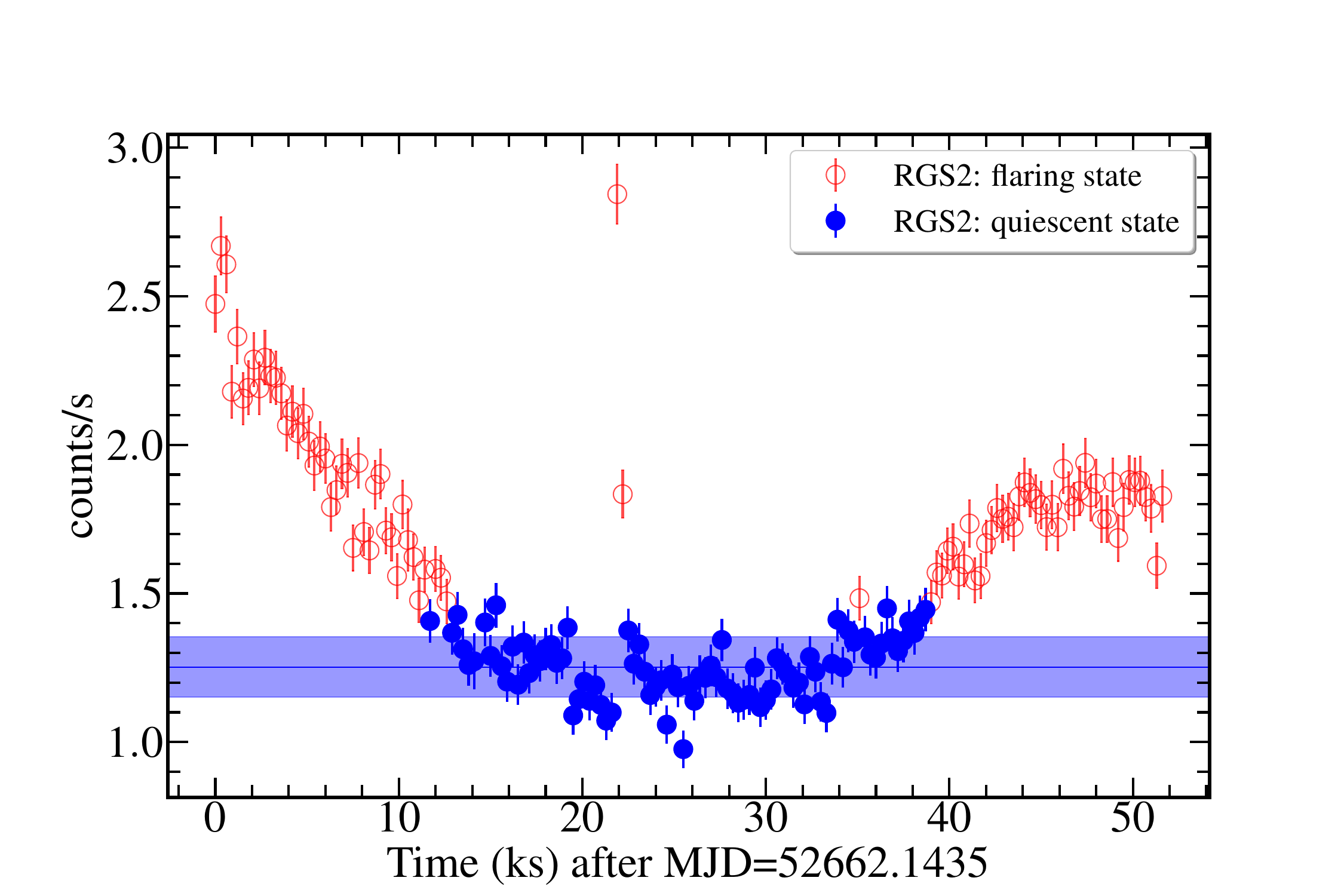}}
\subfigure[Obs ID: 0134522301 (2003-03-30)]{\includegraphics[scale=0.15]{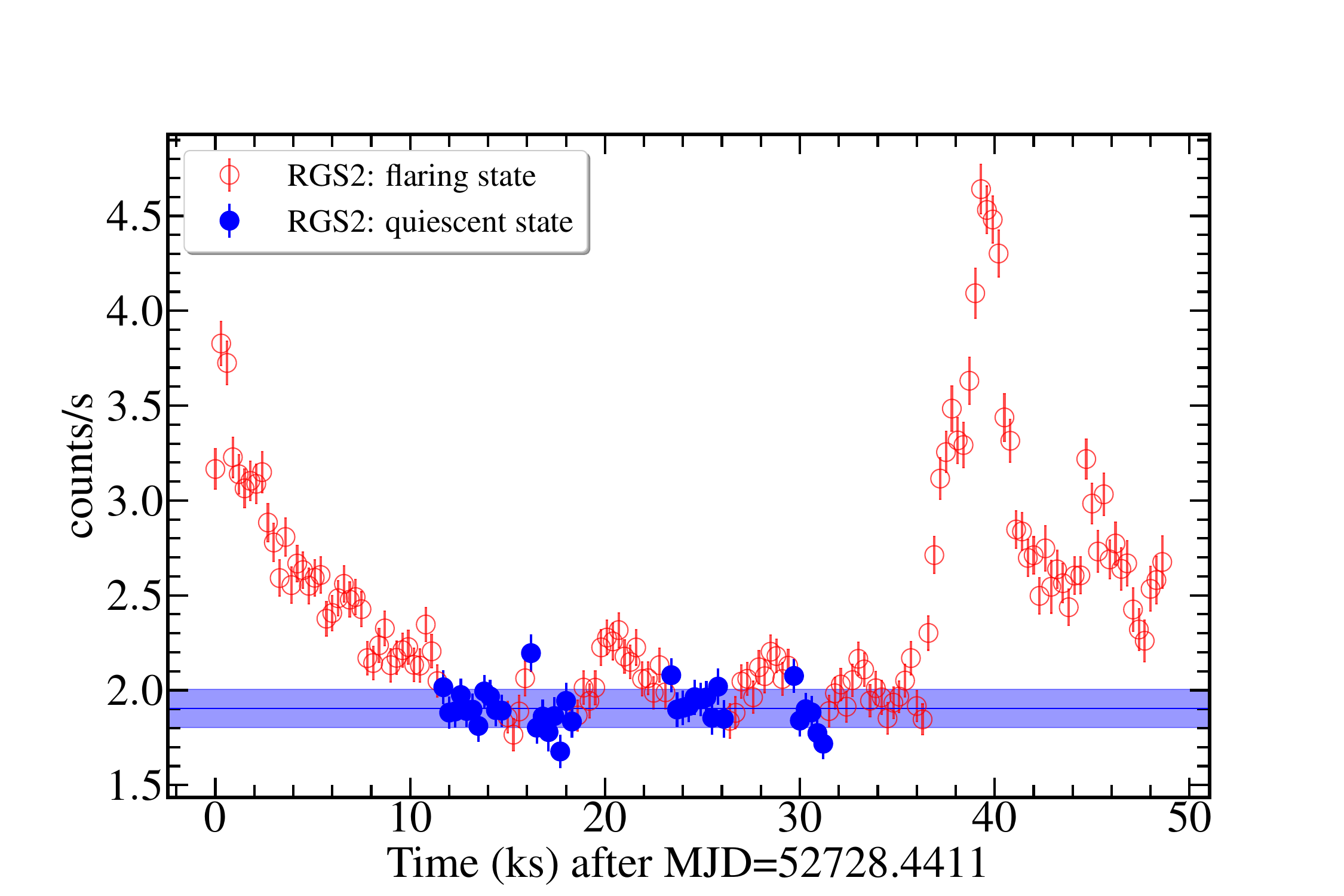}}
\subfigure[Obs ID: 0134522401 (2003-05-31)]{\includegraphics[scale=0.15]{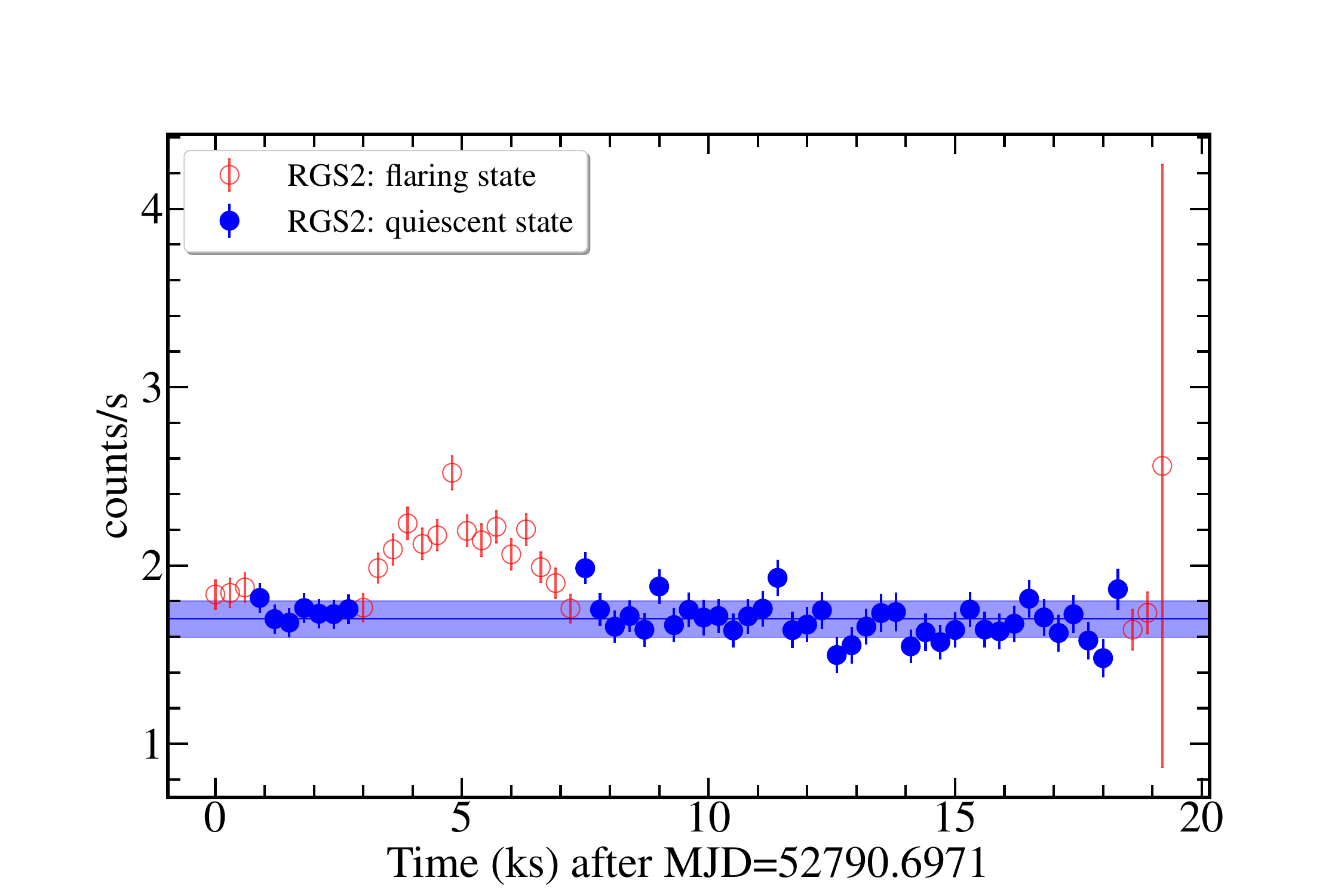}}
\subfigure[Obs ID: 0160362501+0160362601 (2003-08-02)]{\includegraphics[scale=0.15]{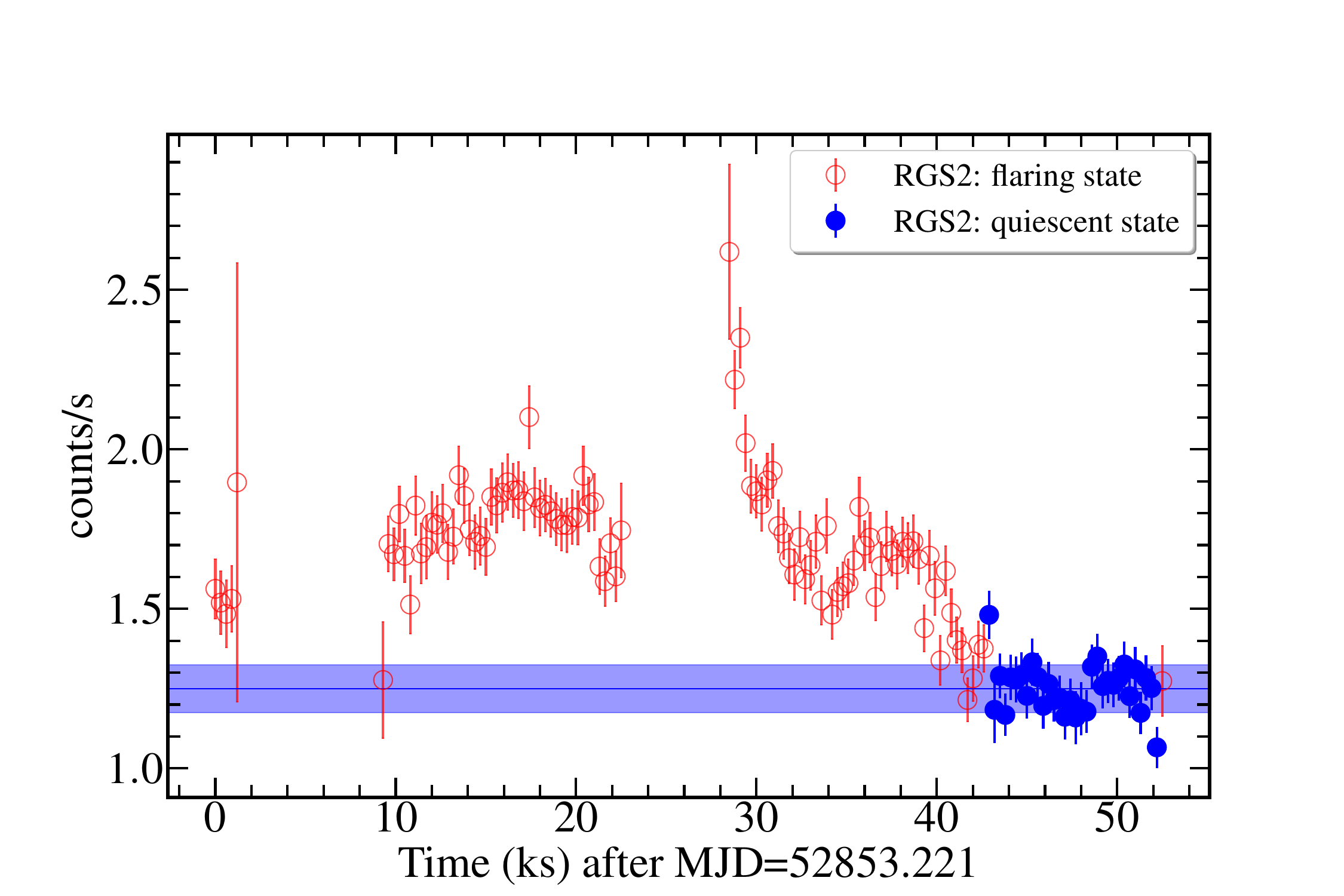}}
\subfigure[Obs ID: 0160362701 (2003-10-23)]{\includegraphics[scale=0.15]{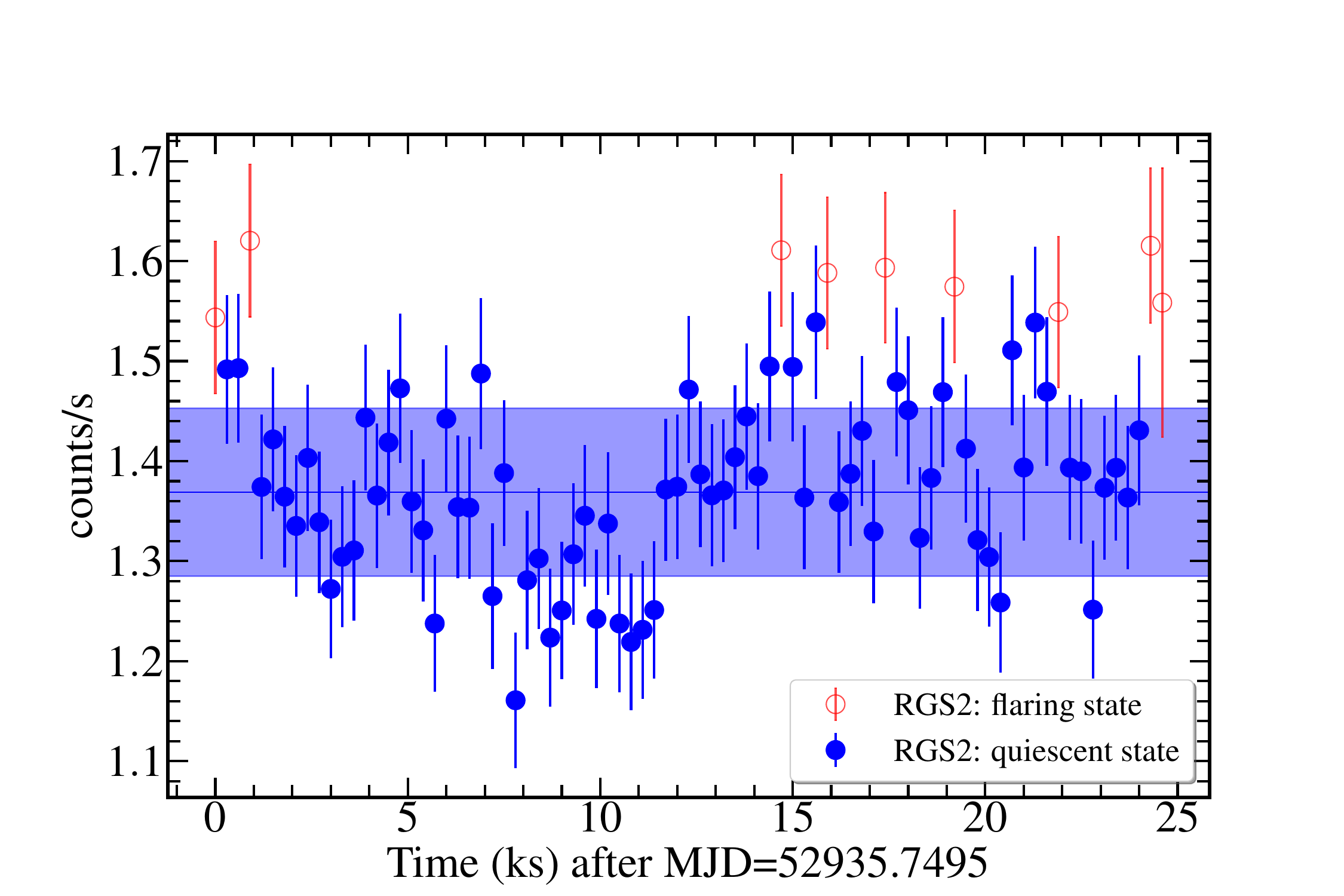}}
\subfigure[Obs ID: 0160362801 (2003-12-08)]{\includegraphics[scale=0.15]{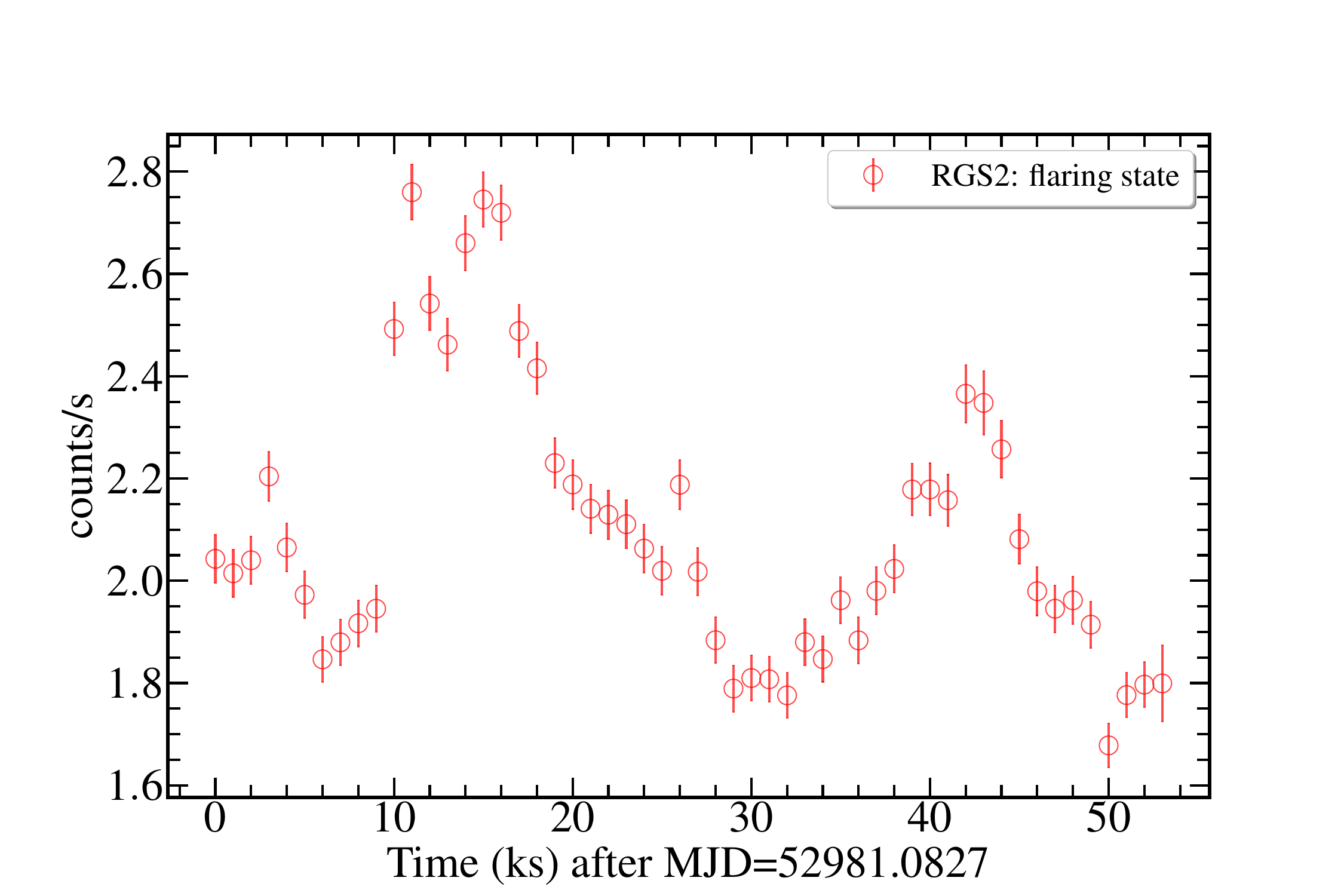}}
\subfigure[Obs ID: 0160362901 (2004-11-27)]{\includegraphics[scale=0.15]{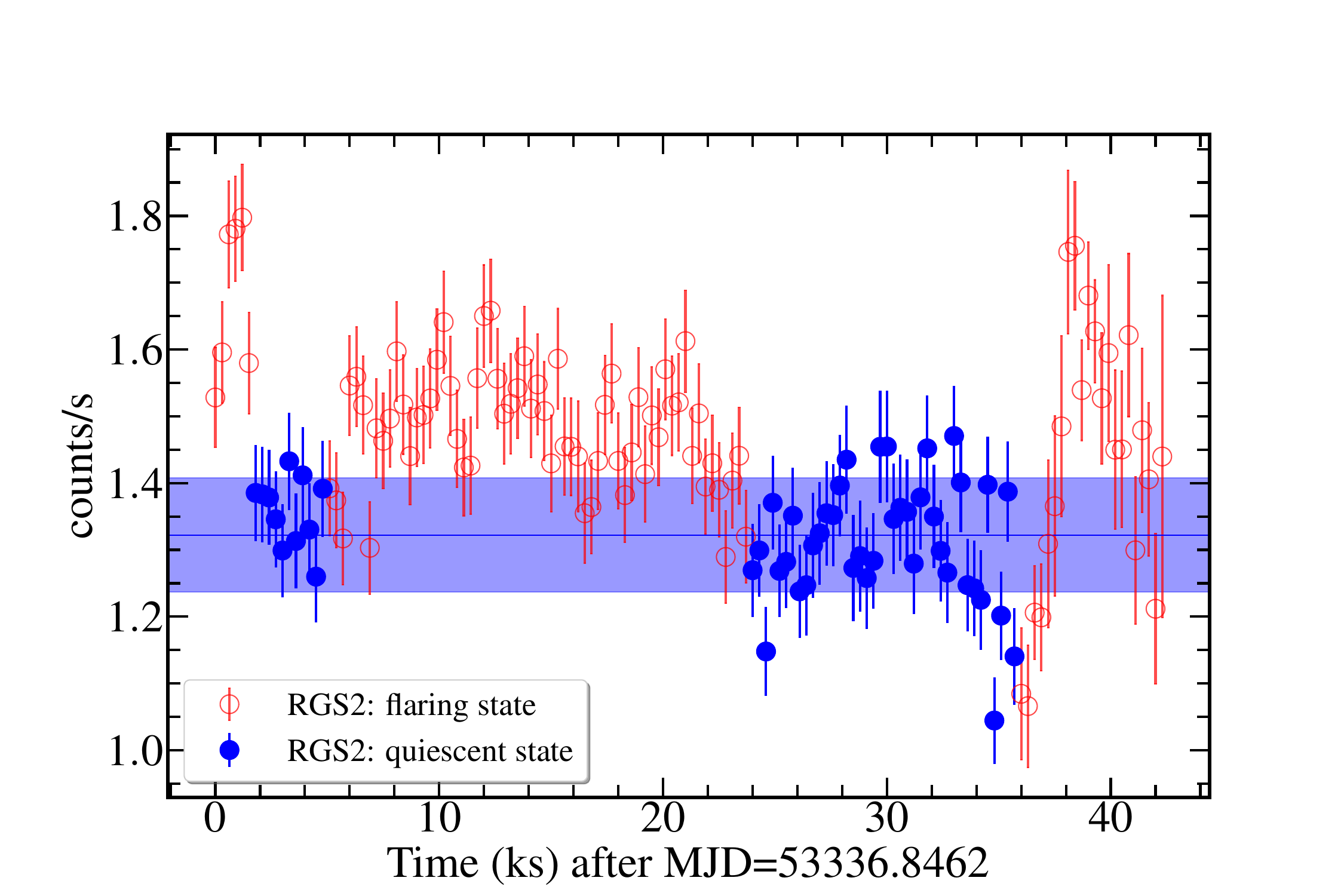}}
\subfigure[Obs ID: 0160363001 (2005-04-18)]{\includegraphics[scale=0.15]{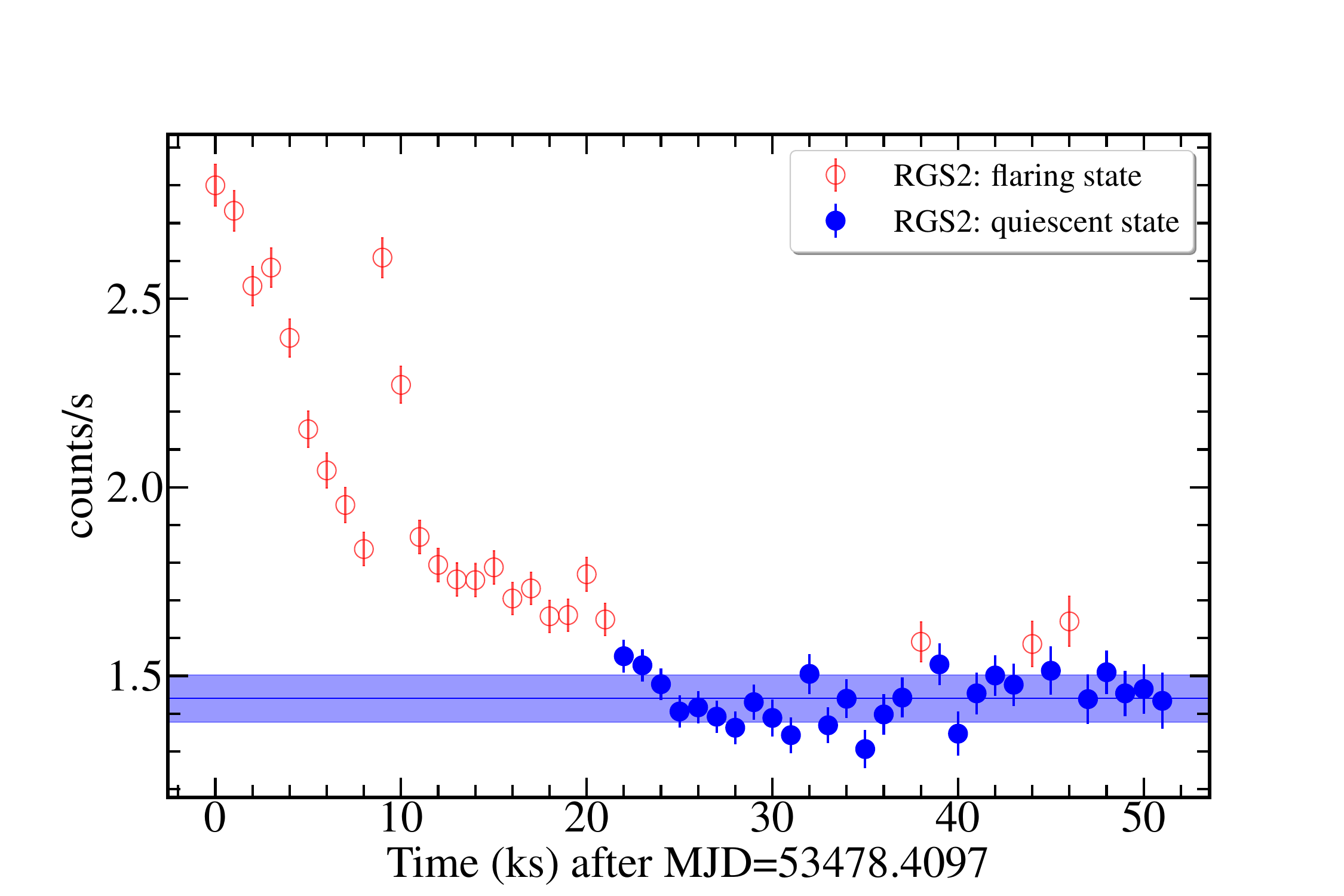}}
\subfigure[Obs ID: 0160363201 (2005-10-16)]{\includegraphics[scale=0.15]{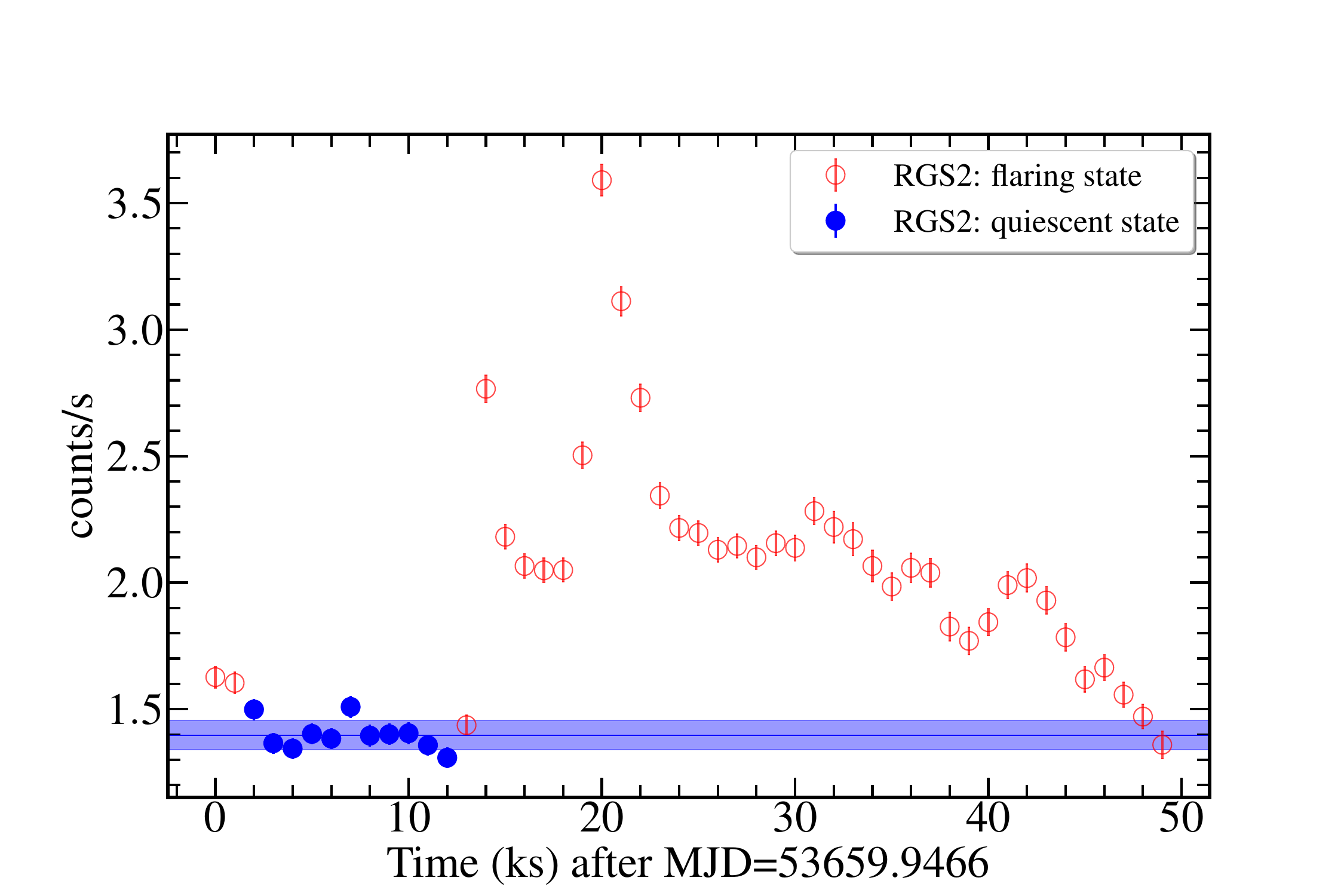}}
\subfigure[Obs ID: 0412580201 (2007-07-19)]{\includegraphics[scale=0.15]{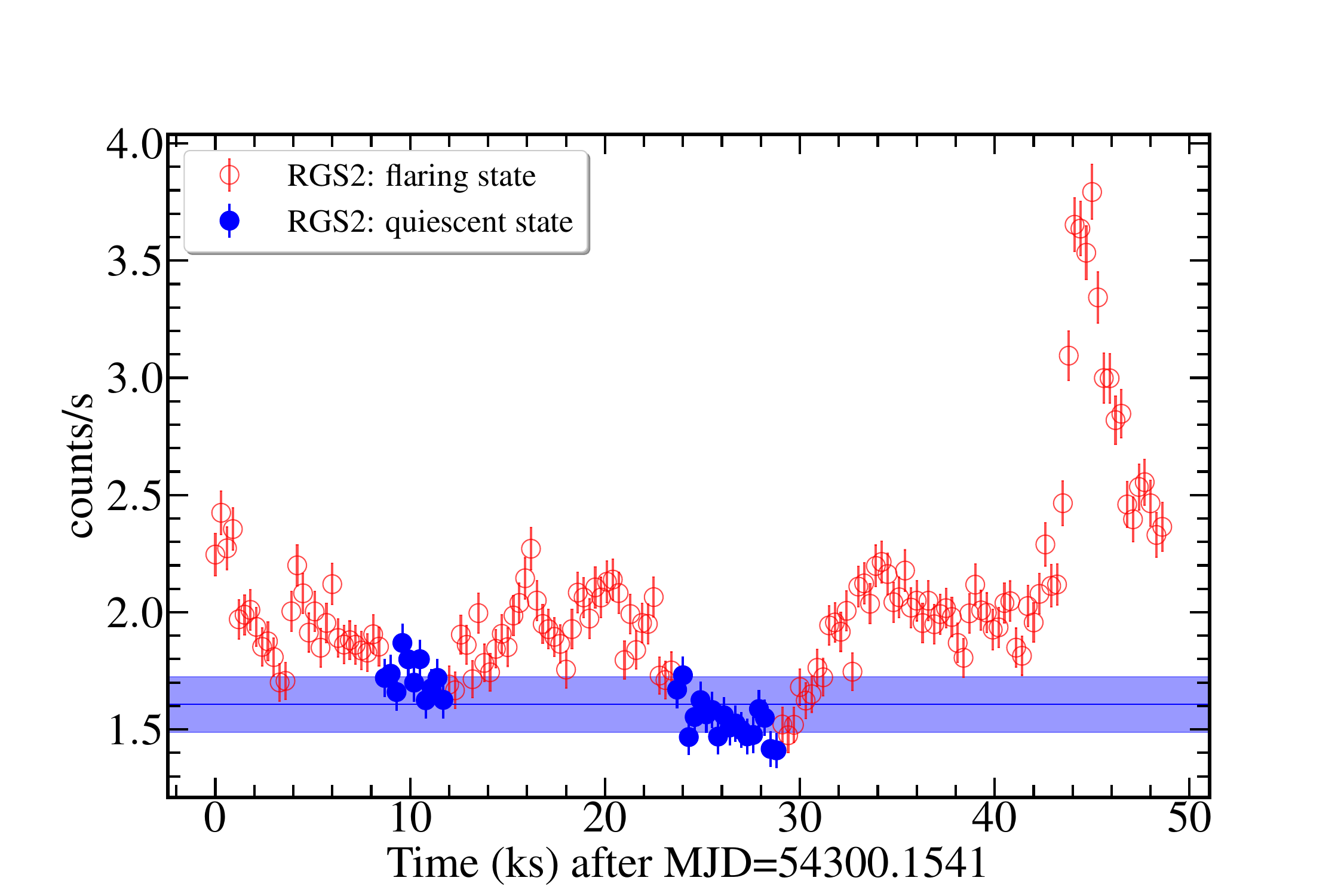}}
\subfigure[Obs ID: 0412580301 (2008-01-03)]{\includegraphics[scale=0.15]{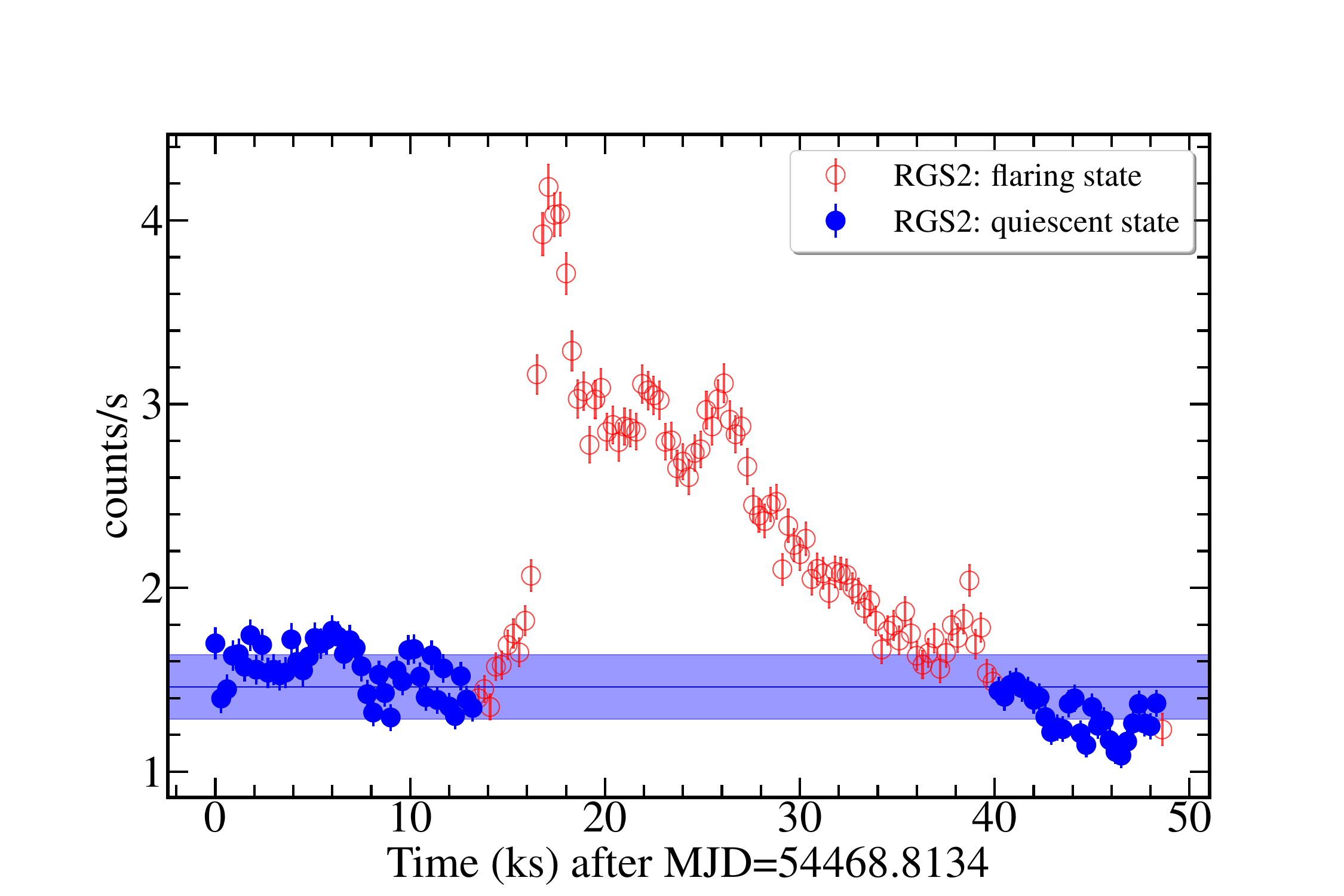}}
\subfigure[Obs ID: 0412580401 (2009-01-04]{\includegraphics[scale=0.15]{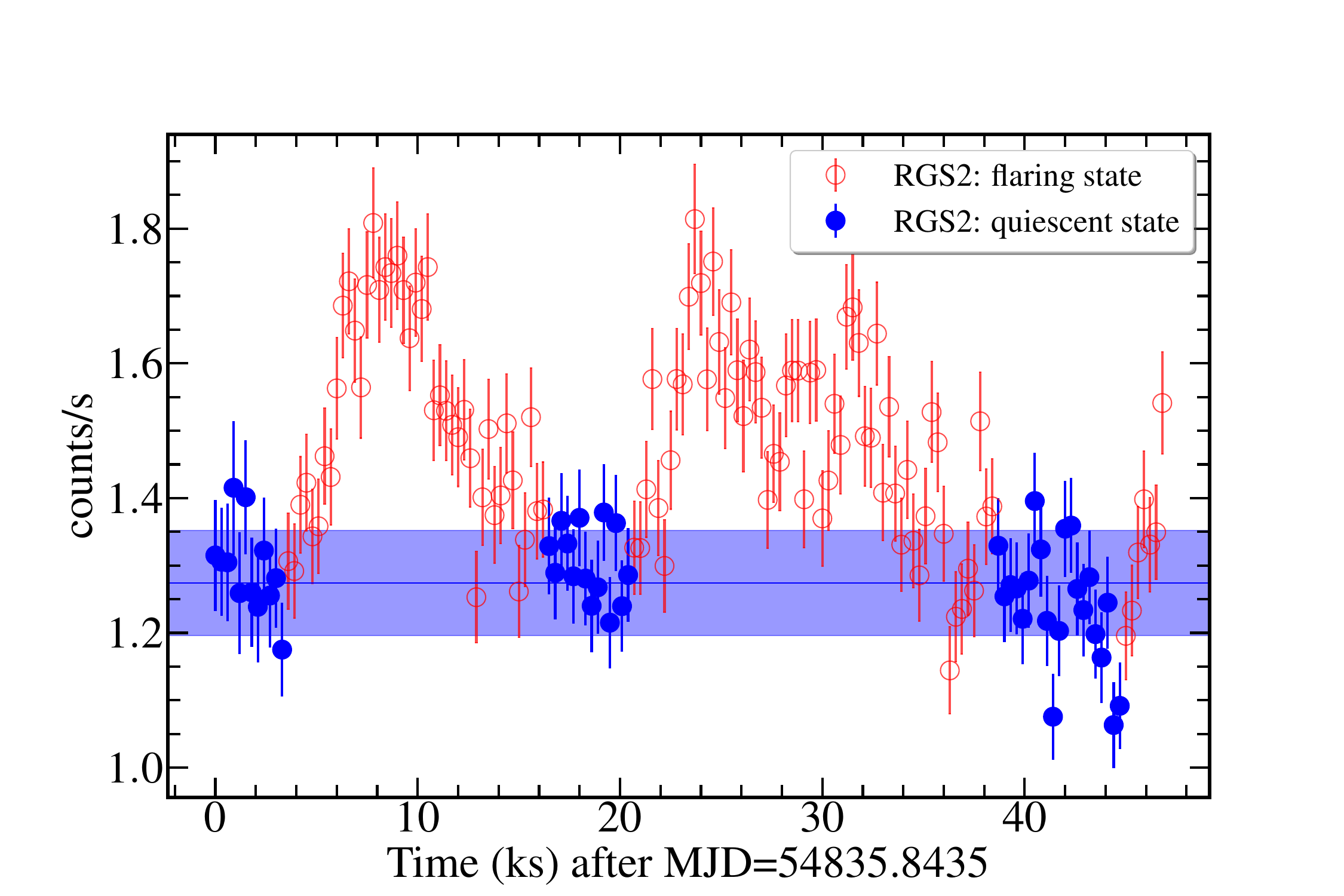}}
\subfigure[Obs ID: 0602240201 (2009-11-25)]{\includegraphics[scale=0.15]{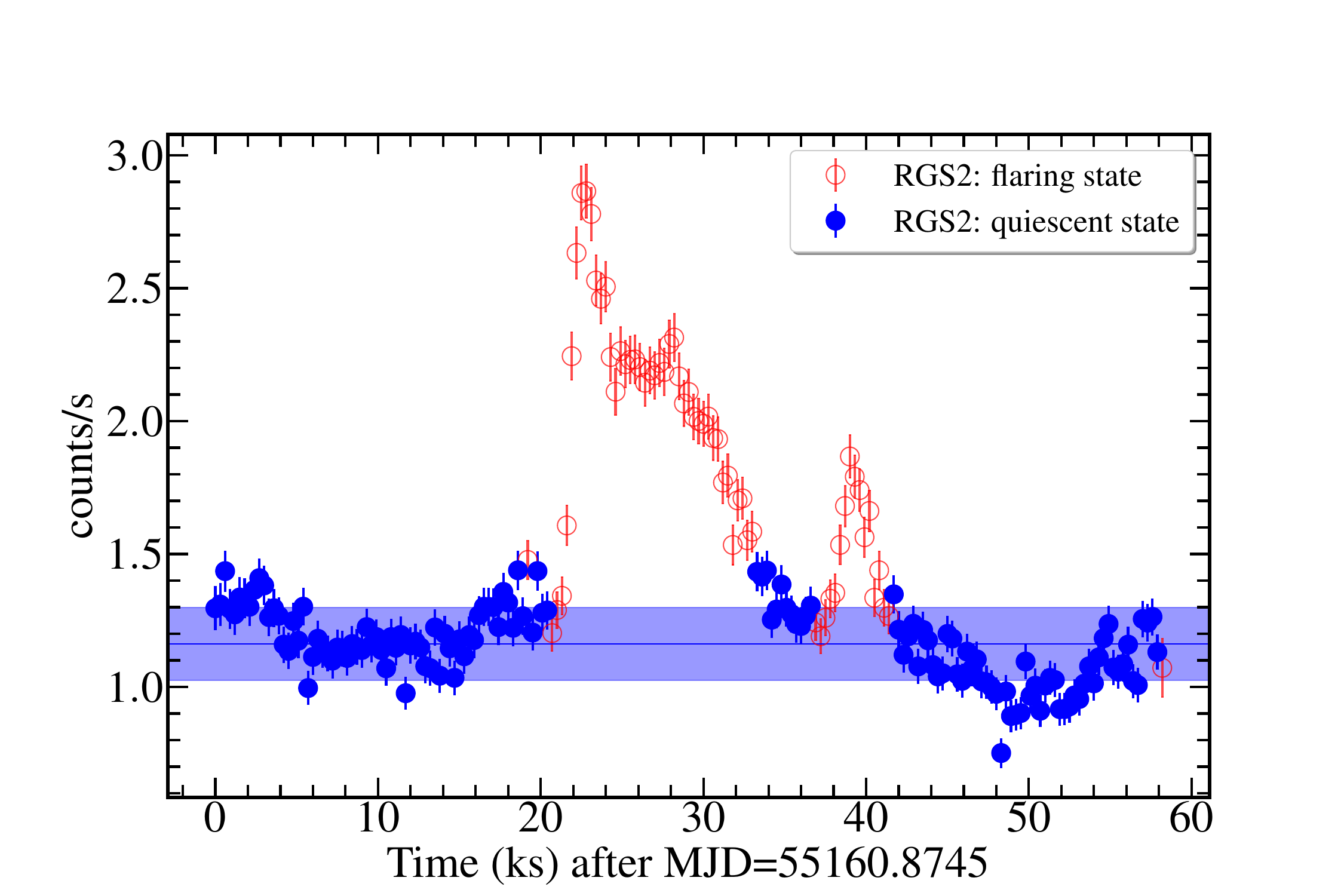}}

\caption{Continued}
    \label{fig:abdor_flare_qui2}
\end{figure*}
\setcounter{figure}{0}  
\begin{figure*}[]
\setcounter{subfigure}{27}
\centering
\subfigure[Obs ID: 0412580601 (2010-01-11)]{\includegraphics[scale=0.15]{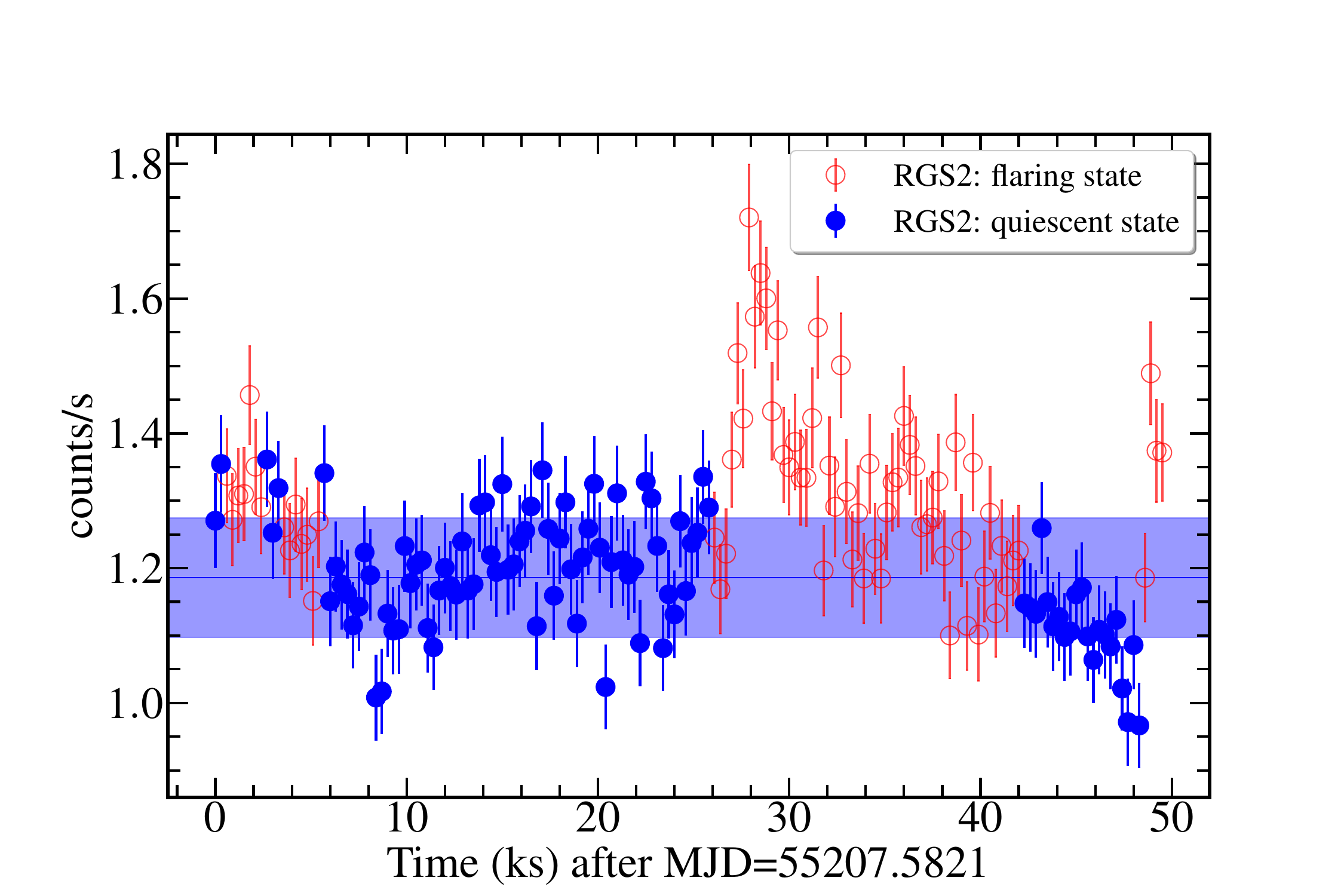}}
\subfigure[Obs ID: 0412580701 (2011-01-02)]{\includegraphics[scale=0.15]{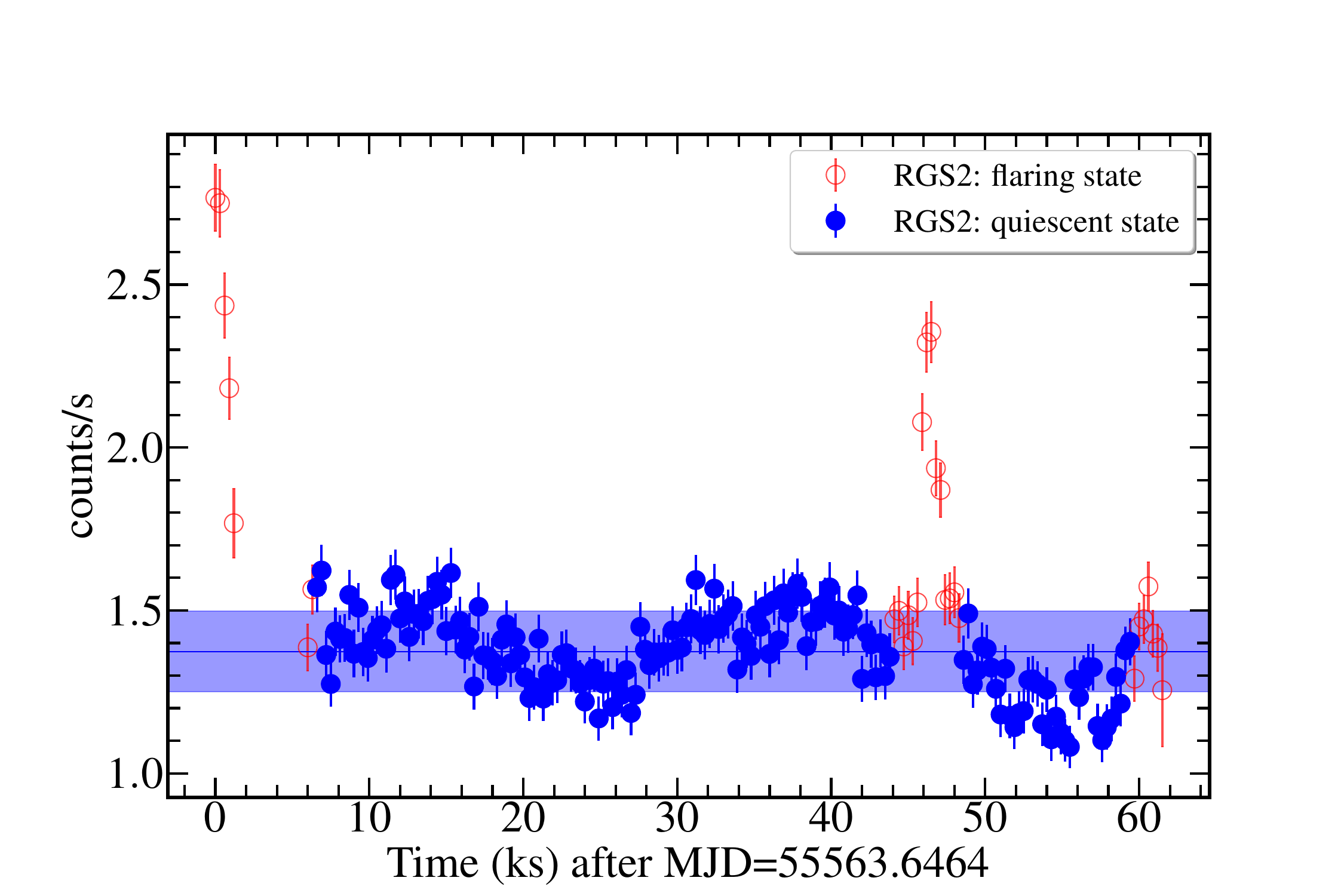}}
\subfigure[Obs ID: 0412580801 (2011-12-31)]{\includegraphics[scale=0.15]{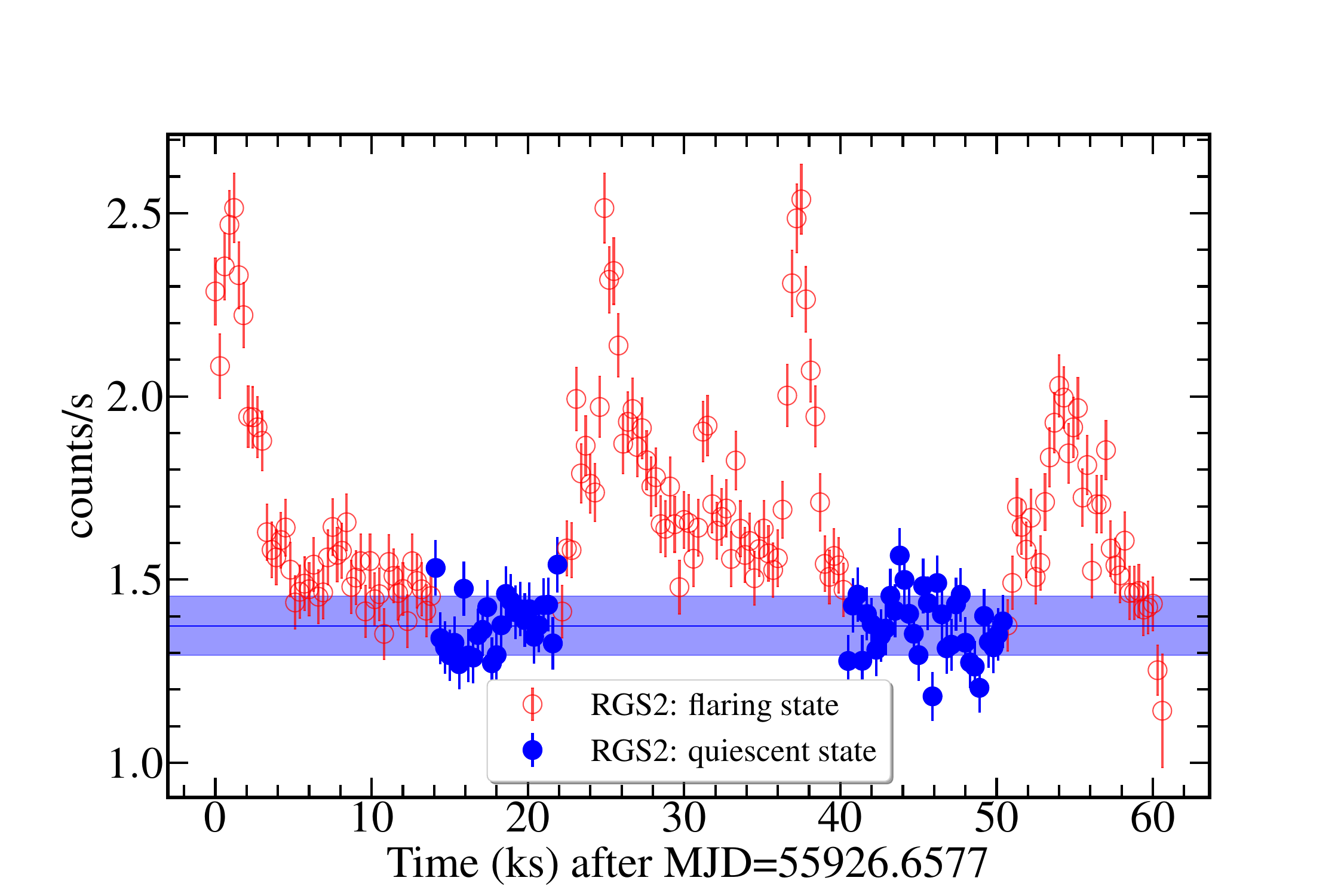}}
\subfigure[Obs ID: 0791980101 (2016-10-07)]{\includegraphics[scale=0.15]{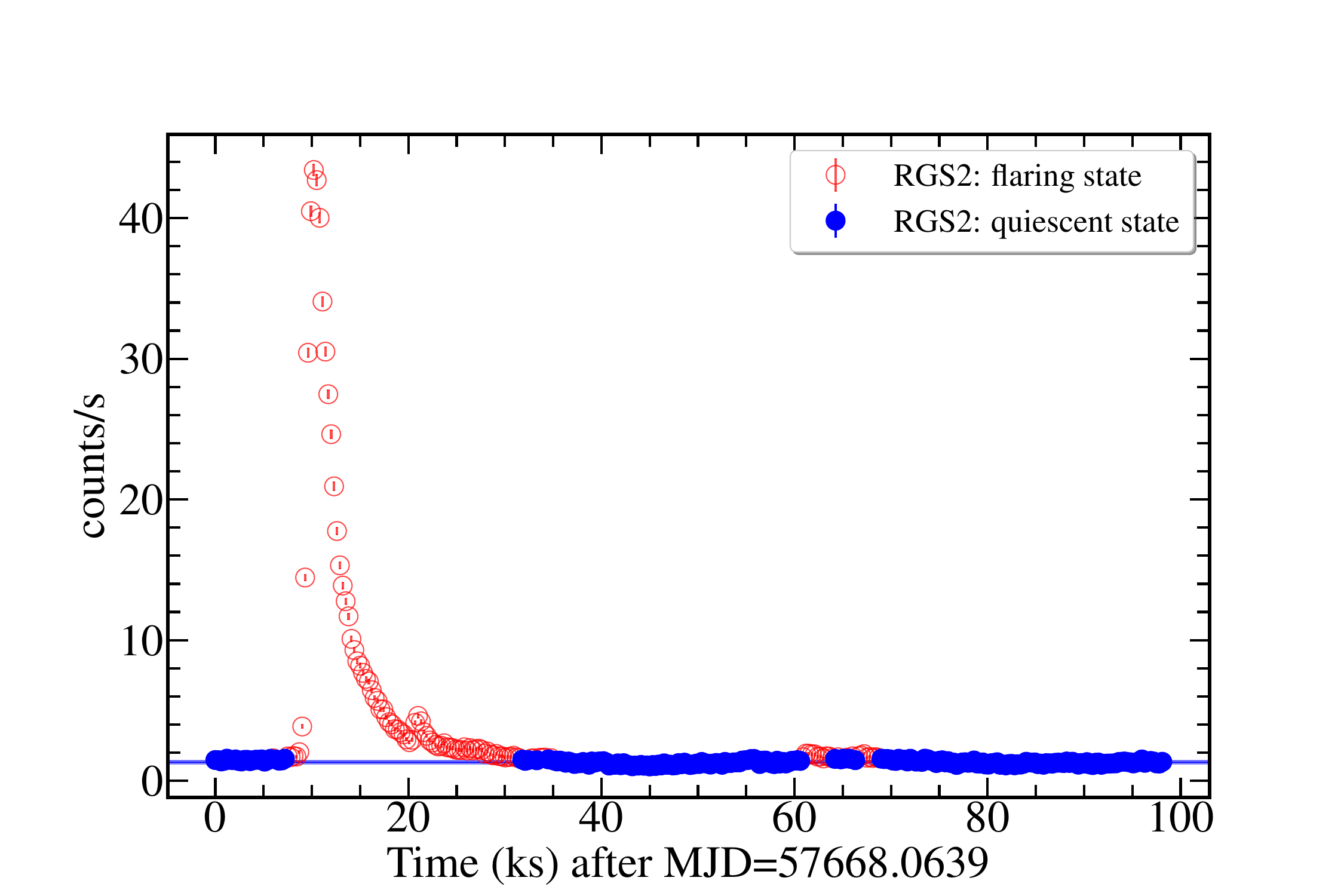}}
\subfigure[Obs ID: 0791980401 (2017-10-10)]{\includegraphics[scale=0.15]{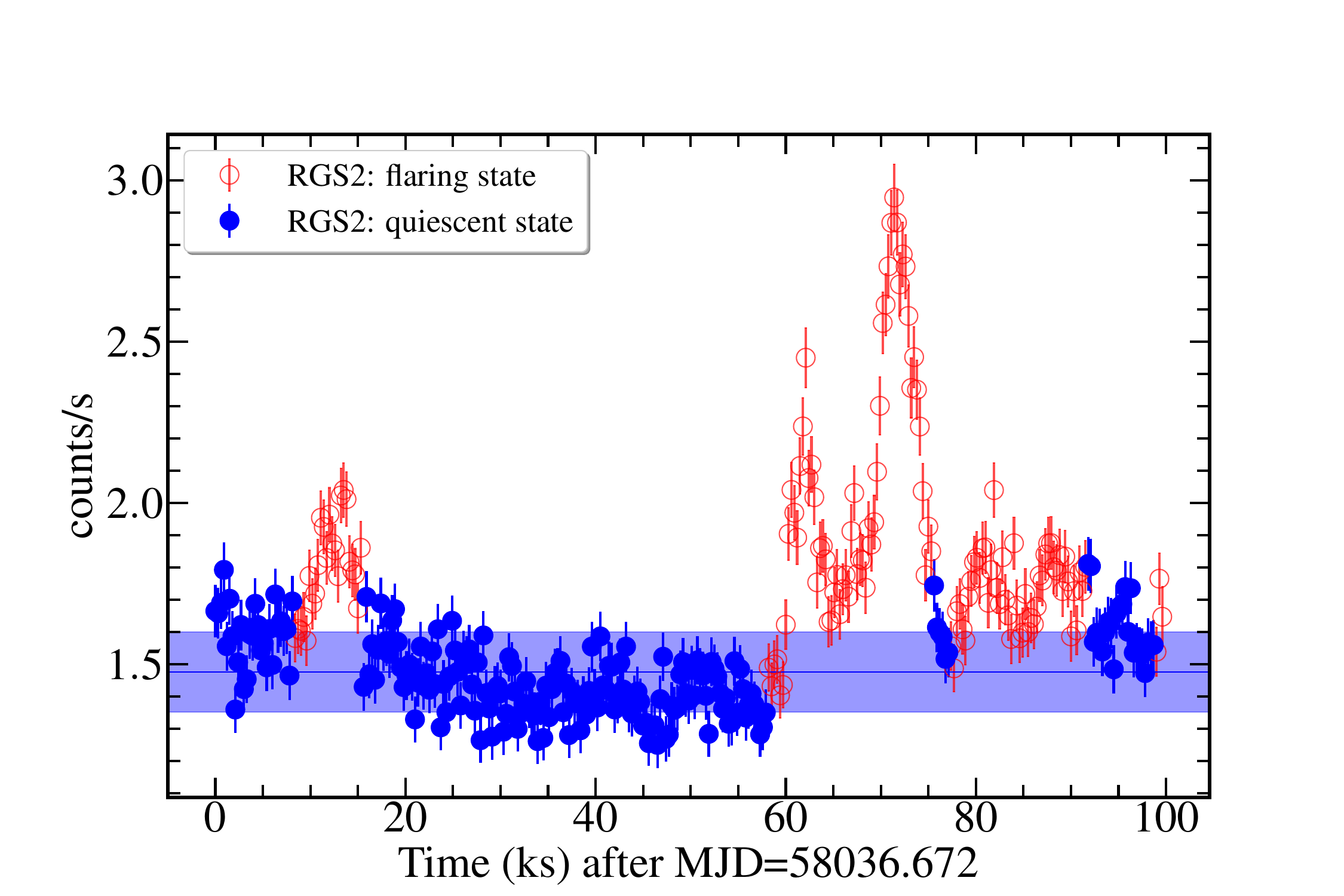}}
\subfigure[Obs ID: 0810850101 (2018-10-02)]{\includegraphics[scale=0.15]{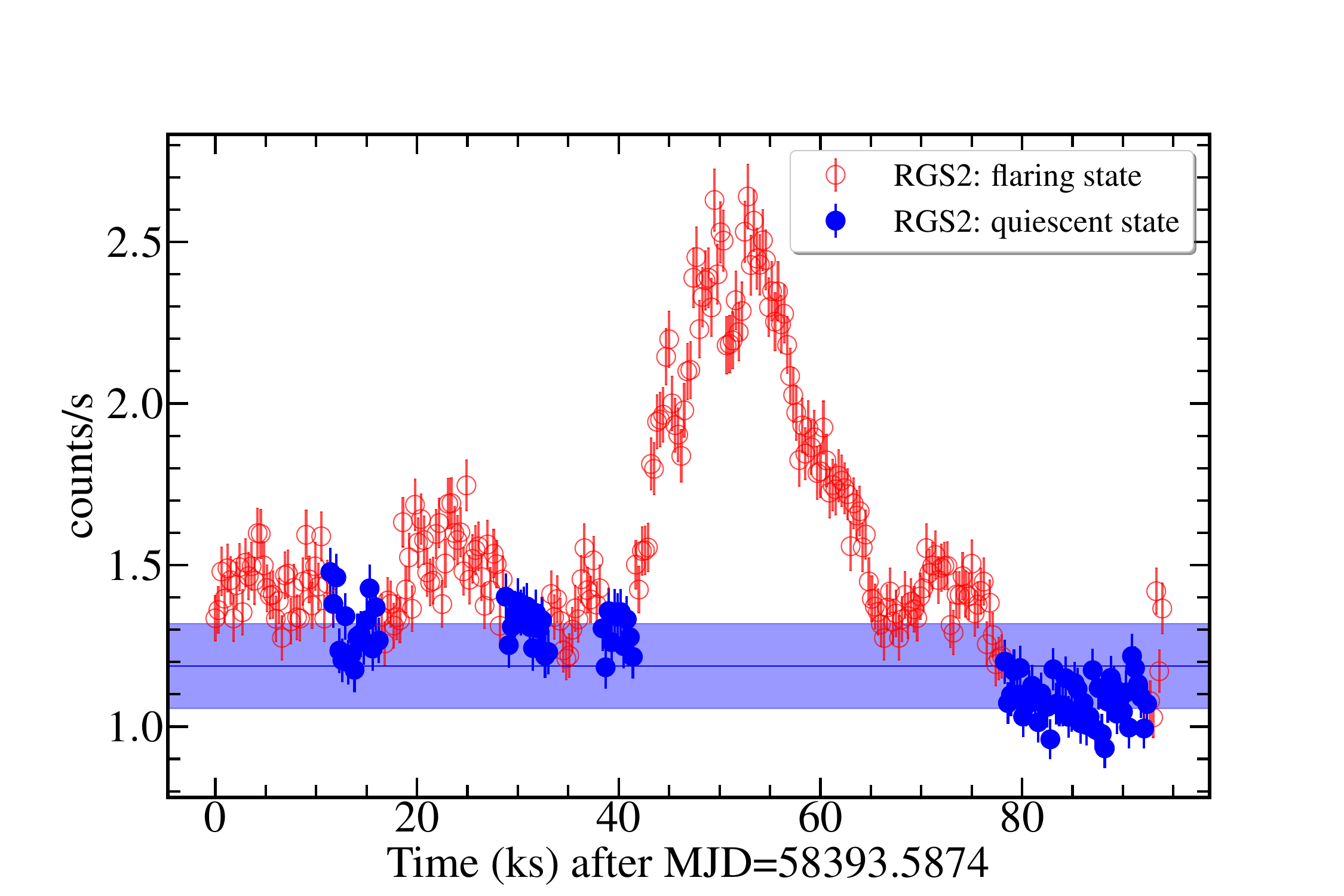}}
\subfigure[Obs ID: 0810850501 (2019-09-30)]{\includegraphics[scale=0.15]{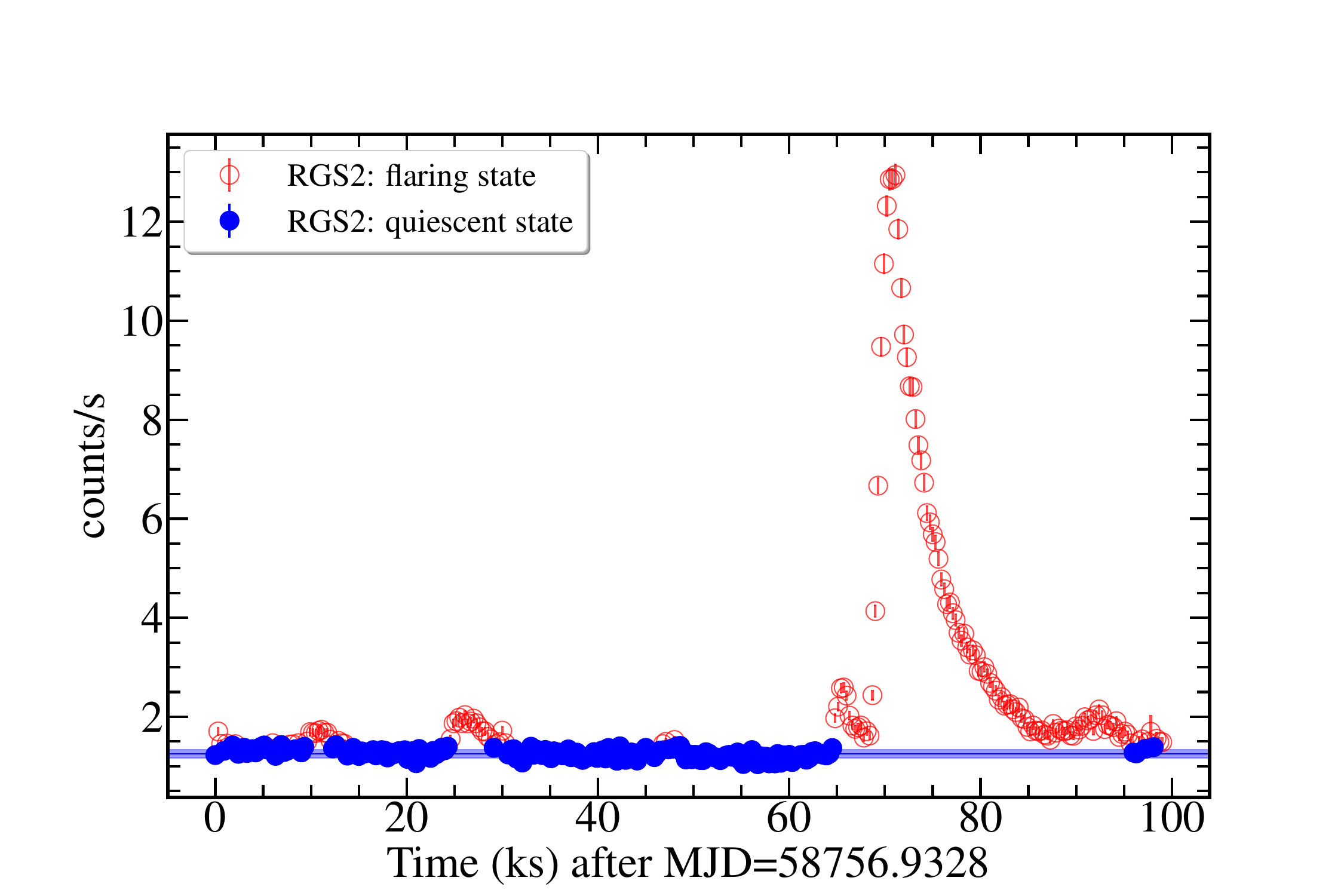}}
\subfigure[Obs ID: 0810850601 (2020-09-29)]{\includegraphics[scale=0.15]{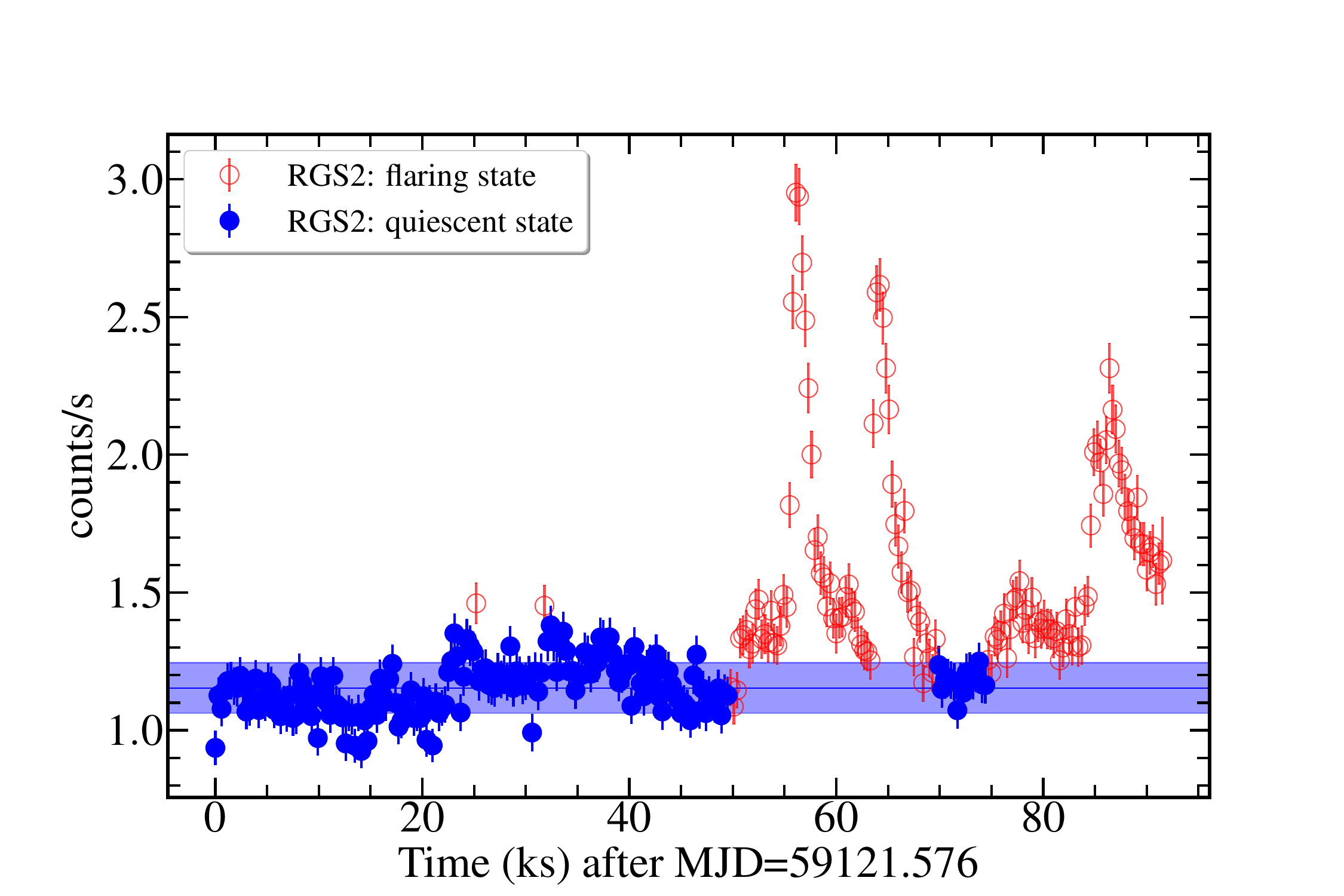}}
\subfigure[Obs ID: 0810850701 (2021-12-04)]{\includegraphics[scale=0.15]{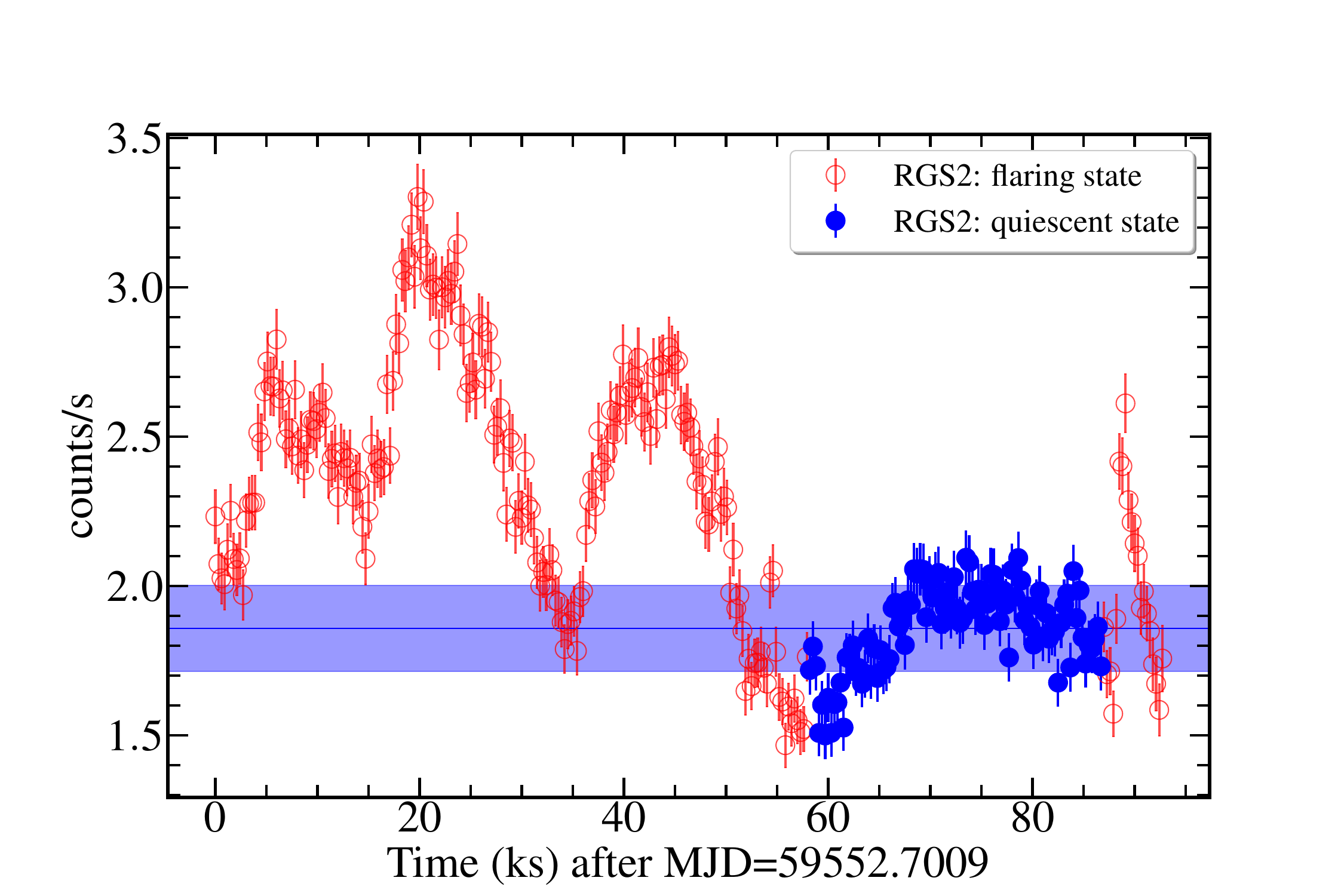}}
\subfigure[Obs ID: 0810850901 (2022-10-19)]{\includegraphics[scale=0.15]{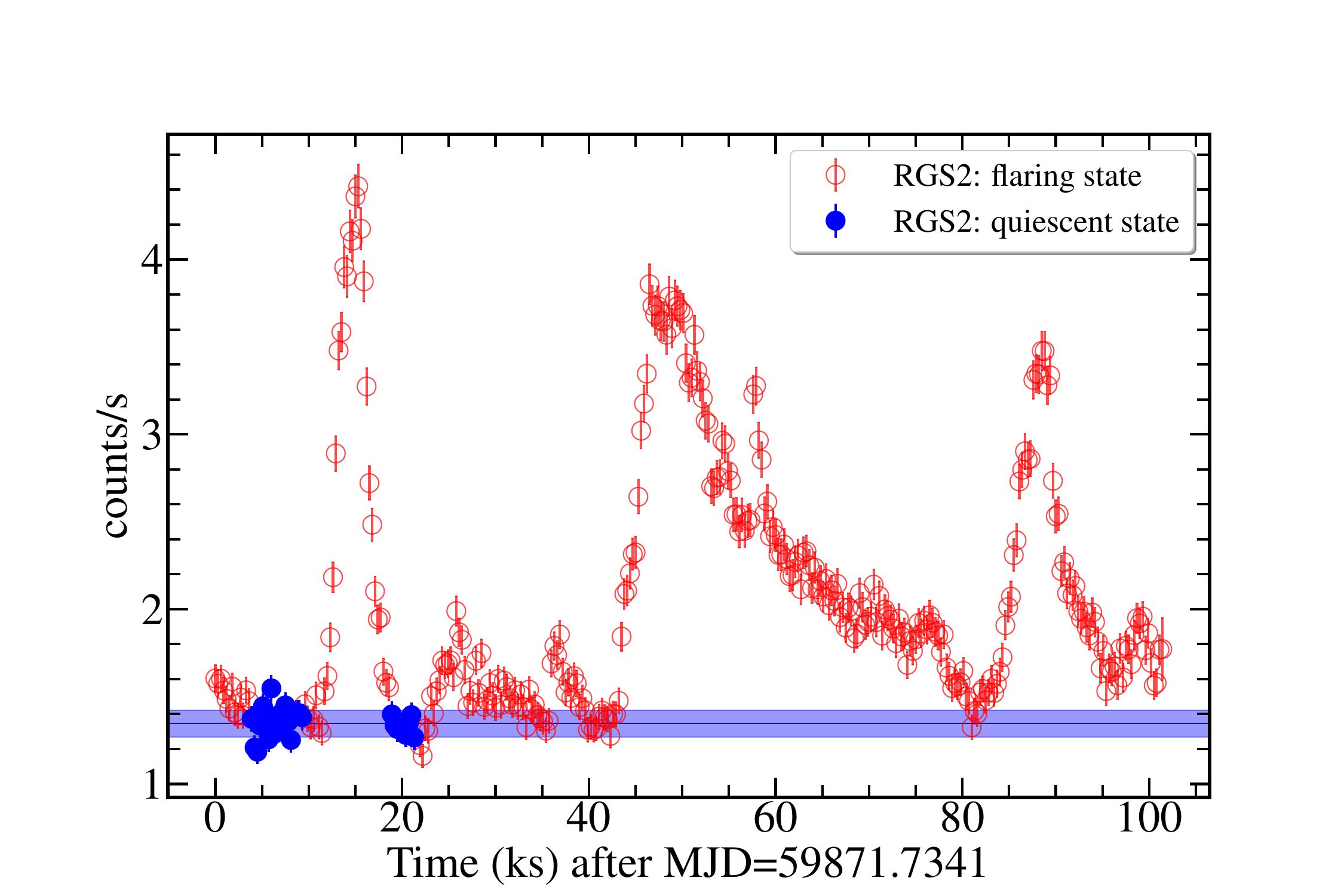}}
\subfigure[EXOSAT (1984-11-12)]{\includegraphics[scale=0.15]{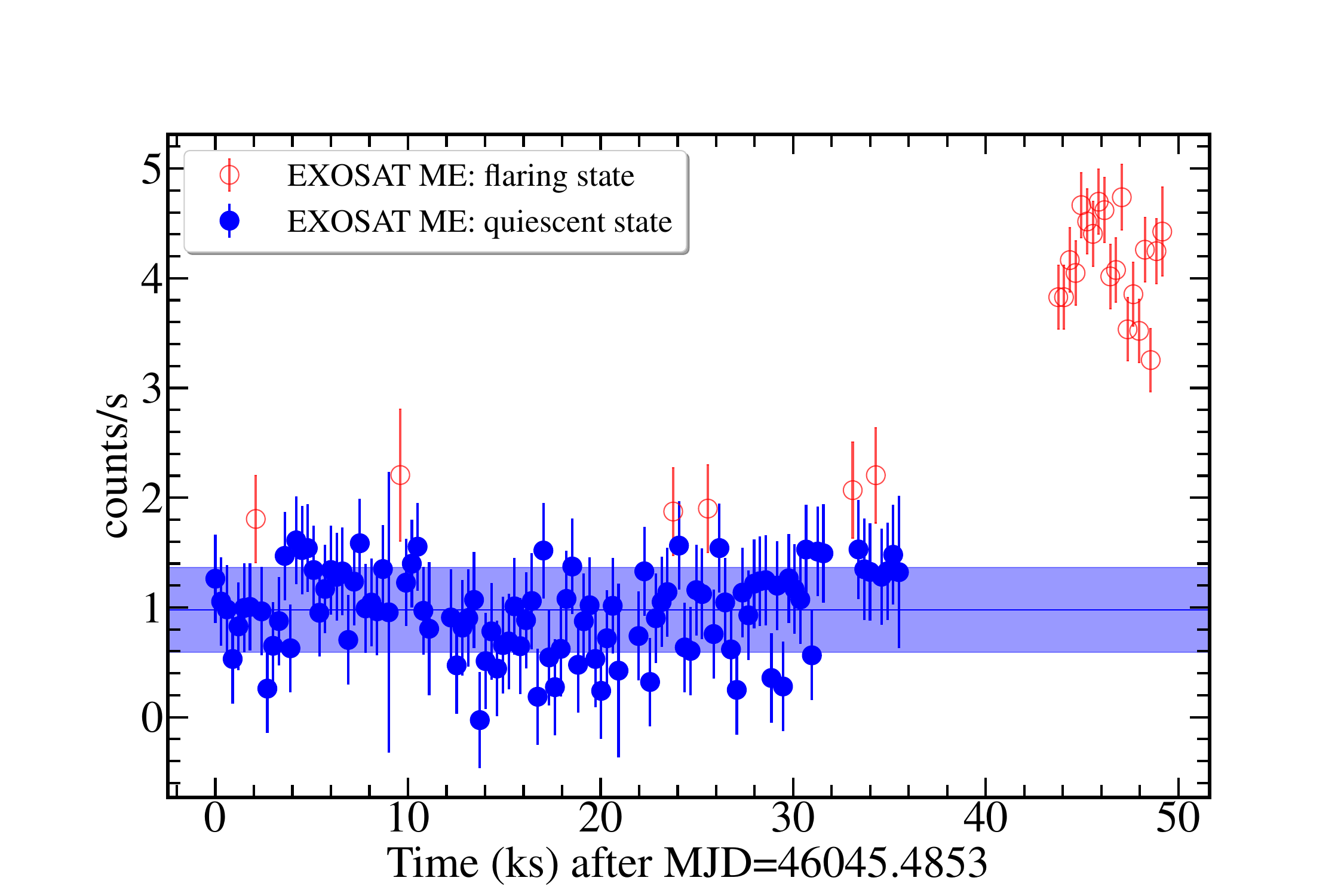}}
\subfigure[EXOSAT (1986-01-21)]{\includegraphics[scale=0.15]{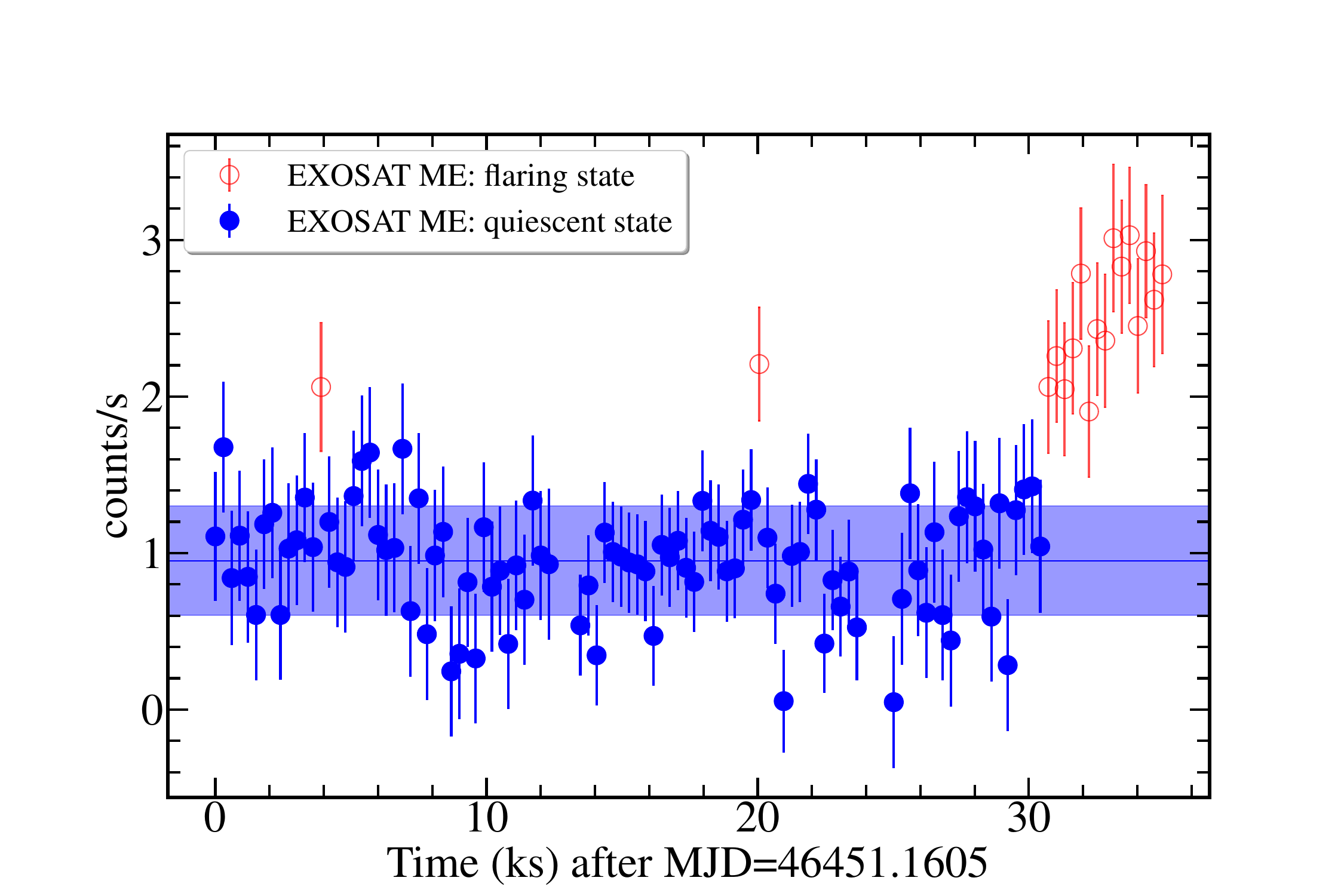}}

\caption{continued}
    \label{fig:abdor_flare_qui3}
\end{figure*}

\clearpage
\newpage
\section{Long-term quiescent X-ray data}
\renewcommand\thetable{\thesection.\arabic{table}}    
\begin{table}[b]
    \centering
    \caption{Yearly averaged X-ray data of AB Dor as observed from various X-ray missions from 1979 to 2022. Here, $\sigma_{CR}$ and $SE_{CR}$ correspond to standard deviation and standard error in count rate, whereas the $\sigma_L$ and $SE_L$ are standard deviation and standard error in Luminosity, respectively. }  \label{tab:abdorlongterm}
    \begin{tabular}{ccccccc}
    \hline
        Telescope & Date & MJD      & Count rate ($CR$) & $\sigma_{CR}/SE_{CR}$ & Luminosity ($L$) & $\sigma_L/SE_L$  \\
                  & YYYY-MM-DD     &          & count/s    &  counts/s   & ($\times$10$^{30}$ erg/s) & ($\times$10$^{30}$ erg/s) \\
        \hline
       Einstein$^1$   &1979-06-01 & 44025.00 & 2.17 & 0.18 / 0.18 & 1.26 & 0.01 / 0.01\\
         EXOSat$^2$   &1984-12-11 & 46045.74 & 0.98 & 0.39 / 0.03 & 2.11 & 0.83 / 0.07\\
                      &1986-01-21 & 46451.36 & 0.95 & 0.35 / 0.03 & 2.06 & 0.76 / 0.07\\
         Ginga$^3$    &1990-10-01 & 48165.00 & 2.02 & 0.57 / 0.18 & 1.37 & 0.38 / 0.12\\
         ROSAT$^4$    &1990-07-02 & 48074.4  & 1.62 & 0.08 / 0.08 & 1.18 & 0.06 / 0.06\\
                      &1991-10-26 & 48555.47 & 1.45 & 0.03 / 0.03 & 1.05 & 0.02 / 0.02\\
                      &1992-03-14 & 48695.88 & 1.72 & 0.02 / 0.02 & 1.26 & 0.01 / 0.01\\
                      &1994-07-06 & 49539.53 & 1.97 & 0.01 / 0.01 & 1.43 & 0.01 / 0.01\\
                      &1995-07-07 & 49905.25 & 1.84 & 0.01 / 0.01 & 1.34 & 0.01 / 0.01\\
                      &1996-07-28 & 50292.34 & 1.85 & 0.01 / 0.01 & 1.35 & 0.01 / 0.01\\
                      &1997-07-12 & 50641.49 & 2.11 & 0.01 / 0.01 & 1.54 & 0.01 / 0.01\\
                      &1998-03-02 & 50874.26 & 1.82 & 0.02 / 0.02 & 1.33 & 0.02 / 0.02\\
         BeppoSAX$^5$ &1997-11-10 & 50762.23 & 0.24 & 0.03 / 0.01 & 2.38 & 0.32 / 0.09\\
                      &1997-11-29 & 50781.92 & 0.28 & 0.05 / 0.01 & 2.80 & 0.49 / 0.1\\
                      &1999-12-09 & 51521.83 & 0.14 & 0.02 / 0.0  & 1.40 & 0.17 / 0.05\\
                      &2000-06-04 & 51699.74 & 0.13 & 0.02 / 0.0  & 1.30 & 0.23 / 0.04\\
        Suzaku$^6$    &2006-12-15 & 54084.50 &  --  & --   / --   & 1.48 & 0.02 / 0.02\\
        XMM-Newton$^7$&2000-08-19 & 51775.82 & 1.43 & 0.05 / 0.003 & 1.39 & 0.05 / 0.003\\
                      &2001-05-30 & 52059.21 & 1.61 & 0.07 / 0.006 & 1.56 & 0.07 / 0.006\\
                      &2002-09-24 & 52541.19 & 1.78 & 0.05 / 0.005 & 1.73 & 0.05 / 0.005\\
                      &2003-07-05 & 52825.46 & 1.55 & 0.04 / 0.004 & 1.5 & 0.04 / 0.004\\
                      &2004-11-28 & 53337.09 & 1.32 & 0.09 / 0.007 & 1.28 & 0.08 / 0.007\\
                      &2005-07-18 & 53569.47 & 1.42 & 0.04 / 0.006 & 1.38 & 0.04 / 0.006\\
                      &2007-07-19 & 54300.44 & 1.61 & 0.12 / 0.009 & 1.56 & 0.12 / 0.009\\
                      &2008-01-04 & 54469.09 & 1.46 & 0.17 / 0.014 & 1.42 & 0.17 / 0.013\\
                      &2009-06-16 & 54998.66 & 1.22 & 0.08 / 0.006 & 1.18 & 0.08 / 0.006\\
                      &2010-01-11 & 55207.87 & 1.19 & 0.09 / 0.007 & 1.15 & 0.09 / 0.007\\
                      &2011-01-03 & 55564.0  & 1.37 & 0.12 / 0.009 & 1.34 & 0.12 / 0.008\\
                      &2012-01-01 & 55927.01 & 1.37 & 0.08 / 0.006 & 1.34 & 0.08 / 0.006\\
                      &2016-10-07 & 57668.63 & 1.32 & 0.14 / 0.008 & 1.28 & 0.13 / 0.007\\
                      &2017-10-11 & 58037.25 & 1.48 & 0.12 / 0.007 & 1.43 & 0.12 / 0.007\\
                      &2018-10-03 & 58394.13 & 1.19 & 0.13 / 0.007 & 1.15 & 0.13 / 0.007\\
                      &2019-10-01 & 58757.51 & 1.25 & 0.09 / 0.005 & 1.22 & 0.08 / 0.005\\
                      &2020-09-30 & 59122.11 & 1.15 & 0.09 / 0.005 & 1.12 & 0.09 / 0.005\\
                      &2021-12-05 & 59553.24 & 1.86 & 0.14 / 0.008 & 1.81 & 0.14 / 0.008\\
                      &2022-10-20 & 59872.32 & 1.35 & 0.08 / 0.004 & 1.31 & 0.07 / 0.004\\

      AstroSat  &$^8$2016-01-23 & 57410.85 & -- & --/-- &  1.37 & 0.1/0.1\\ 
                &$^9$2018-02-26 & 58165.37 & 0.93 & 0.32 / 0.003 & 1.24 & 0.07 / 0.07\\
\hline

    \end{tabular}
    
    ~~\\
            $^1$ \cite{1992ApJS...80..257E}, $^2$ This paper, $^3$\cite{1993A&A...278..467V}, $^4$ ROSAT HRI Catalogue, $^5$ \cite{2002franciosini}, $^6$ \cite{2014PASA...31...21S} ,$^7$ This paper, $^8$ \cite{2023JApA...44...90S}, $^9$ This paper.

\end{table}

\end{document}